\documentclass[twocolumn,prd,nofootinbib,superscriptaddress]{revtex4}
\usepackage{verbatim}
\usepackage{graphicx}
\usepackage{dcolumn}
\usepackage{bm}
\usepackage{color}
\usepackage{url}
\usepackage{amsmath}
\usepackage{adjustbox}
\usepackage{float}
\usepackage{times}
\newcommand\optional[1]{}

\newcommand\qmstateproduct[2]{\left\langle#1|#2\right\rangle}

\newcommand\lnL{ \ln {\cal L}}
\newcommand\lnLmarg{ \ln{\cal L}_{\rm marg}}
\newcommand\unit[1]{{\rm #1}}

\usepackage{color}
\definecolor{amber}{rgb}{1.0, 0.75, 0.0}
\definecolor{orange}{rgb}{1.0, 0.5, 0.0}
\definecolor{amaranth}{rgb}{0.9, 0.17, 0.31}

\newcommand{\Richard}[1]{ {\color{blue}{#1}}}
\graphicspath{{./figures/}}

\def\ltsima{$\; \buildrel < \over \sim \;$}
\def\simlt{\lower.5ex\hbox{\ltsima}}
\def\gtsima{$\; \buildrel > \over \sim \;$}
\def\simgt{\lower.5ex\hbox{\gtsima}}

\begin{document}

\title{A Parameter Estimation Method that Directly Compares Gravitational Wave Observations to Numerical Relativity}
\author{J. Lange}
\affiliation{Center for Computational Relativity and Gravitation, Rochester Institute of Technology, 85 Lomb Memorial Drive, Rochester, NY 14623, USA}
\author{R. O'Shaughnessy}
\affiliation{Center for Computational Relativity and Gravitation, Rochester Institute of Technology, 85 Lomb Memorial Drive, Rochester, NY 14623, USA}
\author{M. Boyle}
\affiliation{Center for Astrophysics and Planetary Science, Cornell University, Ithaca, New York 14853, USA}
\author{J. Calder\'{o}n Bustillo}
\affiliation{Center for Relativistic Astrophysics and School of Physics, Georgia Institute of Technology, Atlanta, GA 30332, USA}
\author{M. Campanelli}
\affiliation{Center for Computational Relativity and Gravitation, Rochester Institute of Technology, 85 Lomb Memorial Drive, Rochester, NY 14623, USA}
\author{ T. Chu}
\affiliation{Department of Physics, Princeton University, Jadwin Hall, Princeton, NJ 08544, USA}
\affiliation{Canadian Institute for Theoretical Astrophysics, University of Toronto, Toronto M5S 3H8, Canada}
\author{ J. A. Clark}
\affiliation{Center for Relativistic Astrophysics and School of Physics, Georgia Institute of Technology, Atlanta, GA 30332, USA}
\author{N. Demos}
\affiliation{Gravitational Wave Physics and Astronomy Center, California State University Fullerton, Fullerton, California 92834, USA}
\author{ H. Fong}
\affiliation{Canadian Institute for Theoretical Astrophysics, University of Toronto, Toronto M5S 3H8, Canada}
\affiliation{Department of Physics, University of Toronto, Toronto M5S 3H8, Canada}
\author{ J. Healy}
\affiliation{Center for Computational Relativity and Gravitation, Rochester Institute of Technology, 85 Lomb Memorial Drive, Rochester, NY 14623, USA}
\author{D. A. Hemberger}
\affiliation{Theoretical Astrophysics 350-17, California Institute of Technology, Pasadena, CA 91125, USA}
\author{I. Hinder}
\affiliation{Max Planck Institute for Gravitational Physics (Albert Einstein Institute), Am M\"{u}hlenberg 1, 14476 Potsdam-Golm, Germany}
\author{K. Jani}
\affiliation{Center for Relativistic Astrophysics and School of Physics, Georgia Institute of Technology, Atlanta, GA 30332, USA}
\author{B. Khamesra}
\affiliation{Center for Relativistic Astrophysics and School of Physics, Georgia Institute of Technology, Atlanta, GA 30332, USA}
\author{ L. E. Kidder}
\affiliation{Center for Astrophysics and Planetary Science, Cornell University, Ithaca, New York 14853, USA}
\author{  P. Kumar}
\affiliation{Canadian Institute for Theoretical Astrophysics, University of Toronto, Toronto M5S 3H8, Canada}
\author{ P. Laguna}
\affiliation{Center for Relativistic Astrophysics and School of Physics, Georgia Institute of Technology, Atlanta, GA 30332, USA}
\author{ C. O. Lousto}
\affiliation{Center for Computational Relativity and Gravitation, Rochester Institute of Technology, 85 Lomb Memorial Drive, Rochester, NY 14623, USA}
\author{ G. Lovelace}
\affiliation{Gravitational Wave Physics and Astronomy Center, California State University Fullerton, Fullerton, California 92834, USA}
\author{ S. Ossokine}
\affiliation{Max Planck Institute for Gravitational Physics (Albert Einstein Institute), Am M\"{u}hlenberg 1, 14476 Potsdam-Golm, Germany}
\author{ H. Pfeiffer}
\affiliation{Canadian Institute for Theoretical Astrophysics, University of Toronto, Toronto M5S 3H8, Canada}
\affiliation{Max Planck Institute for Gravitational Physics (Albert Einstein Institute), Am M\"{u}hlenberg 1, 14476 Potsdam-Golm, Germany}
\affiliation{Canadian Institute for Advanced Research, 180 Dundas St. West, Toronto, ON M5G 1Z8, Canada}
\author{ M. A. Scheel}
\affiliation{Theoretical Astrophysics 350-17, California Institute of Technology, Pasadena, CA 91125, USA}
\author{ D. M. Shoemaker}
\affiliation{Center for Relativistic Astrophysics and School of Physics, Georgia Institute of Technology, Atlanta, GA 30332, USA}
\author{ B. Szilagyi}
\affiliation{Theoretical Astrophysics 350-17, California Institute of Technology, Pasadena, CA 91125, USA}
\affiliation{Caltech JPL, Pasadena, California 91109, USA}
\author{ S. Teukolsky}
\affiliation{Center for Astrophysics and Planetary Science, Cornell University, Ithaca, New York 14853, USA}
\affiliation{Theoretical Astrophysics 350-17, California Institute of Technology, Pasadena, CA 91125, USA}
\author{ Y. Zlochower}
\affiliation{Center for Computational Relativity and Gravitation, Rochester Institute of Technology, 85 Lomb Memorial Drive, Rochester, NY 14623, USA}

\begin{abstract}

We present and assess a  Bayesian method to interpret gravitational wave signals from binary black holes.  Our method directly
compares  gravitational wave data to numerical relativity simulations.  This procedure bypasses  approximations used in semi-analytical models for compact binary coalescence.  In this work,  we use only the full posterior parameter distribution for generic nonprecessing binaries, drawing inferences away from the set of NR simulations used, via interpolation of a single scalar quantity (the marginalized log-likelihood,  $\lnL$) evaluated by comparing data to nonprecessing binary black hole simulations.   We also compare the data to generic simulations, and discuss the effectiveness of this procedure for generic sources. 
We specifically assess the impact of higher order modes, repeating our interpretation with both $l\le2$ as well as $l\le 3$ harmonic modes.  Using the $l\le 3$ higher modes, we gain more information from the signal and can better constrain the parameters of the gravitational wave signal.  We assess and quantify several sources of systematic error that our procedure could introduce, including simulation resolution and duration; most are negligible.  We show through examples that our method can recover the parameters for equal mass, zero spin; GW150914-like; and unequal mass, precessing spin sources.  Our study of this new parameter estimation method demonstrates we can quantify and understand the systematic and statistical error.  This method allows us to use higher order modes from numerical relativity simulations to better constrain the black hole binary parameters. 
\end{abstract}
\maketitle

\begin{widetext}

\tableofcontents
\end{widetext}

\section{Introduction}
On September 14, 2015 gravitational waves (GW) were detected for the first time at the Laser Interferometer
Gravitational Wave Observatory (LIGO) in both Hanford, Washington and Livingston, Louisiana \cite{DiscoveryPaper}.  The LIGO
Scientific Collaboration and Virgo Collaboration (LVC) concluded that the source of the GW signal was a binary black hole (BBH) system with masses
$m_1=26.2^{+5.2}_{-3.8}$ and $m_2=29.1^{3.7}_{-4.4}$ that merged into a more massive black hole (BH) with mass
$m_f=62.3^{+3.7}_{-3.1}$ \cite{O1Paper}.  These parameters were estimated by comparing the signal to state-of-the-art
semi-analytic models \cite{2014PhRvD..89f1502T, 2014CQGra..31s5010P, 2014PhRvL.113o1101H}.   
However, in this mass regime, LIGO is  sensitive to the last few cycles of coalescence, characterized by a strongly nonlinear phase not
comprehensively modeled by analytic inspiral or ringdown models.  In \cite{NRPaper}, the LVC reanalyzed GW150914 with an
alternative method that compares the data directly to numerical relativity (NR), which include aspects of the gravitational radiation omitted by the aforementioned models.  This
additional information led to a shift in some inferred parameters  (e.g., the mass ratio) of the coalescing
binary.

In this work, we  assess the reliability and utility of this novel parameter estimation method in greater detail.  For
clarity and relevance, we apply this method to synthetic data derived from black hole binaries qualitatively similar to GW150914.  
Previous work \cite{NRPaper} demonstrated by example that this method could access information about
GW sources using higher order modes that was not presently accessible by other means.  In this work, we demonstrate the
utility  of this method with a larger set of examples, showing we recover (known) parameters of a synthetic source
more reliably when higher order modes are included. 
 More critically,  we present  a detailed
study of the systematic and statistical parameter estimation errors of this method.  This analysis demonstrates that
these sources of error are under control allowing us to identify source parameters and conduct detailed investigations
into subtle systematic issues, such as  the impact of higher order modes on parameter estimation.  
For simplicity and to best leverage the most exhaustively explored region of binary parameters, our analysis emphasizes
 simulations of nonprecessing black hole binaries as in \cite{NRPaper}, particularly simulations with mass ratios and
 spins that are highly consistent with GW150914.

The paper is outlined as follows.  Section \ref{sec:method} lists the simulations used in the study (both for our template bank and synthetic sources), describes our
method of choice with regards to waveform extraction, and briefly describes the method (see Section III in
\cite{NRPaper}).  Section \ref{sec:diagnostics} describes the diagnostics used in our assessment of the systematics,
illustrating each with concrete examples.  Section \ref{sec:valid} describes several sources of error and their relative impact on our results.
Section \ref{sec:aligned}
presents 3 end-to-end runs, $q=1$ zero spin; $q=1.22$ anti-aligned (GW150914-like); and $q=1.23$ short precessing,
including both $l\le2$ and $l\le 3$ (for the GW150914-like) results.  Section \ref{sec:conclusion} summarizes our findings. Appendix
\ref{sec:appendixA} includes more end-to-end studies that use intrinsically different sources to explore more of the
parameter space using our method.   
For context,  the same method used to analyze GW150914 has also been applied to synthetic data using numerical
relativity simulations \cite{LIGO-Puerrer-NR-LI-Systematics}.   

\section{Methods and inputs}
\label{sec:method}
\subsection{Numerical relativity simulations}
\label{subsec:NR:Sims}
\begin{figure}
\includegraphics[width=\columnwidth]{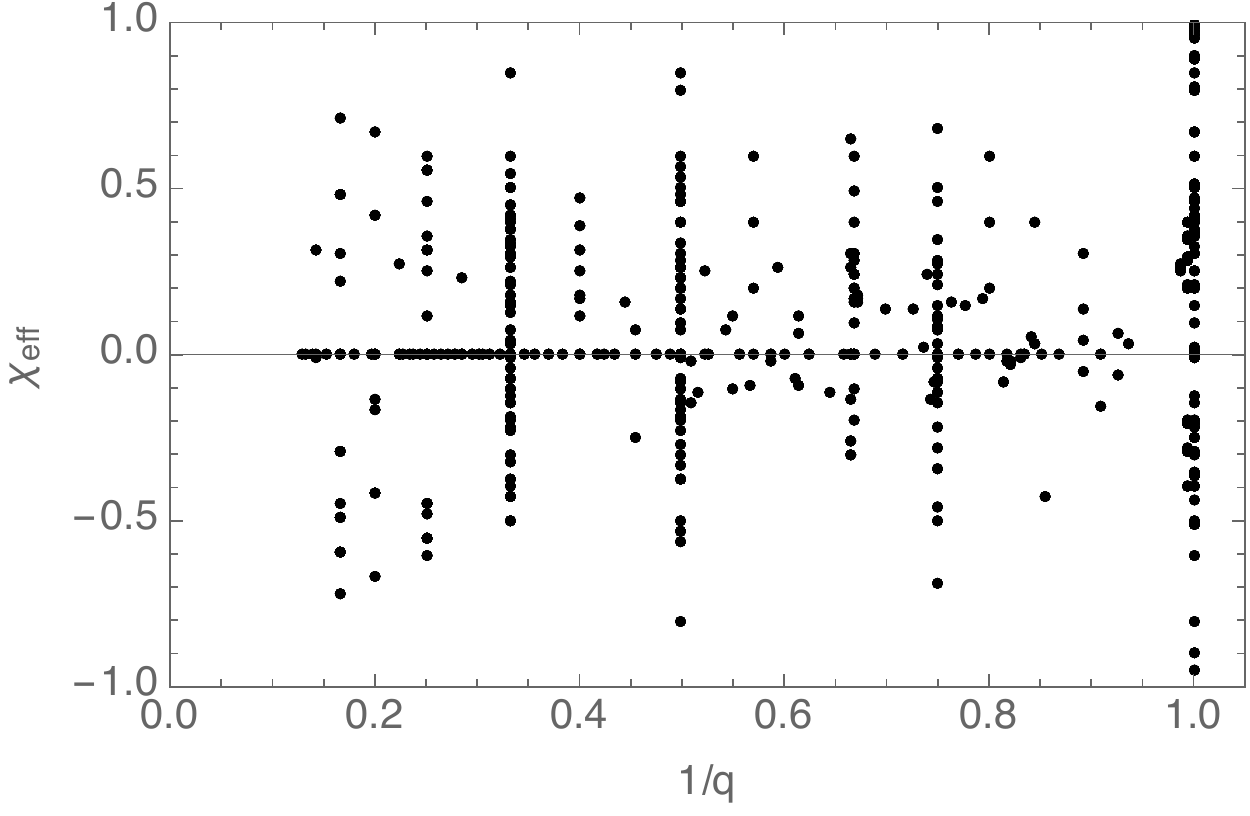}
\caption{\textbf{NR template bank}: An illustration of all the simulations used in this study in the 2D space of $1/q$ and $\chi_{\rm
    eff}$ [Eq. (\ref{eq:chieff})]. Combined with our interpolation methods, the wide range of mass ratios and spins
  represented in this illustration allow us to reproduce binary parameters for much of the parameter space. }
\label{fig:simulations}
\end{figure}

A numerical relativity (NR) simulation of a coalescing compact binary can be completely characterized by its intrinsic
parameters, namely its individual masses and spins.  We parameterize the binary using the mass ratio $q=m_{1}/m_{2}$
with the convention $q\ge 1$ ($m_{1}\ge m_{2}$) and the dimensionless spin parameters
\begin{equation}
\label{eq:chii}
\bm{\chi_{i}}=\bm{S_{i}}/m_{i}^{2}. 
\end{equation}
where $i=1,2$ indexes the component black holes in the binary. 
With regard to spin, we define another dimensionless parameter that is a combination of the spins \cite{2001PhRvD..64l4013D,2008PhRvD..78d4021R,2011PhRvL.106x1101A}:
\begin{equation}
\label{eq:chieff}
\chi_{\rm eff}=(\bm{S_{1}}/m_{1}+\bm{S_{2}}/m_{2})\cdot\hat{L}/M.
\end{equation}
Figure \ref{fig:simulations} illustrates our NR template bank, with each simulation represented as a point in the $\chi_{\rm
  eff},q$ plane.  Finally we quantify the duration of each simulation signal by a dimensionless parameter $M\omega_0$, corresponding to the dimensionless starting binary frequency measured at infinity. 

For a given simulation, the GW strain $h(t,r,\hat{n})$ can be characterized by a spin-weighted
spherical harmonic decomposition at large enough distance:
$h(t,r,\hat{n})=\Sigma_{l\ge2}\Sigma_{m=-l}^{l}h_{lm}(t,r)_{-2}Y_{lm}(\hat{n})$.   In this expression, $\hat{n}$ is
characterized by polar angles $\iota,-\phi_{\rm ref}$; see \cite{gwastro-PE-AlternativeArchitectures}.
For the majority of sources,  the ($2,\pm 2$) mode dominates the
summation and can adequately characterize the observationally-accessible radiation in any direction to a relative good
approximation; however, other higher modes can often contribute in a significant way to the overall signal
\cite{2008CQGra..25k4047S}.  
More exotic sources  (i.e. high mass ratio and/or precessing, high spins) have significant power in higher modes
\cite{2007PhRvD..76f4034B,2008PhRvD..77d4031S,2010PhRvD..82j4006O,1995PhRvD..52..821K,2014LRR....17....2B}.
\begin{table*}
\centering
\begin{adjustbox}{max width=\textwidth}
\begin{tabular}{l|cccccccccc|c}
Group & Param & $M/M_{\odot}$ & q & $s_{1,x}$ & $s_{1,y}$ & $s_{1,z}$ & $s_{2,x}$ & $s_{2,y}$ & $s_{2,z}$ & $\imath$ & Label\\ \hline
Sequence-RIT-Generic & \verb|D12.25_q0.82_a-0.44_0.33_n120|& 70 & 1.22 & - & - & 0.330 & - & - & -0.440 & $0,\frac{\pi}{6},\frac{\pi}{4},\frac{\pi}{3},\frac{\pi}{2},\frac{3\pi}{4}$ & RIT-1a\\
Sequence-RIT-Generic & \verb|D12.25_q0.82_a-0.44_0.33_n110|& 70 & 1.22 & - & - & 0.330 & - & - & -0.440 & $\frac{\pi}{4}$ & RIT-1b\\
Sequence-RIT-Generic & \verb|D12.25_q0.82_a-0.44_0.33_n100|& 70 & 1.22 & - & - & 0.330 & - & - & -0.440 & $\frac{\pi}{4}$ & RIT-1c\\
Sequence-RIT-Generic & \verb|DD_D10.99_q2.00_a-0.8_n100|& 70 & 2.0 & - & - & -0.801 & - & - & -0.801 & $\frac{\pi}{4}$ & RIT-2\\
Sequence-RIT-Generic & \verb|U0_D9.53_q1.00_a0.0_n100| & 70 & 1.0 & - & - & - & - & - & - & $0,\frac{\pi}{6},\frac{\pi}{4},\frac{\pi}{3},\frac{\pi}{2},\frac{3\pi}{4}$ & RIT-3\\
Sequence-RIT-Generic & \verb|D21.5_q1_a0.2_0.8_th104.4775_n100| & 70 & 1.0 & - & - & 0.200 & 0.775 & 0 & -0.200 & $0,\frac{\pi}{6},\frac{\pi}{4},\frac{\pi}{3},\frac{\pi}{2},\frac{3\pi}{4}$ & RIT-4\\
Sequence-RIT-Generic & \verb|D11_q0.50_a0.0_0.0_n100| & 70 & 2.0 & - & - & - & - & - & - & $\frac{\pi}{4}$ & RIT-5\\
Sequence-SXS-All & \verb|1| & 70 & 1.0 & - & - & - & - & - & - & $0,\frac{\pi}{6},\frac{\pi}{4},\frac{\pi}{3},\frac{\pi}{2},\frac{3\pi}{4}$ & SXS-1\\
Sequence-SXS-All & \verb|Ossokine_0233| & 70 & 1.23 & - & - & 0.320 & - & - & -0.580 & $0,\frac{\pi}{6},\frac{\pi}{4},\frac{\pi}{3},\frac{\pi}{2},\frac{3\pi}{4}$ & SXS-0233\\
Sequence-SXS-All & \verb|Ossokine_0234v2| & 70 & 1.23 & 0.0943 & 0.0564 & 0.322 & 0.266 & 0.213 & -0.576 & $0,\frac{\pi}{6},\frac{\pi}{4},\frac{\pi}{3},\frac{\pi}{2},\frac{3\pi}{4}$ & SXS-0234v2\\
Sequence-SXS-All & \verb|BBH_SKS_d14.4_q1.19_sA_0_0_0.420_sB_0_0_0.380| & 70 & 1.19 & - & - & 0.420 & - & - & 0.380 & $\frac{\pi}{4}$ & SXS-$\chi_{\rm eff}$0.4\\
Sequence-SXS-All & \verb|BBH_SKS_d12.8_q1.31_sA_0_0_0.962_sB_0_0_-0.900| & 70 & 1.31 & - & - & 0.962 & - & - & -0.900 & $\frac{\pi}{4}$ & SXS-high-antispin\\
\end{tabular}
\end{adjustbox}
\caption{\textbf{Synthetic sources}: A list of the synthetic sources used in our mismatch studies and end-to-end
  runs. These are done at different inclinations and with higher order modes.  All synthetic sources are performed using
  the same SNR (20) and the same extrinsic parameters: GPS time $10^9 \unit{s}$; RA=0; DEC=3.1; and line of sight
  relative to the NR simulation characterized by Euler angles $\iota,\phi,\psi$   with $\iota$ provided in the table and
  $\phi=\psi=0$. 
}
\label{tab:simulations}
\end{table*}

\subsection{Simulations used}
In this work, we use a wide parameter-range of NR simulations similar to the set used in \cite{NRPaper}.  We use all of
the 300 public and 13 non-public SXS simulations for a total of 313 \cite{2013PhRvL.111x1104M}.  From the RIT group, we use all 126 public and 281
non-public simulations to bring the total contribution up to 407 \cite{2017arXiv170303423H}.  We  also use  a total of
282 simulations provided from the GT group \cite{2016CQGra..33t4001J}.  Including all the contributions from these three groups, we have a total NR template bank of 1002 simulations.  Figure \ref{fig:simulations} shows all the NR simulations in the 2D parameter space of $\chi_{\rm eff}$, as defined in Eq. (\ref{eq:chieff}), vs $1/q$ i.e. the mass ratio. All these simulations have already been published and were produced by one of three familiar procedures, see Appendix A in \cite{NRPaper} for more details for each particular group.

From these simulations, we selected 12 simulations to focus on as candidate synthetic sources.  Table
\ref{tab:simulations} shows the specific simulations used, specifying the mass ratio ($q>1$), component spins of each
BH, and total mass.  To simplify the process of referring to these heterogeneous simulations, in the last column
  we assign a shorthand label to each one.  These candidates have a variety of mass ratios and spins including zero,
  aligned, and precessing systems from different NR groups.   The first three simulations (RIT-1a,-1b, and -1c) have
   identical initial conditions/parameters, carried out with different simulation numerical resolution.   In many of the validation studies, RIT-1a is used; this is a GW150914-like simulation with comparable masses and anti-aligned spins. We use this simulation for its relative simplicity (higher order modes start to become important at the total mass we'll scale the simulation to, namely $70M_\odot$) and to relate it to our similar work done on the real event GW150914.

In this paper, we present 3 end-to-end studies of our parameter estimation method using data from synthetic sources. We
use: a zero spin q=1.0 NR simulation (SXS-1) to show that the method recovers the parameters for the most basic source,
an aligned spin GW150914-like simulation (SXS-0233) to show that higher order modes and therefore NR is needed to
optimally recover the parameters even with aligned spin cases, and a precessing source (SXS-0234v2) to show our method
arrives at reasonable conclusions for any heavy, comparable-mass  binary system with generic spins.

\begin{widetext}
\subsection{Extracting asymptotic strain from $\psi_4(r,t)$ }
From our large and heterogeneous set of simulations, we need to consistently and reproducibly estimate  $r h_{lm}(t)$.
Many general methods for strain estimation exist; see the review in \cite{2016arXiv160602532B}.  
The method adopted here must be robust, using the minimal subset of all groups' output; function with all simulations, precessing
or not; and rely on only knowledge of asymptotic properties, not (gauge-dependent) information about dynamics.  
For these reasons, we implemented our own strain reconstruction and extrapolation algorithm, which as input requires
only $\psi_{4,lm}(t)$ on some (known) code extraction radius.   
This method combines two standard tools --  perturbative
extrapolation \cite{2015PhRvD..91j4022N} and the fixed-frequency integration method \cite{2011CQGra..28s5015R} -- into a
single step.

Specifically, we extract $r h(t)$ at infinity from $\psi_4(r,t)$ at finite radius using a perturbative extrapolation technique based
on Eq. (29) in \cite{2015PhRvD..91j4022N},
implemented in the fourier domain and using a low-frequency cutoff  \cite{2011CQGra..28s5015R}.   Specifically, if $f_{\rm min}$ is identified as the minimum frequency content for
the mode,  we construct the gravitational wave strain from $\psi_4$ at a single finite radius from
\begin{eqnarray}
{} r \tilde{h}_{lm}(f) = \frac{\tilde{\psi}_{4,lm}}{(i\omega)^2}(1 - 2 M/r) [1 - \frac{(\ell-1)(\ell+2)}{2 r}\frac{1}{i    \omega}
   + \frac{(\ell-1)(\ell+2)(\ell^2+\ell-4)}{8 r^2}\frac{1}{(i \omega)^2}]
\nonumber
 \\
  + \frac{\tilde{\psi}_{4,l+1,m}}{(i\omega)^2}\frac{2i a}{(\ell+1)^2} 
  \sqrt{   \frac{(\ell+3)(\ell-1)(\ell+m+1)(\ell-m+1)}{(2l+1)(2l+3)}}
  [i\omega - \frac{\ell(\ell+3)}{r}]
\nonumber \\
  - \frac{ \tilde{\psi}_{4,l-1,m}}{(i \omega)^2}\frac{2i a}{(\ell)^2} \sqrt{
    \frac{(\ell+1)(\ell-2)(\ell+m)(\ell-m)}{(2l-1)(2l+1)}}[i \omega - \frac{(\ell-2)(\ell+1)}{r}]
\label{eq:PTExtract}
\end{eqnarray}
where the effective frequency is implemented as
\begin{eqnarray}
i\omega = i 2\pi \text{sign}(f) \text{max}(|f|, f_{\rm min})
\end{eqnarray}
and where $a$ is an estimate for the final black hole spin.  
This method nominally introduces an obvious obstacle to practical calculation: the last two terms manifestly require an estimate of
  $a$ and are tied to a frame in which the final black hole spin is aligned with our coordinate axis.
In practice, the two spin-dependent terms are small and can be safely omitted in most practical calculations; moreover,
each group provides a suitable estimate for the final state.  We will
clearly indicate when these terms are incorporated into our analysis in subsequent discussion.

When implementing this procedure numerically, we first clean $\psi_{4,lm}$ using pre-identified simulation-specific
criteria to eliminate junk radiation at early and late times, tapering the start and end of the signal to avoid
introducing discontinuities.  For example, for many simulations and for all modes, any content in $\psi_{4,lm}$ prior to
$t \le r + t_{0}$  was set to zero, for some suitable $t_0$ (fixed for all modes); subsequently, to eliminate the
discontinuity this choice introduces,
each mode was multiplied by a Tukey window chosen to cover 5\% of the remaining waveform duration.
Similarly, all data after a mode-dependent time $t_{e}$ was set to zero, where the time $t_e$ was identified via the
first time (after the time where $|\psi_{4,22}|$ is largest)  where $r |\psi_{4,lm}|$ fell below a fixed, mode-independent
threshold.  To smooth discontinuity, a cosine taper was applied at the end, with duration the larger of either 15 M or 10\% of the
remaining post-coalescence duration, whichever is larger.  

The Fourier transform implementation includes additional interpolation/resampling and padding.  First, particularly to
enable non-uniform time-sampling, each mode is interpolated and resampled to a uniform grid, with spacing set by the
time-sampling rate of the underlying simulation.   In carrying out this resampling, the waveform is padded to cover a
duration $2T+100 M$, where $T$ is the remaining duration of the (2,2) mode after the truncation steps identified above.
To simplify subsequent visual interpretation and investigation, the padding is aligned such that the peak of the (2,2) mode occurs near the center of the
interval ($t=0$).   

Finally, the characteristic frequency $M f_{\rm min, (l,m)}$ is identified from the starting frequency of each  $\psi_{4,lm}$.
In cases where the starting frequency cannot be reliably identified (e.g., due to lack of resolution), the frequency is
estimated from the minimum frequency of the 22 mode as $|m| f_{min, (2,2)}/2$.\footnote{This fallback approximation is not always
  appropriate for strongly precessing systems. However, for strongly precessing systems, the relevant starting frequency
can be easily identified.}  In Section \ref{sub:rextr} we will demonstrate the reliability of this procedure to extract $h(t)$ from $\psi_4$.

\end{widetext}

\subsection{Framework for directly comparing simulations to observations I: Single simulations}
\label{sub:framework}

 In this section, we briefly review the methods introduced in \cite{gwastro-PE-AlternativeArchitectures} and \cite{NRPaper} to infer compact binary parameters from GW data.  All analyses of the data begin with the likelihood of the data given noise, which always has the form (up to normalization)
\begin{equation}
\label{eq:lnL}
\ln {\cal L}(\lambda ;\theta )=-\frac{1}{2}\sum\limits_{k}\langle h_{k}(\lambda ,\theta )-d_{k} |h_{k}(\lambda ,\theta )-d_{k}\rangle _{k}-\langle d_{k}|d_{k}\rangle _{k},
\end{equation}
where $h_{k}$ are the predicted response of the k$^{th}$ detector due to a source with parameters ($\lambda$, $\theta$) and
$d_{k}$ are the detector data in each instrument k; $\lambda$ denotes the combination of redshifted mass $M_{z}$ and the
numerical relativity simulation parameters needed to uniquely specify the binary's dynamics; $\theta$ represents the
seven extrinsic parameters (4 spacetime coordinates for the coalescence event and 3 Euler angles for the binary's
orientation relative to the Earth); and $\langle a|b\rangle_{k}\equiv
\int_{-\infty}^{\infty}2df\tilde{a}(f)^{*}\tilde{b}(f)/S_{h,k}(|f|)$ is an inner product implied by the k$^{th}$ detector's
noise power spectrum $S_{h,k}(f)$. 
In all calculations, we adopt the fiducial O1 noise power spectra associated with data near GW150914 \cite{DiscoveryPaper}.
In practice we adopt a low-frequency cutoff f$_{\rm min}$ so all inner products are modified to
\begin{equation}
\label{eq:overlap}
\langle a|b\rangle_{k}\equiv 2 \int_{|f|>f_{\rm min}}df\frac{\tilde{a}(f)^{*}\tilde{b}(f)}{S_{h,k}(|f|)}.
\end{equation}
The joint posterior probability of $\lambda ,\theta$ follows from Bayes' theorem:
\begin{equation}
p_{\rm post}(\lambda ,\theta)=\frac{ {\cal L}(\lambda ,\theta)p(\theta)p(\lambda)}{\int d\lambda d\theta {\cal L}(\lambda ,\theta)p(\lambda)p(\theta)},
\end{equation}
where $p(\theta)$ and $p(\lambda)$ are priors on the (independent) variables $\theta ,\lambda$. For each $\lambda$, we evaluate the marginalized likelihood
\begin{equation}
 {\cal L}_{\rm marg}\equiv\int  {\cal L}(\lambda ,\theta )p(\theta )d\theta
\end{equation}
via direct Monte Carlo integration, where $p(\theta)$ is uniform in 4-volume and source orientation.  
To evaluate the likelihood in regions of high importance, we use an adaptive Monte Carlo as described in
\cite{gwastro-PE-AlternativeArchitectures}.  We will henceforth refer to the algorithm to ``integrate over extrinsic
parameters'' as ILE.   
The marginalized
likelihood is a way to quantify the similarity of the data and template.  If we integrate out all the parameters except total mass, we get a curve that looks like Figure \ref{fig:ILE-Ex}. Having $\lnL$ in this form is the most useful for our purposes, and plots involving $\lnL$ will be as a function of total mass.

\subsection{Framework for directly comparing simulations to observations II: Multidimensional fits and posterior distribution}
\label{sub:framework2}

The posterior distribution for intrinsic parameters, in terms of the marginalized likelihood and assumed prior $p(\lambda)$ on intrinsic parameters like mass and spin, is
\begin{equation}
\label{eq:post}
p_{\rm post}=\frac{{\cal L}_{marg}(\lambda )p(\lambda)}{\int d\lambda {\cal L}_{\rm marg}(\lambda ) p(\lambda )}.
\end{equation}
As we  demonstrate by concrete examples in this work, using a sufficiently dense grid of intrinsic parameters,
Eq. (\ref{eq:post}) indicates that we can reconstruct the full posterior parameter distribution via interpolation or
other local approximations. The reconstruction only needs to be accurate near the peak. If the marginalized likelihood ${\cal L}_{\rm marg}$ can be
approximated by a d-dimensional Gaussian, with (estimated) maximum value  ${\cal L}_{\rm max}$,  then we anticipate only configurations $\lambda$ with 
\begin{equation}
\label{eq:chi}
\ln {\cal L}_{\rm max}/{\cal L}_{\rm marg}(\lambda)>\chi^{2}_{d,\epsilon}/2
\end{equation}
contribute to the posterior distribution at the 1-$\epsilon$ creditable interval, where $\chi^{2}_{d,\epsilon}$ is the
inverse-$\chi^{2}$ distribution.
[The practical significance of this threshold will be more apparent in Section \ref{sub:ILE}, which implicitly
  illustrates it using one dimension.]  
Since the mass of the system can be trivially rescaled to any value, each NR simulation is represented by particular values for the seven intrinsic parameters (mass ratio and the three components of the spin vectors) and is represented by a one-parameter family of points in the 8-dimensional parameter space of all possible values of $\lambda$.
Given our NR archive, we evaluate the natural log of the marginalized likelihood as a function of the redshifted mass
$\ln {\cal L}_{\rm marg}(M_{z})$. As in \cite{NRPaper}, our first-stage result is this  function,  explored almost
continuously in mass and discretely as our fixed simulations permit.   This information alone is sufficient to estimate
what parameters are consistent with the data: for example, using a cutoff such as  Eq. (\ref{eq:chi}), we identify the
masses that are most consistent for each simulation.   

As demonstrated first in \cite{NRPaper} and explored more systematically here, this likelihood is smooth and broad extending over many
NR simulations' parameters.   As a result, even though our function exploration is a restricted to a discrete grid of NR
simulation values, we can interpolate between simulations to reconstruct the entire likelihood and hence entire
posterior. We can do this because of the simplicity of the signal, which for the most massive binaries involves only a
few cycles.   More broadly, our method works because many NR simulations produce very similar radiation, up to an
overall mass scale; as a result, as has been described previously in other contexts \cite{2014PhRvD..89d2002K},
surprisingly few simulations have been needed to  explore the model space (e.g., for nonprecessing binaries). 

Finally, as we demonstrate repeatedly below by example,  $\lnLmarg$ is often well approximated by a simple
low-order series, typically just a quadratic.   Moreover, for the short GW150914-like signals here, many nonprecessing
simulations fit both observations and even precessing simulations fairly well.  As a result, we employ a quadratic approximation to $\lnLmarg$ near the peak under the restrictive approximation that all angular momenta are parallel using information from only nonprecessing
binaries.   Using this fit, we can estimate $\lnLmarg$ for all masses and aligned spins and therefore estimate
the full posterior distribution.  %
Section IV B in \cite{NRPaper} gives the results of this
method based on the LIGO data containing GW150914.   In this work, we apply this method to a larger set of examples.

\section{Diagnostics}
\label{sec:diagnostics}

Many steps in our procedure to compare NR simulations to GW observations  can introduce systematic error into our
inferred posterior distribution.  Sources of error include  the numerical simulations' resolution; waveform extraction; finite duration; Monte Carlo integration error; the
finite, discrete, and sparsely spaced simulation grid; and our fit to said grid.   In the following sections, we describe tools
to characterize the magnitude and effect of these systematic errors.
First and foremost, we introduce the broadly-used \emph{match}, a   complex-valued inner
product which arises naturally in data analysis and parameter inference applications.   Following many previous studies \cite{2007PhRvD..76j4018C}, we review how systematic error
shows up as a mismatch and parameter bias.  
Second, we describe an analogy to the match which uses our full multimodal infrastructure and is more directly
connected to our final posterior distribution: the marginalized likelihood versus mass $\lnLmarg(M)$, or equivalently
(one-dimensional) posterior distribution implied by assuming the data must be drawn from a specific simulation up to
overall unknown mass and orientation.  
Due to systematic error, the inferred one-dimensional distribution (or match versus mass) may change, both globally and
through any concrete confidence interval (CI) derived from it. 
To appropriately quantify the magnitude of these effects, we introduce two measures to compare similar distributions.
On the one hand, any change in the 90\% CI provides a simple and easily-explained measure of how much an error changes
our conclusions.  
On the one hand, the KL divergence $(D_{\rm KL}$) gives a simple, well-studied, theoretically appropriate, and numerical measure of the
difference between two neighboring distributions.    
In this section we describe these diagnostics and illustrate them using concrete and extreme examples to illustrate how
a significant error propagates into our interpretation. 

\subsection{Inner products between waveforms: the mismatch  }
\label{sub:match}
The match is a well-used and data-analysis-driven tool to compare two candidate GW signals in an
idealized setting.   Unlike most discussions of the match, which derive them from the response of a single idealized
instrument, we
follow  \cite{gwastro-mergers-HeeSuk-FisherMatrixWithAmplitudeCorrections} and work with the response of an idealized
\emph{two}-detector instrument, with both co-located identical interferometers oriented at $45^o$ relative to one another, and the
source located directly overhead this network.\footnote{Equivalently, we work in the limit of many identical detectors,
  such that the network has equal sensitivity to both polarizations for all source propagation directions.}
As is well-known, the match arises naturally in the likelihood of a candidate signal, given known
and noise-free data --
or, in the notation of this work, from  Eq. (\ref{eq:lnL}) restricted to this idealized network, setting $d$ to $h_{0}=h(t,\lambda_{0})$ and $h(\lambda, \theta)=h$:
\begin{equation}
\begin{split}
\label{eq:lnLratio}
\ln {\cal L}&=-\frac{1}{2}\{\langle h-h_{0}|h-h_{0}\rangle-\langle h_{0}|h_{0}\rangle\}\\
                &=-\frac{1}{2}\{\langle h|h\rangle-2 \Re \langle h_{0}|h\rangle\},
\end{split}
\end{equation}
\noindent
where $\Re$ is the real part.  Again $\langle a|b\rangle$ is the complex overlap (inner product) between two waveforms
for a single detector as shown in Eq. (\ref{eq:overlap}); the GW strain $h=h_+-i h_\times$ contains two
polarizations, and is assumed to propagate from directly overhead the network; the likelihood reflects the response
of both detectors' antenna response and noise.
Eq. (\ref{eq:lnLratio}) is slightly different than the the likelihood obtained in Eq. (17) of \cite{gwastro-mergers-HeeSuk-FisherMatrixWithAmplitudeCorrections} by an overall constant. What we use, described in \cite{gwastro-mergers-HeeSuk-CompareToPE-Aligned}, is the likelihood ratio (divided by the likelihood of zero signal). If we add this constant back into the equation, we recover Eq. (17) from \cite{gwastro-mergers-HeeSuk-FisherMatrixWithAmplitudeCorrections}:
\begin{equation}
\label{eq:lnL2}
\ln {\cal L}_{\rm single}=-\frac{1}{2}\{\langle h_{0}|h_{0}\rangle+\langle h|h\rangle -2\Re\langle h_{0}|h\rangle\}.
\end{equation}
This single-detector likelihood depends  on the parameters $\lambda,\theta$ of $h$ and $\lambda_o,\theta_0$ of $h_0$.
For the purposes of our discussion, we will include ``systematic error'' parameters that enhance or change the model
space in $\lambda$ (e.g., changes in simulation resolution).   

The parameters which maximize the likelihood identify the configuration of parameters that make $h$ most similar to
$h_0$.   For a fixed emission direction from the source, three key parameters in $\theta$ dominate how $h$ can be
changed to maximize the likelihood: the event time $t_{\rm event}$; the source luminosity distance $D_L$; and  the polarization
angle $\psi$, characterizing rotations of the source (or detector) about the line of sight connecting the source and
instrument.  In terms of these parameters, 
\begin{eqnarray}
h = e^{-2i\psi} \frac{D_{\rm L,ref}}{D_{\rm L}}h_{\rm ref}(t -t_{\rm event}|\lambda,\theta_{\rm rest})
\end{eqnarray}
where $h_{\rm ref}$ is the value of $h$ at $D_{\rm L}=D_{ \rm L,ref},t_{\rm event}=0$, and $\psi=0$ and $\theta_{\rm rest}$
denotes the four remaining extrinsic parameters besides these three.    
As noted in \cite{gwastro-mergers-HeeSuk-FisherMatrixWithAmplitudeCorrections}, a change of the polarization angle
$\psi$ corresponds to a rotation of the argument of the complex strain function, $h(\psi)=e^{-2i\psi}h(\psi=0)$.
As a result, maximizing the likelihood versus $\psi$ corresponds to choosing a phase angle so $\qmstateproduct{h}{h_0}$
is purely real: 
\begin{equation}
\label{eq:max}
{\rm max}_{\psi}\langle h_{0}|h\rangle=|\langle h_{0}|h\rangle|.
\end{equation}
Similarly  maximizing the likelihood versus distance, the likelihood becomes
\begin{equation}
\max_{\psi,D_L} \ln {\cal L}_{\rm single}=-\rho^{2}(1-P_*).
\end{equation}
where in this expression  $\rho^{2}=\langle h_{0}|h_{0}\rangle=\langle h|h\rangle$ and the function $P$ is 
\begin{equation}
\label{eq:p}
P_*(h_{0},h)\equiv {\rm max}_{\psi}\frac{|\langle h_{0}|h\rangle|}{\sqrt{\langle h_{0}|h_{0}\rangle\langle h|h\rangle}},
\end{equation}
This partially-maximized likelihood depends strongly on the event time.  If we furthermore maximize over event time, we
find the final and important relationships
\begin{align}
\label{eq:final}
\ln {\cal L}_{\rm single,max} &=\max_{\psi,D_L,t_{\rm event}} \ln {\cal L}_{\rm single}=-\rho^{2}(1-P), \\
P(h_{0},h) &\equiv {\rm  max}_{\psi,t_{\rm event}}\frac{|\langle h_{0}|h\rangle|}{\sqrt{\langle h_{0}|h_{0}\rangle\langle h|h\rangle}}.
\end{align}

In the rest of this paper, we will use the mismatch $\mathcal{M}$ between two signals:
\begin{equation}
\mathcal{M}(h_{0},h)=1-P(h_{0},h).
\end{equation}
Because of its form -- an inner product -- the mismatch identifies differences between the two candidate signals;
substituting this expression into the maximized ideal-detector likelihood [Eq.  (\ref{eq:final})] yields:
\begin{equation}
\label{eq:snr-lnL}
\ln {\cal L}_{\rm single,max}=-\rho^{2}\mathcal{M}.
\end{equation}

As the above relationships make apparent, a candidate signal $h$ which has a significant mismatch cannot be scaled to
resemble $h_0$ and therefore must be unlikely.  This relationship has been used to motivate simple criteria to
characterize when two signals $h,h_0$ are indistinguishable (or, conversely, distinguishable); working to order of
magnitude [cf. Eq. (\ref{eq:chi})], two signals are indistinguishable if \cite{1994PhRvD..49.2658C,2008PhRvD..78l4020L,2009JPhCS.189a2024M,2009PhRvD..79l4033R}
\begin{equation}
\label{eq:accuracy}
 \mathcal{M}\le\frac{1}{\rho^{2}}.
\end{equation}
In this work, we apply the match criteria to assess when two simulations of the same or similar parameters (or the same
simulation at a different mass) can be distinguished from a reference configuration. 

As a concrete example, discussed at greater length in Section \ref{sec:sub:ex1}, the top-right panel in Figure \ref{fig:Ex1} shows two plots of mismatch  versus total mass. 
In the black curve,  we calculate the match of two identical waveforms from the RIT-1a simulation: one set at a fixed
total mass $M=70 M_{\odot}$ while the other changes over a given mass range.  At the true total mass, the mismatch
goes to zero.  For comparison, the red curve in that figure shows the mismatch between another simulation $h$ and a fixed
RIT-1a ($h_0$), versus total mass for $h$.    As illustrated in the top-left panel of Figure \ref{fig:Ex1}, the two
simulations are not identical; hence,  the mismatch in the top-right panel  between $h$
and $h_0$ never reaches zero.  Moreover, due to differences in the source $h_o$ and template family $h$, the location of
the minimum mismatch and hence best fit occurs at a different, offset total mass, close to $50 M_\odot$.  

  As the reader will see in subsequent
sections, we can also calculate the mismatch as a function of particular properties of NR simulations to see how much
error is introduced, see Section \ref{sec:valid}.

\subsection{Marginalized likelihood versus mass}
\label{sub:ILE}
\begin{figure*}
\includegraphics[width=\columnwidth]{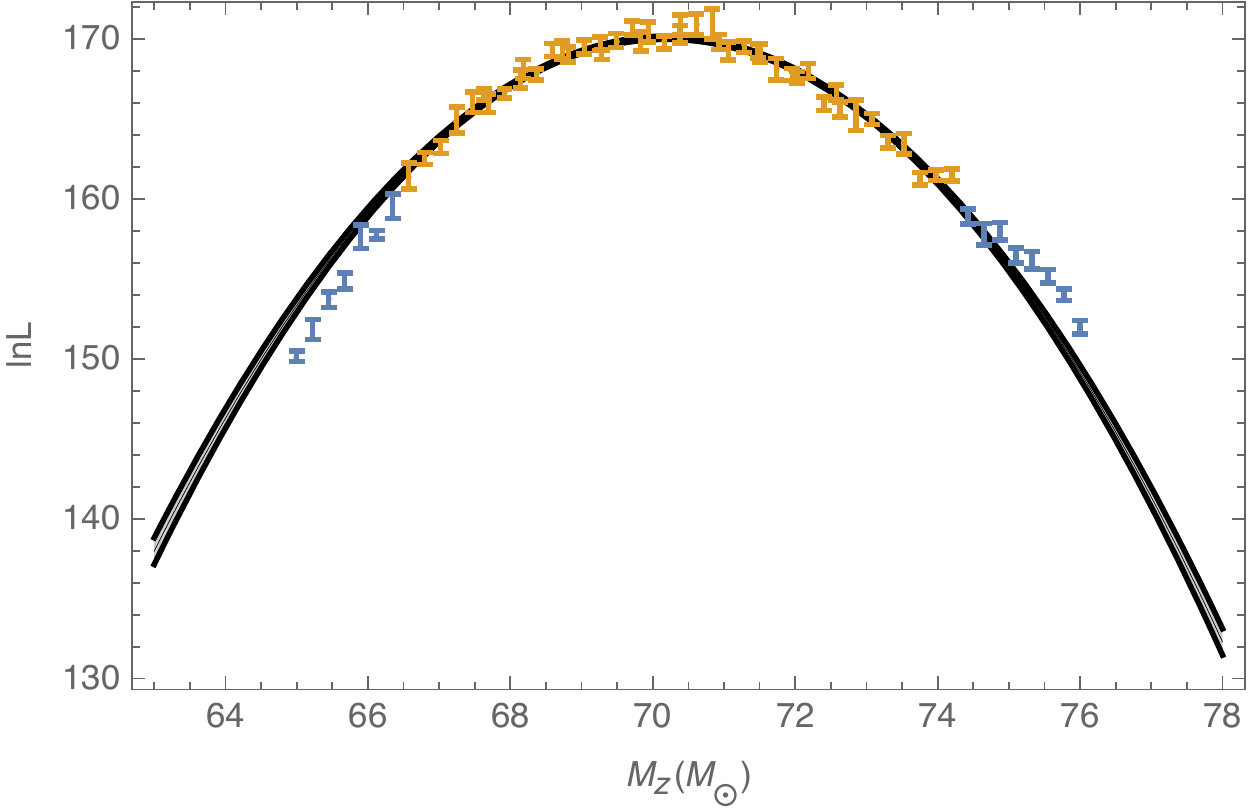}
\includegraphics[width=\columnwidth]{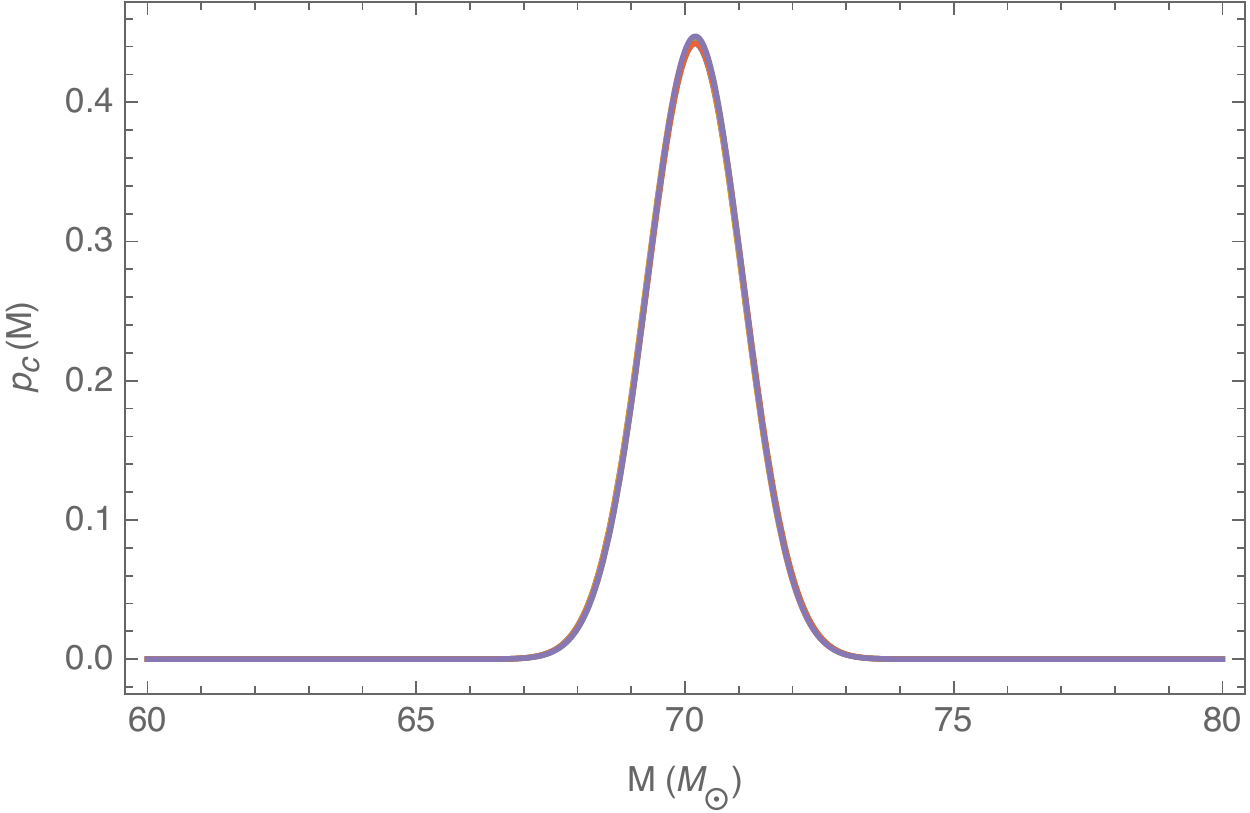}
\caption{\textbf{Example of $\lnLmarg(M)$: comparing a simulation to itself}: \emph{Left panel}: Blue and yellow points
  (with error bars) show results of evaluating  $\lnLmarg(M)$ with RIT-1a as a source compared to itself.  
  The shaded region is derived by fitting a quadratic to these data via least-squares [Eq. (\ref{eq:leastsquares})],
  providing a mean and confidence interval (shown).   The reference
  source has  total mass $M=70 M_{\odot}$ and an
  inclination $\imath=0.785$; all calculations are carried out using  $f_{\rm min}=30 \unit{Hz}$.
This curve will be duplicated as a
  black curve in  the right
panel from Figure \ref{fig:Ex1} and left panel from Figure \ref{fig:Ex2ILE}.   
\emph{Right panel}:  Nominal one-dimensional posterior distributions [Eq. (\ref{eq:1d})] derived from the fit to left.
This figure shows five examples, randomly drawn from the fit coefficient distribution derived by least squares, drawn to
exemplify the propagated systematic uncertainties due to Monte Carlo integration error. 
For studies similar to this one (i.e., high-mass investigations where direct comparison to numerical relativity is most appropriate), this figure suggests
that  Monte Carlo error is much smaller than the posterior width (i.e., has  little relevance given the substantial statistical uncertainty introduced by the limited
number of GW cycles available for comparison from short NR simulations).}
\label{fig:ILE-Ex}
\end{figure*}
Another simple diagnostic is the result $\lnLmarg (M)$ for a single simulation on some reference
data (e.g., the simulation itself, or a signal with comparable physical origin).    
This function enters naturally into our full parameter estimation calculation; therefore, it allows us to test all of the quantities that influence our principal result
directly including NR resolution, extraction radius, etc. as described below.
For simplicity, as computed for the purposes of this
test, this function depends on part (only $l\le2$ modes) of the NR
radiation and the data.    Figure \ref{fig:ILE-Ex} shows a null example run
with RIT-1a, a GW150914-like simulation, as a source compared against itself.  As previous work from both real LIGO and synthetic data has suggested, $\lnL(M)$ can be well-approximated by a locally quadratic fit (see Section \ref{sub:null} for a more in-depth discussion of this example).

\subsection{Probability Density Function/KL Divergence}
\label{sub:PDF}
To quantitatively assess whether two given versions of
$\ln {\cal L}(M)$  are demonstrably different, we employ an observationally-motivated diagnostic to prioritize
agreement in regions with significant posterior support.    Motivated by the applications we perform when comparing results of
this kind, we translate $\ln{\cal L}(M) $ into a probability distribution (i.e., assuming all other parameters are fixed):
\begin{equation}
\label{eq:1d}
p_c(M)  = \frac{1}{\int dM e^{\lnL}}e^{\lnL}.
\end{equation}
In practice, this distribution is always extremely well approximated by a gaussian, so we can further simplify by
characterizing any 1d distribution by its mean $M_*$ and variance $1/\Gamma_{MM} = \sigma_*^2$.  Using this ansatz, we
can therefore define a quantity to assess the difference between any pair of results for $\ln {\cal L}(M)$.  In this
work, we use  the KL divergence between these two approximately-normal distributions:
\begin{align}
\label{eq:dkl}
D_{KL}(p_{*}|p)& =\int dx p(x) \ln p(x)/p_*(x)  \nonumber \\
 &=\ln\frac{\sigma}{\sigma_{*}}-\frac{1}{2}+\frac{(\bar{x}-\bar{x}_{*})^{2}+\sigma_{*}^{2}}{2\sigma^{2}}.
\end{align}
We also will plot the derived PDF
$p_c(M)$ and evaluate the implied 1D 90\% CI derived from it.

The implications of a significant disagreement for this diagnostic -- already illustrated via high mismatch in Figure
\ref{fig:Ex1} -- can be clearly seen in the 1D posterior distributions derived from the fit of $\lnLmarg(M)$ as shown in Figure \ref{fig:Ex1ILE} and Figure \ref{fig:Ex2ILE}.  Loosely following  the work in \cite{2007PhRvD..76j4018C} for estimating parameter errors due to mismatch, we expect the parameter error will be a significant fraction of the statistical error.  Using the notation above and approximating $P \simeq 1 - \frac{1}{2}\bar{\Gamma}_{xx} \delta x^2$ for some nominal perturbed parameter $x$, we estimate the statistical error to be $\sigma_{x,stat} \simeq 1/\rho \sqrt{\Gamma_{xx}}$.  Conversely, balancing mismatch and parameter biases,  similar changes in likelihood occur when
\begin{equation}
\label{eq:changeM}
\delta x \simeq \frac{1}{\bar{\Gamma}_{xx}^{1/2}} {\cal M}^{1/2};
\end{equation}
however, much more detailed calculations is presented in \cite{2007PhRvD..76j4018C}. The above relationship illustrates how a
high mismatch causes a deviation in the $\lnLmarg(M)$ curve as well as its corresponding posterior distribution.  Figure \ref{fig:Ex1ILE} show a comparison between two waveforms from RIT-1a and RIT-2 (red curve). With
significantly different parameters (see Table \ref{tab:simulations}), the mismatch is significantly high. This causes
a radical shift in the $\lnLmarg(M)$ result as well as its corresponding PDF compared to to it's true value.  
This example will be described in greater detail in Section \ref{sec:sub:ex1}.

\subsection{Example 0: Null test/Impact of Monte Carlo Error}
\label{sub:null}
To illustrate the use of these diagnostics, we first apply them to the special case where the data contains the response
due to a known source.  In this case, by construction, the match will be
unity when using the same parameters.  Following a similar procedure  to that we would apply if we didn't know
the source mass, we can also plot the mismatch $\qmstateproduct{h_A(M)}{h_A(M_*)}/||h_A(M)||||h_A(M_*)||$.  Referring to
the notation in Eq. (\ref{eq:p}), we assign the RIT-1a waveform to $h_{0}=h_{RIT-1a}$(source) and again the RIT-1a
waveform to $h=h_{RIT-1a}$ (template).  This plot can be seen in any of the following examples as the black curve
(top-right panels from Figure \ref{fig:Ex1} and Figure \ref{fig:Ex2}).  It has a peak value of unity (not plotted) and
rapidly falls as one moves away from the mass corresponding to the peak match value.  The left panel of Figure
\ref{fig:ILE-Ex} shows the log likelihood $\lnLmarg$ provided by \textit{ILE} as a function of mass.  From here we fit a
local quadratic to the $\lnLmarg$ close to the peak.  Using the fit, we generate five random samples and use them for
subsequent calculations (i.e. 1D distributions).  We derived a 1D distribution using Eq. (\ref{eq:1d}).

First and foremost, these figures illustrate the relationships between the three diagnostics.  As suggested by Eq. (\ref{eq:snr-lnL}), the match and log likelihood $\lnLmarg$ are nearly proportional up to an overall constant.  Second, as required by Eq. (\ref{eq:1d}), the one-dimensional posterior is proportional to ${\cal L}_{\rm marg}$.  This visual illustration corroborates our earlier claim implicit in the left panel of Figure  \ref{fig:ILE-Ex}:  only the part of $\lnLmarg$ within a few of its the peak value contributes in any way to the posterior distribution and to any conclusions drawn from it (e.g., the 90\% CI).

Each evaluation of the Monte Carlo integral has limited accuracy, as indicated in Figure \ref{fig:ILE-Ex}.  By taking
advantage of many evaluations of this integral, we dramatically reduce the overall error in the fit.  
To estimate the impact of this uncertainty, we use standard frequentist polynomial fitting techniques \cite{Ivezic} to estimate the
best fit parameters and their uncertainties (i.e., of a quadratic approximation to $\lnL$ near the peak): if $\lnLmarg=\sum_\alpha \lambda_\alpha F_\alpha(M_z)$ and $\gamma_{kk} = 1/\sigma_k^2$ is an inverse covariance matrix characterizing our measurement errors, then the best-fit estimate for $\lnLmarg$ and its variance is

\begin{subequations}
\label{eq:leastsquares}
\begin{eqnarray}
\lnLmarg {}_{\rm ,est} = F (F^T\gamma F)^{-1} \gamma  y \\
\label{eq:fiterrors}
\Sigma(x) = F_\alpha(x)  [(F^T\gamma F)^{-1}]_{\alpha \beta} F_{\beta}(x)
\end{eqnarray}
\end{subequations}
where $y$ is an array representing the $\lnLmarg$ estimates at the data points and $F$ is a matrix representing the
values of the basis functions on the data points: $F_{\alpha}(x_k)$.  The left panel of Figure  \ref{fig:ILE-Ex} shows the 90\% CI derived from this fit, assuming gaussian errors. %

\begin{table}
\centering
\begin{tabular}{l|lccc}
sample & $D_{KL}$ & CI (90\%)\\
1 & 0 & (68.71 - 71.66)\\
2 & 2.5e-4 & (68.71 - 71.68\\
3 & 1.2e-4 & (68.71 - 71.68)\\
4 & 7.2e-4 & (68.71 - 71.67)\\
5 & 2.3e-4 & (68.70 - 71.68)\\
\end{tabular}
\caption{\textbf{KL Divergence and 90\% CI between different samples from the null test fit}: This
  table shows the $D_{KL}$ and 90\% CI for five different sample PDFs.  The $D_{KL}$ was calculated
  comparing the 1D distributions to the first sample (i.e. $D_{KL}$ for sample 1 is zero).  The CI are also given to show the change between them.   Both diagnostics suggest the distributions are nearly indistinguishable.
}
\label{tab:Ex0}
\end{table}

To translate these uncertainties into changes in the one-dimensional posterior distribution $p_c$, we  generate random
draws from the corresponding approximately multinomial distribution for fit parameters; and thereby generate random
samples and hence one-dimensional distributions for $p_c(M)$ consistent with different realizations of the Monte Carlo
errors.  The right panel of Figure \ref{fig:ILE-Ex} shows five random samples from the fit in the left panel.  This
figure demonstrates this level of Monte Carlo error, by design, has negligible impact on the posterior distribution.  To
quantify the impact of Monte Carlo error on the posterior, we calculate the KL Divergence from Eq. (\ref{eq:dkl}).  In
all cases, the KL divergence was small, of order  $10^{-4}$, see Table \ref{tab:Ex0} for more details on $D_{KL}$ and
the 90\% CI.   In Section \ref{sub:MC}, we further verify this conclusion by repeating our analysis many times.

\subsection{Example 1: Two NR simulations with different parameters/Illustrating how sensitively parameters can be
  measured}
\label{sec:sub:ex1}
\begin{figure*}
\includegraphics[width=\columnwidth]{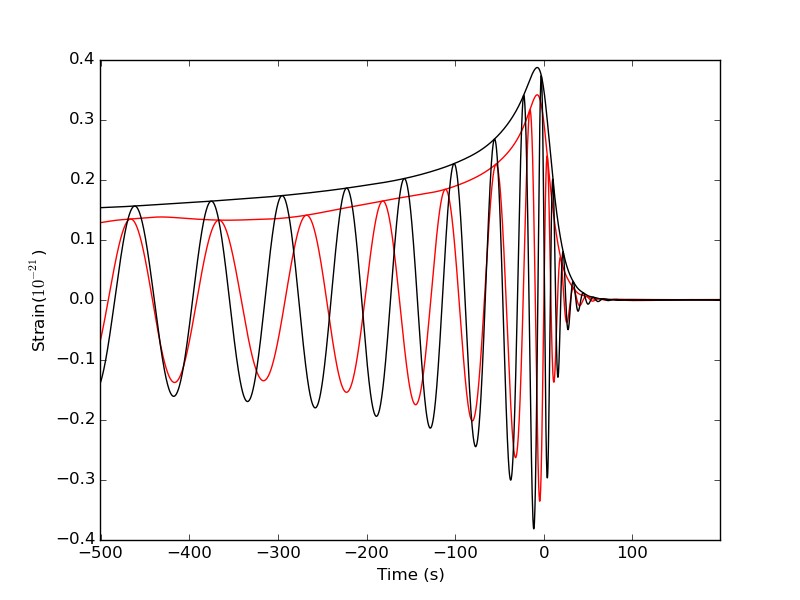}
\includegraphics[width=\columnwidth]{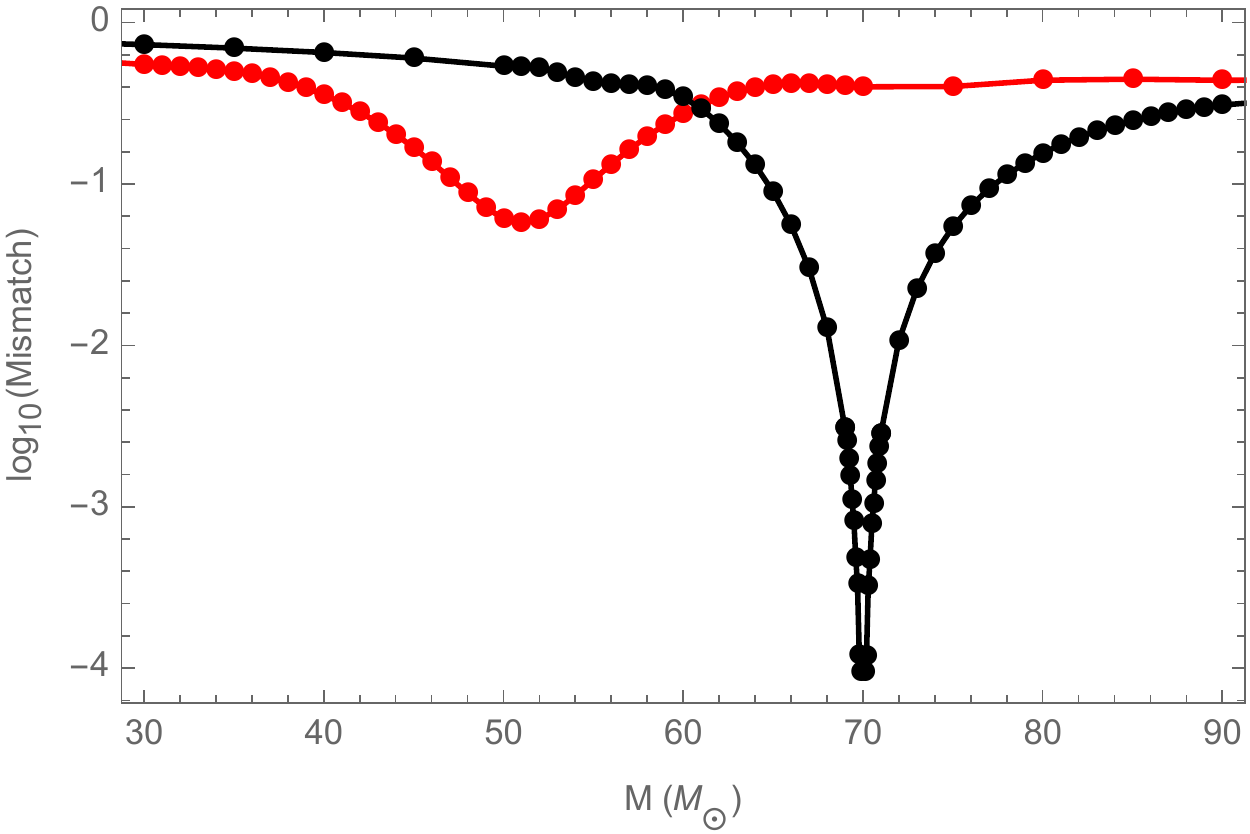}
\includegraphics[width=\columnwidth]{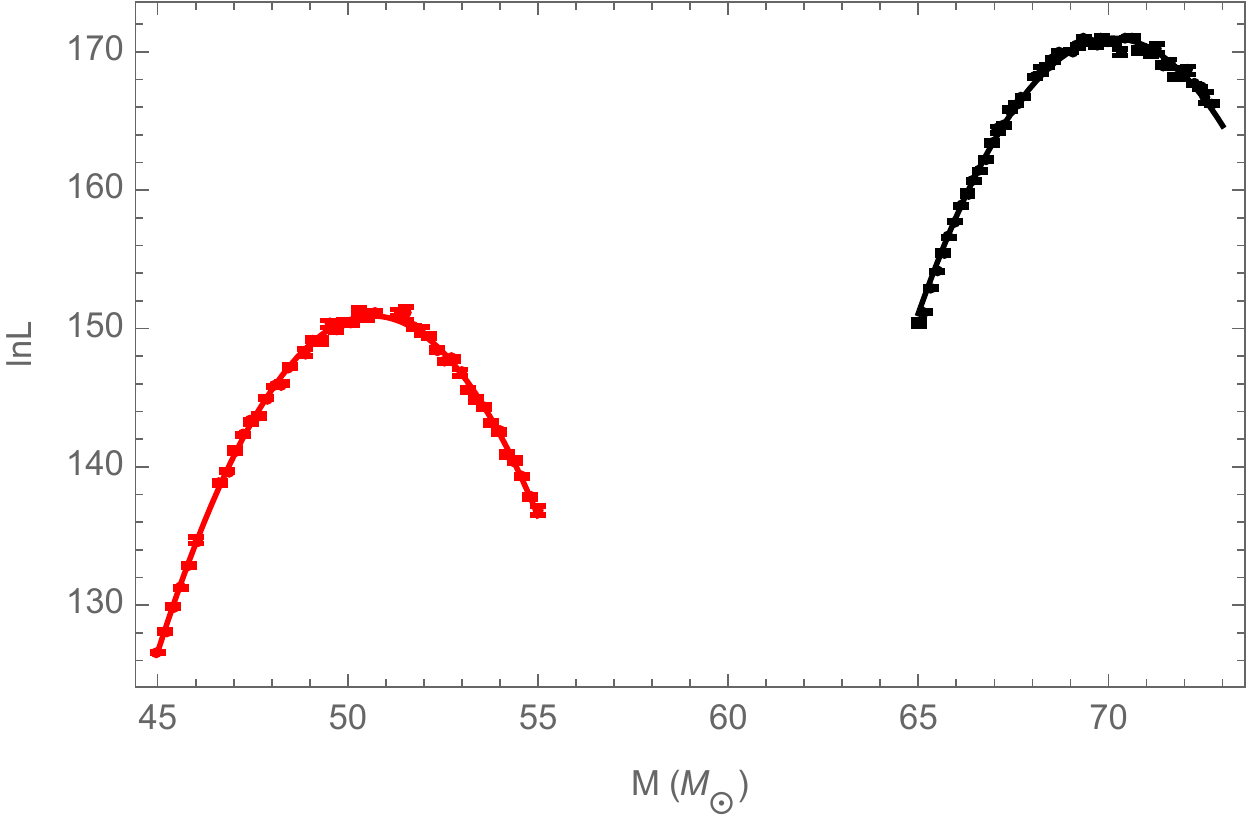}
\includegraphics[width=\columnwidth]{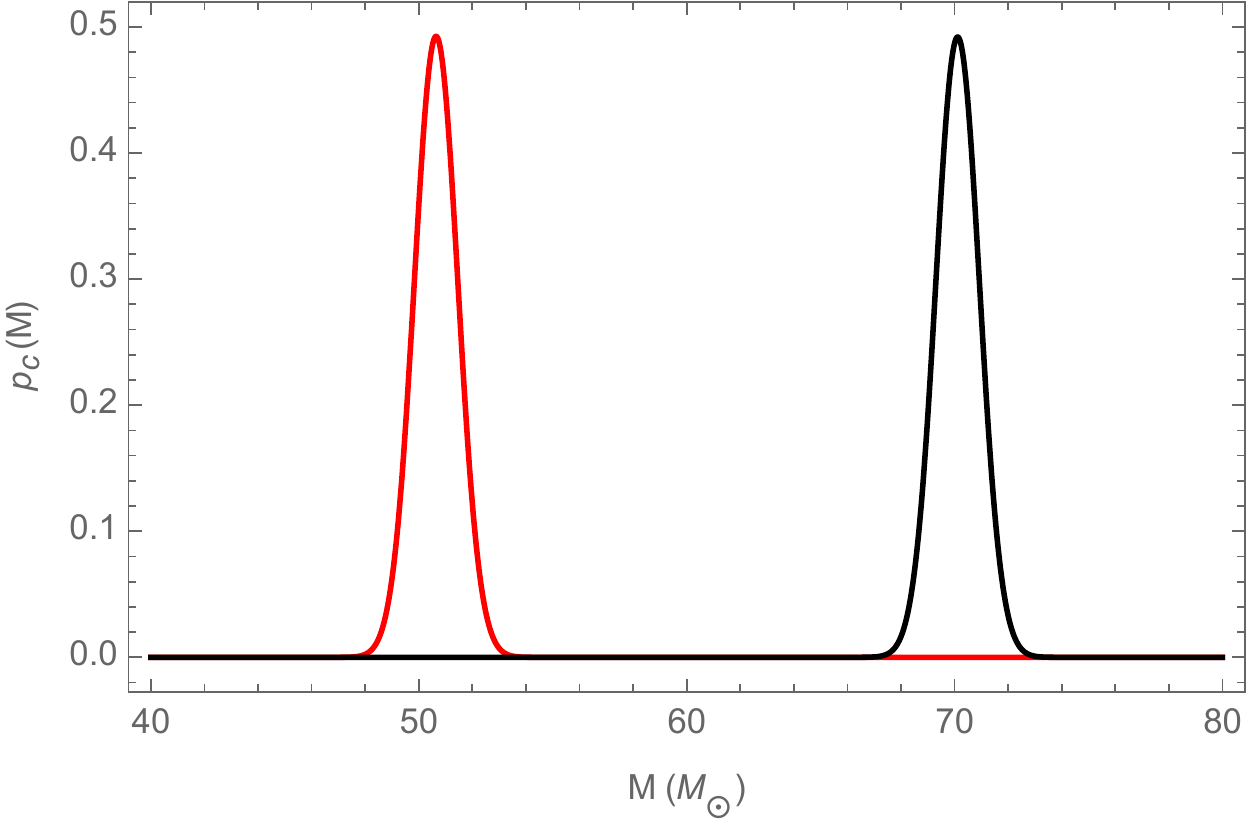}
\caption{\label{fig:Ex1}\label{fig:Ex1ILE}\small\textbf{Example 1-Assessing differences between two NR simulations with different parameters}:  Two representations of
  the different predictions of RIT-1a and RIT-2, which are aligned spin binaries with mass ratios $q=1.22$ and $q=2.0$ respectively,
  illustrating how dramatic differences propagate into our diagnostics.
  \emph{Top-left panel}: The strain along a line of sight inclined at $\iota=0.785$ and evaluated for a
  total mass $M=70M_\odot$, with RIT-1a in
  black and RIT-2 in red.  
\emph{Top-right panel}: The mismatch between synthetic data and candidate templates as a function of the template's mass.
In both cases, the RIT-1a  simulation is used as the template (i.e., as $h$ in Eq. (\ref{eq:p})).  For the black curve,
RIT-1a for a $70M_\odot$ binary is also used as the source (i.e., $h_0=h_{\rm RIT-1a}$).   For the red curve, the source
is  RIT-2 set at M=70 $M_{\odot}$.
 while RIT-1a has a changing mass
 \emph{Bottom-left panel}: Points show the  marginalized likelihood versus total mass calculated
  by applying the same template simulation (RIT-1a) to two different sources: RIT-1a in black and  RIT-2 in red.   Each
  source  has fixed mass $M=70 M_{\odot}$ and inclination $\imath=0.785$; as in Figure \ref{fig:ILE-Ex}, we evaluate
  ${\cal L}$ using a low-frequency cutoff  $f_{\rm min}=30\unit{Hz}$.  For context, red and black solid curves show a
  corresponding quadratic least-squares fit to these data. 
\emph{Bottom-right panel}: The corresponding one-dimensional posteriors $p_c(M)$ [Eq. (\ref{eq:1d})].  
Both bottom panels illustrate how an ill-suited simulation with  large mismatch (i.e., the red curve) correlates with a drastic shift in parameters (here, total mass) relative to
the true best-fit solution (here, the black curve), [see Eq. (\ref{eq:changeM})].  Also, the ill-matched simulation cannot
recover all the information available to the true solution, so the peak $\ln{\cal L}_{\rm marg}$  for the red curve is substantially
lower ($\simeq 20$) than the peak of the black curve. 
}
\end{figure*}

In this example we compare two NR simulations with significantly different parameters to demonstrate how our diagnostics
handle waveforms of extreme contrast. The two NR simulations used are RIT-1a and RIT-2.  As shown in Table
\ref{tab:simulations}, these simulations are both aligned spin with different magnitudes with $q=1.22$ and $q=2.0$
respectively.  To illustrate the extreme differences between the radiation from these two systems, the top-left panel of
Figure \ref{fig:Ex1} shows the two simulations' $r h(t)$. %

Our three diagnostics equally reveal the substantial differences between these two signals.  To be concrete,
  since these diagnostics treat data and models asymmetrically, we operate on synthetic data containing RIT-1a with inclination $\imath=\pi/4$ in these
applications.
First, the top-right panel of Figure \ref{fig:Ex1} shows the results of our mismatch calculations.  The black curve is the same null test mismatch
calculation as in the top-right panel of Figure \ref{fig:Ex2}: it has a narrow minimum (of zero) at the true binary mass
($70 M_\odot$).  For the red curve, we calculate the mismatch while
holding RIT-2 at a fixed mass and changing the mass of RIT-1a.  Using the notation in Eq. (\ref{eq:p}), we assign the RIT-2 waveform to $h_{0}=h_{\rm RIT-2}$(fixed mass at $M=70 M_{\odot}$) and the RIT-1a waveform to $h=h_{\rm RIT-1a}$(changing mass).  In this case, the match does not reach unity,
  differing by a few percent, while  the peak value occurs at significantly offset parameters (here, in total mass).
Second, the bottom-left panel of Figure \ref{fig:Ex1ILE}  shows the results for $\lnLmarg(M)$, using these two NR simulations to look at the same stretch of synthetic data including our local quadratic fit to them.  Third, the bottom-right panel of Figure \ref{fig:Ex1ILE} shows the implied one-dimensional posterior distribution derived from our fits.
There is a clear shift in total mass with the null test again peaking around $70 M_{\odot}$ and this example's peak
around $50 M_{\odot}$. There are also orders of magnitude difference between the $\lnLmarg$ of the two cases.  These
diagnostics show something that could be seen just by looking at the waveforms; however, we now have some idea on how
major differences propagate through our diagnostics and how the error in each diagnostic relate to each other.  For
completeness, we also include the $D_{KL}$ and CI for these two waveforms in Table \ref{tab:Ex1}. The $D_{KL}$ as well
as the CI are both considerably offset, as expected given the two significantly different simulations involved.

\begin{table}
\centering
\begin{tabular}{l|lccc}
\textit{ILE} run (source/template) & $D_{KL}$ & CI (90\%)\\
RIT-1a/RIT-1a & 0.0 & (68.8 - 71.4)\\
RIT-2/RIT-1a & 288.8 & (49.3 - 52.0)\\
\end{tabular}
\caption{\textbf{KL Divergence and 90\% CI between two NR simulations with different parameters}: This table shows the $D_{KL}$ and 90\% CI between: RIT-1a/RIT-1a and RIT-1a/RIT-2.  The $D_{KL}$ was calculated comparing the 1D distributions to RIT-1a/RIT-1a distribution (notice its $D_{KL}$ is zero i.e. they're identical).  The CI are also given to show the difference between these two distributions.} %
\label{tab:Ex1}
\end{table}

Finally, the parameter shift seen above is roughly consistent in magnitude with what we would expect for
  such an extreme mismatch error, given the SNR and match: we expect using Eq. (\ref{eq:changeM})  $\delta M \simeq \sigma_{M} \rho {\cal M}^{1/2}
  \simeq 5 \sigma_M \simeq 5 M_\odot$ (using ${\cal M} = 6\times 10^{-2},\rho=20$ and $\sigma_M = 1.1 M_\odot$), or a
  shift in best fit of several standard deviations and many solar masses.   While noticeably smaller than our actual
  best-fit shift, our result from Eq. (\ref{eq:changeM}) provides a valuable sense of the order-of-magnitude biases
  incurred by specific level of mismatch in general.  Moreover, this example is a concrete illustration of the critical need to have
${\cal M}\le 1/\rho^2$ to insure that any systematic parameter biases are small and under control.

\subsection{Example 2: Different  physics: SEOB vs NR/Illustrating the value of numerical relativity}
\label{sub:Ex2}
\begin{figure*}
\includegraphics[width=\columnwidth]{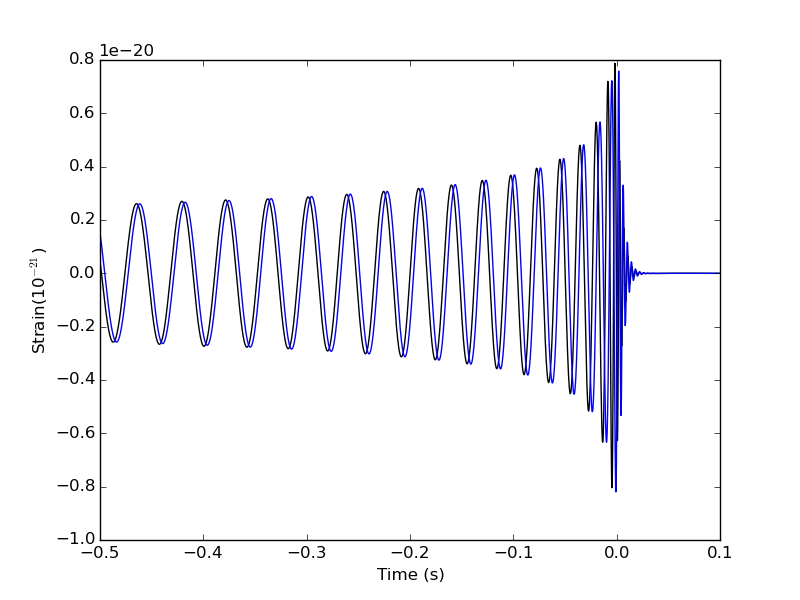}
\includegraphics[width=\columnwidth]{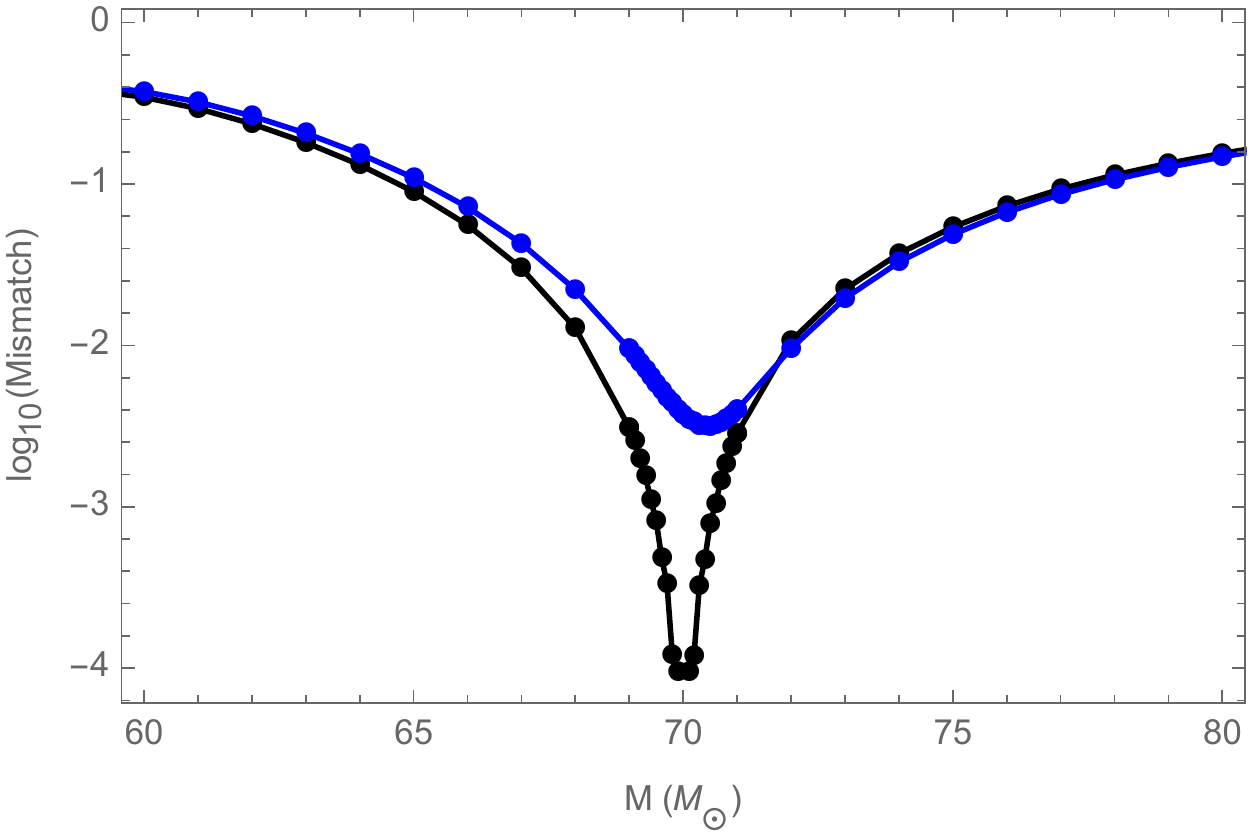}
\includegraphics[width=\columnwidth]{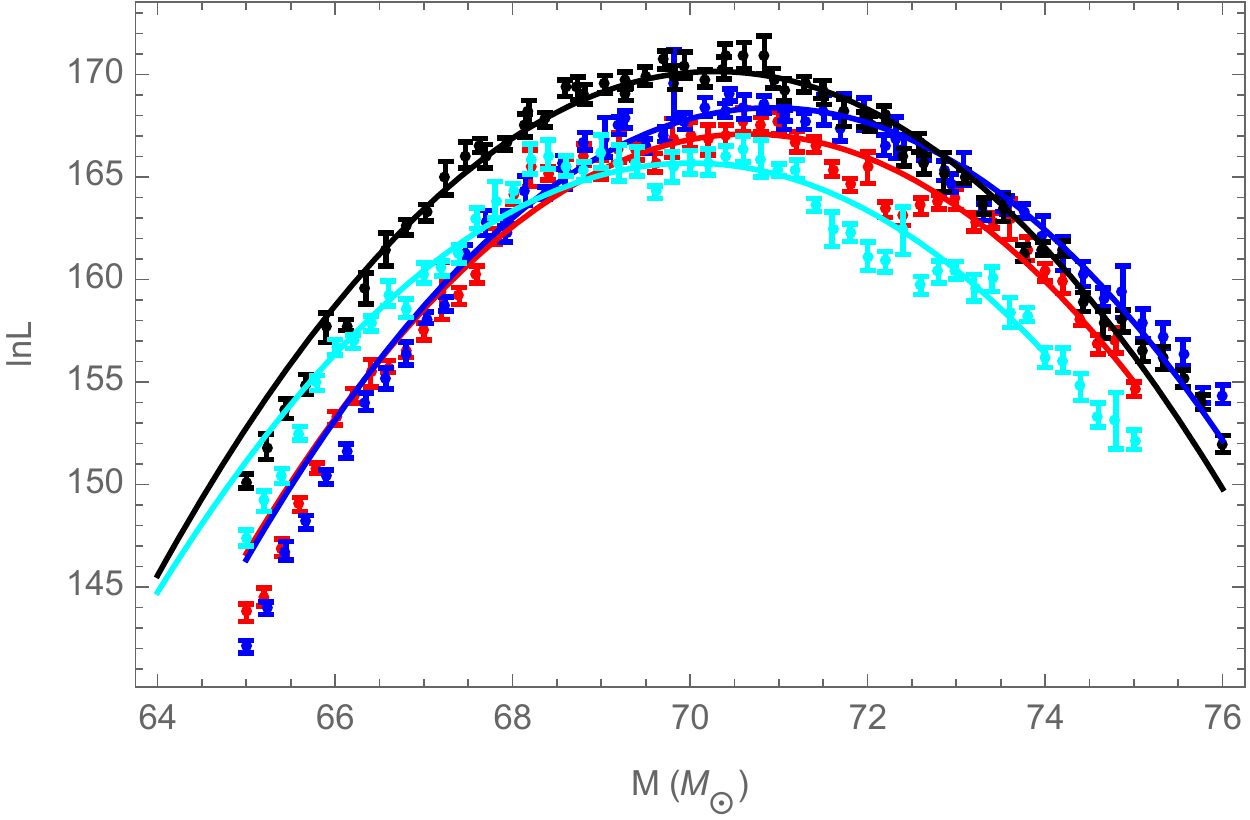}
\includegraphics[width=\columnwidth]{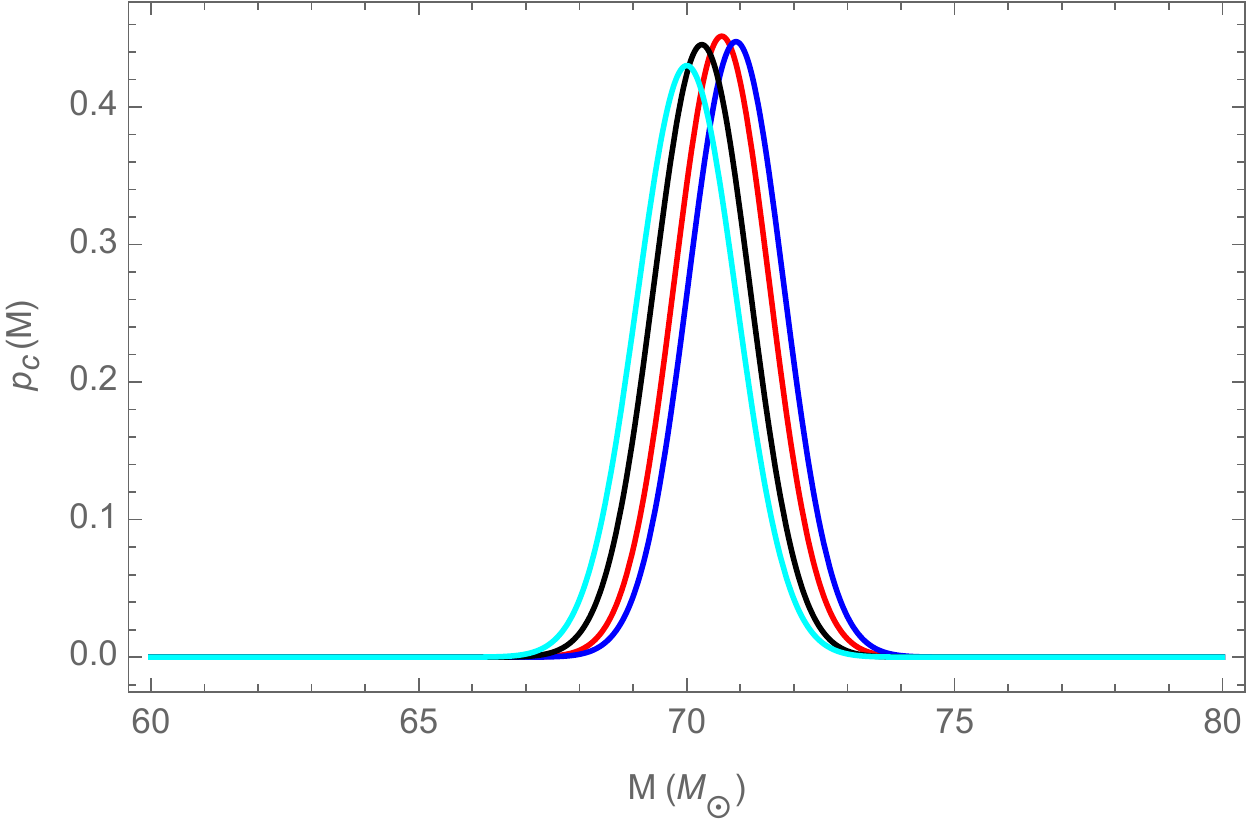}
\caption{\label{fig:Ex2ILE}\small \textbf{Example 2-Assessing differences in SEOB and NR waveforms that have the same parameters}: This figure  shows how subtle differences between an NR solution and an approximation to GR (here, EOB) can
  propagate into mismatch and parameter estimation.  These two companion figures follow the pattern of Figures \ref{fig:Ex1}.
\emph{Top-left panel}: The black and blue curves show the strain evaluated from RIT-a and SEOBNRv2, respectively, for a
source with identical parameters.   Source parameters and strain results for the black curve are identical to Figure
\ref{fig:Ex1} (e.g., $\iota=\pi/4$).
\emph{Top right-panel}: Following the top-right panel of Figure \ref{fig:Ex1}, this figure shows the match between the two
waveforms on the top-left with the corresponding template from RIT-1a.
\emph{Bottom-left}: The marginalized likelihood $\lnLmarg$ for the two waveforms shown above, evaluated using
\emph{both} RIT-1a and SEOBNRv2 as templates:   NR source compared to same NR template in black;  the SEOBNRv2 source to
a SEOBNRv2 template in red; the SEOBNRv2 source to a NR (RIT-1a) template in blue; and  the NR (RIT-1a) source to a
SEOBNRv2 template in cyan. 
\emph{Bottom-right}: The one-dimensional posterior distributions $p_c(M)$ derived from the quadratic fits shown in the bottom-left.
Both bottom panels show  a clear change along the total mass for SEOBNRv2 sources.  The NR/NR comparison  has the
highest $\lnLmarg$ with with a corresponding total mass $\sim70 M_{\odot}$. The NR/SEOBNRv2 template curve correctly
finds the total mass $\sim70 M_{\odot}$; however, the $\lnLmarg$ is orders of magnitudes different than the null
example. The differences between NR simulations and the SEOBNRv2 model is significant for parameter estimation.
}
\label{fig:Ex2}
\end{figure*}

Several studies have previously demonstrated the critical need for numerical relativity, since even the best models do
not yet capture all available physics \cite{gwastro-mergers-nr-SXS_RIT-2016,2015PhRvD..92j2001K}.  For example, these
models generally omit higher-order modes, whose omission will impact inferences about the source
\cite{2014PhRvD..90l4004V,gwastro-Varma-2016,2016PhRvD..93h4019C}.  

To illustrate the value of NR in the context of this work, we compare parameter estimation with NR and with an analytic
model.  In this particular example, we use NR simulation RIT-1a including the $l\le2$ modes (see Table \ref{tab:simulations}) evaluated along an
inclination $\iota=\pi/4$.   Using this line of sight and our fiducial mass ($M=70 M_\odot$), higher harmonics play a
nontrivial role.   For our analytical model, we use an Effective-One-Body model with spin (SEOBNRv2), described in
\cite{gw-astro-EOBspin-Tarrachini2012}, which was one of the models used in the parameter estimation of GW150914
\cite{2016PhRvL.116x1102A} and which was recently compared to this simulation \cite{gwastro-mergers-nr-SXS_RIT-2016}.
 The top-left panel of Figure \ref{fig:Ex2} shows the time-domain strains from the NR simulation
and SEOBNRv2 with the same parameters. %
To better quantify the small but visually apparent difference in the two waveforms, we use the diagnostics described earlier on these two waveforms.

One way to characterize the differences in these waveforms is the mismatch [Eq. (\ref{eq:p})].   In the top-right panel of Figure
\ref{fig:Ex2}, we calculate the mismatch by holding the SEOBNRv2 waveform at a fixed mass while changing the mass of the NR waveform shown in
blue.  Referring to the notation in Eq. (\ref{eq:p}), we assign the SEOBNRv2 waveform to $h_{0}=h_{\rm SEOBNRv2}$ and  the RIT-1a waveform to $h=h_{\rm RIT-1a}$.  For comparison, a mismatch calculation was done with the null test from Section \ref{sub:null} (RIT-1a compared to
itself) shown here in black.   Two differences between the two curves are immediately apparent.  First, the blue curve does not go to zero;
the mismatch is a few times $10^{-3}$, significantly in excess of the typical accuracy threshold
[Eq. (\ref{eq:accuracy}), evaluated at $\rho=25$].  Second, the minimum occurs at offset parameters.   The best-fit offset and mismatch are
qualitatively consistent with the naive estimate presented earlier:  a
high mismatch yields a high change in total mass [see Eq. (\ref{eq:changeM})].  This simple calculation
illustrates how mismatch could propagate directly into significant biases in parameter estimation. 

Another and more observationally relevant way to characterize the differences between these two waveforms is by
carrying out a full \textit{ILE}
based parameter estimation calculation.  We carry out four comparisons:  the null test (a NR
source compared to same NR template (black));  the SEOBNRv2 source compared to a
SEOBNRv2 template (red); the NR source compared to a SEOBNRv2 template (cyan);  and an SEOBNRv2 source compared to a NR
template (blue).  The bottom panels of Figure \ref{fig:Ex2ILE} shows both the underlying $\lnLmarg(M)$ results; our quadratic approximations
to the data; and our implied one-dimensional posterior distributions [Eq. (\ref{eq:1d})].
All \textit{ILE} calculations were carried out with $f_{\rm min}=30\unit{Hz}$.  
All four likelihoods $\lnLmarg$ and posterior distributions $p_c$ are manifestly different, with generally different
peak locations and widths.   
Table \ref{tab:Ex2} quantifies the differences between  the possible four configurations, using  $D_{KL}$ and 90\% CI.
The $D_{KL}$ was always calculated by comparing one of them to the NR/NR case. 
These systematic differences exist even without higher modes, whose neglect will only exacerbate the biases seen here.

Keeping in mind the two figures adopt a comparable color scheme,  the shift in
peak value and location between  the black and blue curves seen in the bottom panels of Figure \ref{fig:Ex2ILE} can be traced back to the top-right of Figure \ref{fig:Ex2}: to a first approximation, systematic errors identified by the mismatch ($\cal M$) show up in the
marginalized likelihood ($\lnLmarg$).   Again, based on calculations using Eq. (\ref{eq:changeM}), we expect the change in mass
location of order unity holding all other things equal, comparable to the observed offset. 

In many ways, one-dimensional biases shown in the bottom-right panel \emph{understate} the differences between these signals:
that comparison explicitly omits the peak value of $\lnLmarg$, which occurs not only at a different location but also
with a different value for all four cases. 
As we
would expect, the NR/NR case has the highest $\lnLmarg$ with a peak near the true total mass 70$M_{\odot}$. The
NR/SEOB case can also produce a peak near 70$M_{\odot}$; however, the $\lnLmarg$ is orders of magnitude lower, which
translates to a lower likelihood that this was in fact the correct template.   
When performing a full multidimensional fit, template-dependent biases in the peak value of $\lnLmarg$ can also impact
our conclusions.  

To summarize, we have shown that using SEOBNRv2 in place of a more precise solution of Einstein's equations introduces
non-negligible systematic errors, of a magnitude comparable to the statistical error for plausible sources, and that it can impact astrophysical conclusions.

\begin{table}
\centering
\begin{tabular}{l|lccc}
\textit{ILE} configuration (source/template) & $D_{KL}$ & CI (90\%)\\
SEOB/SEOB & 0.086 & (69.2 - 72.1)\\
SEOB/RIT-1a & 0.25 & (69.4 - 72.4)\\
RIT-1a/RIT-1a & 0 & (68.8 - 71.8)\\
RIT-1a/SEOB & 0.050 & (68.5 - 71.5)\\
\end{tabular}
\caption{\textbf{KL Divergence and 90\% CI between SEOB and NR}: This table shows the $D_{KL}$ and 90\% CI for the four different configurations using SEOBNRv2 and NR as sources and templates.  The $D_{KL}$ was calculated comparing the 1D distributions to the NR/NR case (notice its $D_{KL}$ is zero i.e. they're identical).  The CI also given to show the change between them.  Based on the $D_{KL}$ results, the 1D posteriors are similar but not exactly the same distribution.  These nontrivial differences affect our parameter estimation results and also change our astrophysical conclusions about the source.}
\label{tab:Ex2}
\end{table}

\subsection{Example 3: Signal duration and cutoff frequency/Illustrating the impact of simulation duration with SEOB}
  \label{sub:Ex3}
  \begin{figure*}
\includegraphics[width=\columnwidth]{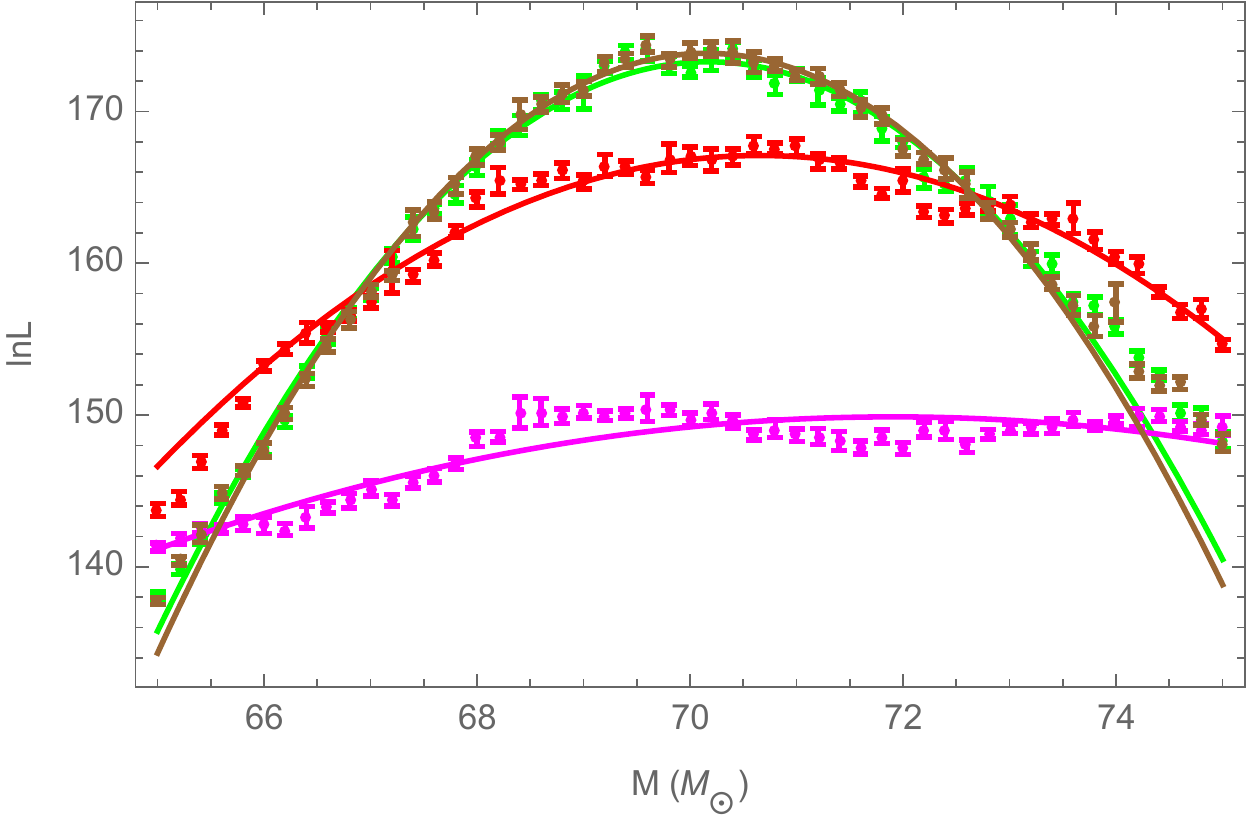}
\includegraphics[width=\columnwidth]{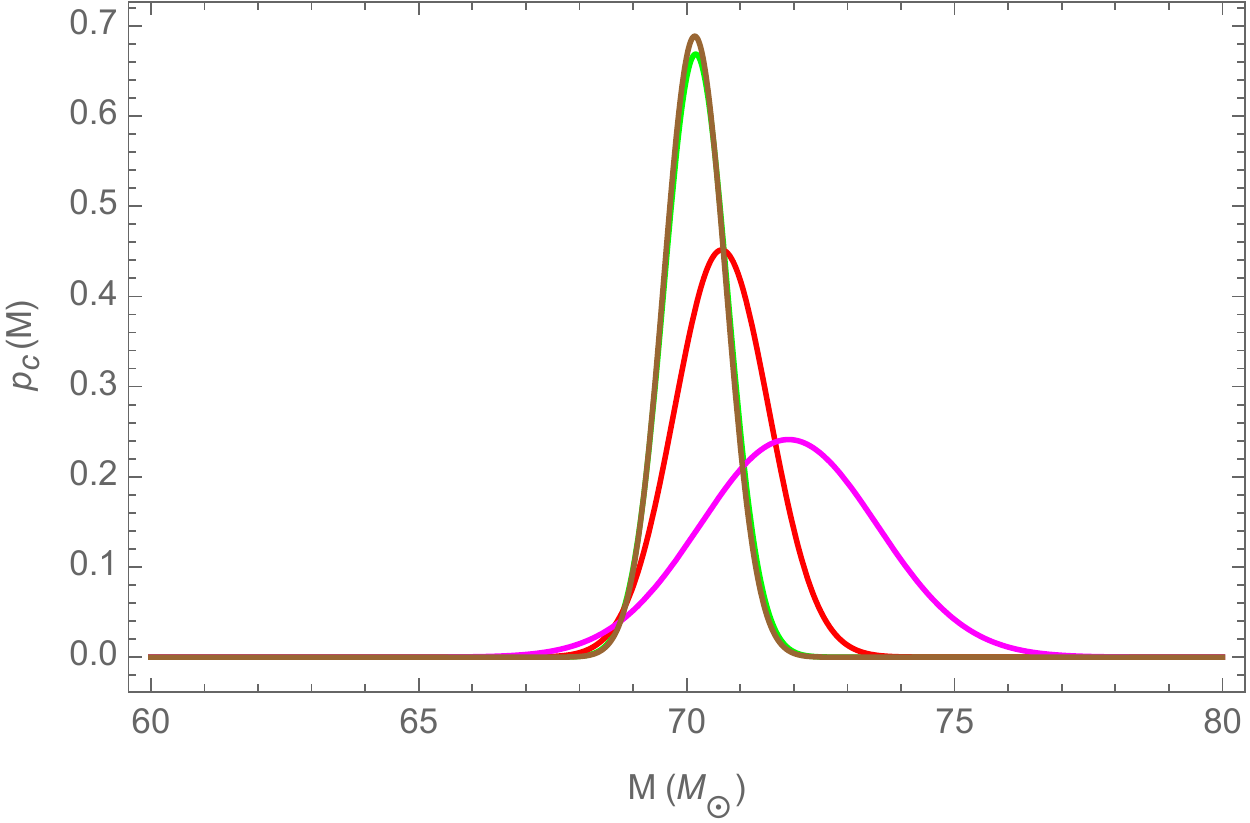}
\caption{\textbf{Example 3-Quantifying the impact of the low-frequency cutoff}:  Using analytic SEOBNRv2 templates with
  user-specified starting frequency and length, this figure quantifies the impact of our choice of low-frequency cutoff
  on parameter estimation.
\emph{Left panel}: Plot of $\lnLmarg$ versus total mass evaluated using SEOBNRv2 templates with different starting frequencies
with $f_{\rm min}=10\unit{Hz}$ (brown), $f_{\rm min}=20\unit{Hz}$ (green), $f_{\rm min}=30\unit{Hz}$ (red), and
$f_{\rm min}=40\unit{Hz}$ (magenta).   In all cases, the source signal is also SEOBNRv2 using the same parameters as
RIT-1a, but starting frequency $f_{\rm min}=5\unit{Hz}$.  
\emph{Right panel}: The one-dimensional posteriors $p_c(M)$ [Eq. (\ref{eq:1d})] implied by the results to left.    As
you increase the low frequency cutoff, the $\lnLmarg$ decreases significantly, and both the posterior and $\lnLmarg$ are
wider and offset from the true parameters. 
}
\label{fig:Ex3}
\end{figure*}

Numerical relativity simulations have finite duration.  Until hybrids \cite{2008PhRvD..77d4020H,2008PhDT.......315B,2008PhRvD..77j4017A,2013PhRvD..87b4009M} are ubiquitously available, these finite
  duration cutoffs will  impair the utility of direct comparison between data and multimodal NR simulations.   To assess
  this impact of finite simulation duration, we adopt  a contrived but easily-controlled approach, using an analytic
  model where we can freely adjust signal duration.    While our specific numerical conclusions depend on the noise
  power spectrum adopted, as it sets the required low-frequency cutoff, the general principles remain true for advanced instruments.

\begin{table}
\centering
\begin{tabular}{l|lccc}
$f_{\rm min}$ for \textit{ILE} run (Hz) & $D_{KL}$ & CI (90\%)\\
10 & 0.0 & (69.2 - 71.1)\\
20 & 1.3e-3 & (69.2 - 71.1)\\
30 & 0.62 & (69.2 - 72.1)\\
40 & 7.1 & (69.2 - 74.6)\\
\end{tabular}
\caption{\textbf{KL Divergence and 90\% CI of PDFs derived from SEOB sources with different low frequency cutoffs}: This table shows the $D_{KL}$ and 90\% CI for the four different configurations using SEOBNRv2 source with a set duration of $5 \unit{Hz}$ and compared against SEOBNRv2 templates with different low frequency cutoffs.  The $D_{KL}$ was calculated comparing the 1D distributions to the $f_{\rm min}=10 \unit{Hz}$ case (notice its $D_{KL}$ is zero i.e. they're identical).  The CI also given to show the change between them.  Based on the $D_{KL}$ results, the 1D posteriors of $f_{\rm min}=10, 20 \unit{Hz}$ seem to be the same distribution; however, they differ significantly to $f_{\rm min}=30, 40 \unit{Hz}$.}
\label{tab:Ex3}
\end{table}

In this example, we plot $\lnLmarg$  for a fiducial SEOBNRv2 source versus itself using different choices for the
low-frequency cutoff  (and, equivalently, different initial orbital frequencies for the binary).  The left panel of
Figure \ref{fig:Ex3} shows $\lnLmarg$ versus $M$.  In this figure, the $\lnLmarg$ curves for $f_{\rm min}= 10\unit{Hz}$ and $20\unit{Hz}$ (brown and green) are significantly narrower and higher compared to the $\lnLmarg$ curves for $f_{\rm min}=30 \unit{Hz}$ or
$40\unit{Hz}$(red and magenta).  As described in \cite{NRPaper}, even though very little signal power is associated with
very low frequencies for this combination of detector and source, a significant
amount of information about the total mass is available there with all other parameters of the system perfectly known.
These differences are immediately apparent in our one-dimensional diagnostics  $\lnLmarg(M)$ and $p_c(M)$, which are
both narrower and more informative when more information is included (i.e., for lower $f_{\rm min}$).  
That said, our PSD does not provide access to arbitrarily low frequencies, and the lowest two frequencies have nearly identical posterior distributions, as measured by KL divergence, see Table \ref{tab:Ex3}.
This investigation strongly suggests our analysis could  be sharper with longer simulations or hybrids.  
That said, \cite{NRPaper} demonstrated  this procedure will, for GW150914-like data and noise, arrive at similar results to an analysis which includes these
lower frequencies.   As noted in \cite{NRPaper}, this virtue leverages a fortuitous degeneracy in astrophysically
relevant observables: the limitations of our high-frequency analysis are mostly washed out due to strong  degeneracies between mass, mass ratio, and spin.

\section{Validation studies}
\label{sec:valid}
In this section we self-consistently assess our errors in $h(t)$ and $\lnL$.  Using the diagnostics described above, via
targeted one-dimensional studies, we
systematically assess the impact of Monte Carlo error; waveform extraction error; simulation resolution; and limited
access to low frequency content.  We will show via our diagnostics that the effects from these potential sources of
error can be either ignored or mitigated (e.g., by a suitable choice of operating point for our analysis procedure, such
as a high enough extraction radius).  For each potential source of error, we use  the KL divergence $D_{KL}$
[Eq. (\ref{eq:dkl})] to quantify small differences in one-dimensional posterior distributions $p_c(M)$
[Eq. (\ref{eq:1d})] derived from $\lnLmarg$.   We will relate our results to familiar mismatch-based measures of
error.  
To be concrete, we will employ a target signal amplitude (SNR) $\rho=25$, similar to GW150914.  For similarly-loud
sources, the mismatch criteria [Eq. (\ref{eq:accuracy})] suggests any parameters with mismatch below
$\log_{10}(\mathcal{M})=-2.8$ will lead to ``statistical errors'' (associated with the width of the posterior) will be
smaller than systematic biases.

\subsection{Impact of Monte Carlo error}
\label{sub:MC}

  \begin{figure*}
\includegraphics[width=\columnwidth]{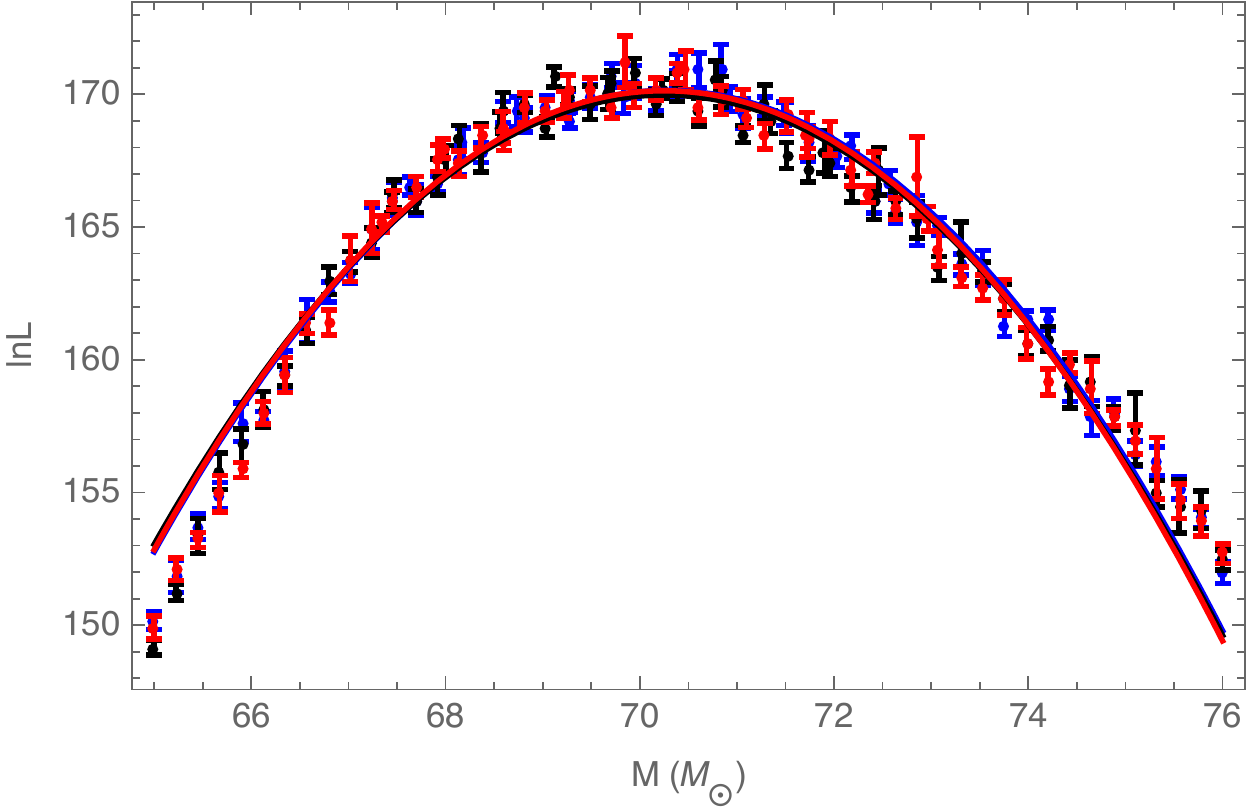}
\includegraphics[width=\columnwidth]{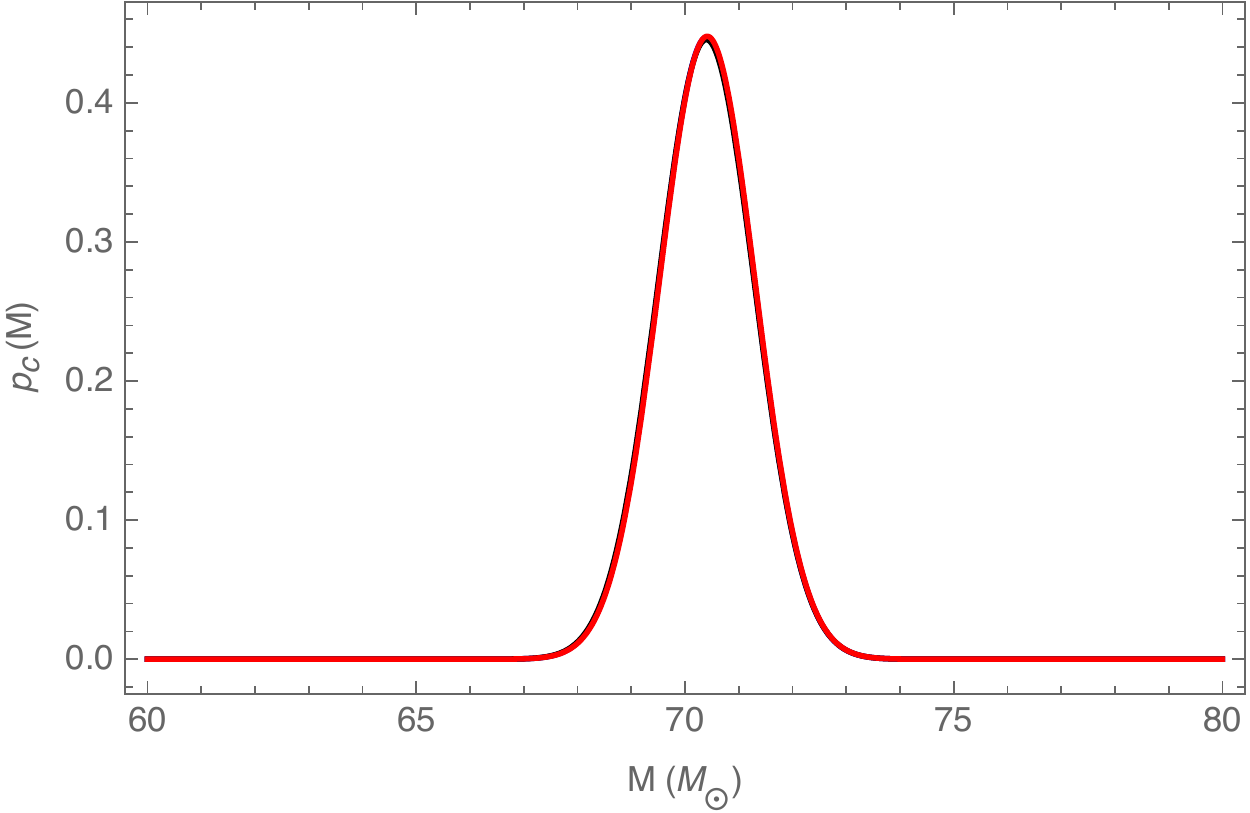}
\caption{\textbf{Monte Carlo error revisited: Repeating the fitting process multiple times}: This figure shows several
  repeated, independent end-to-end calculations of $\lnLmarg$ (left panel) and $p_c(M)$ (right panel), shown in
  different colors.  The calculation performed is identical to the calculation
described for Figure \ref{fig:ILE-Ex}.  This figure demonstrates we understand and have control over our Monte Carlo errors.
}
\label{fig:MC}
\end{figure*}
\begin{table}
\centering
\begin{tabular}{l|lccc}
Trial & $D_{KL}$ & CI (90\%)\\
v1 & 0 & (68.9 - 71.9)\\
v2 & 4.8e-5 & (68.9 - 71.9)\\
v3 & 5.6e-5 & (68.9 - 71.9)\\
\end{tabular}
\caption{\textbf{KL Divergence and 90\% CI between different runs of the same null test.}: This table shows the $D_{KL}$, calculated using Eq. (\ref{eq:dkl}) and 90\% CI for three different runs of the same configuration as described in Section \ref{sub:null}.  The $D_{KL}$ was calculated comparing the 1D distributions to Trial v1 (notice its $D_{KL}$ is zero i.e. they're identical).  The CI also given to show the change between them.  Based on the $D_{KL}$ results, the 1D posteriors of these different trials are identical.}
\label{tab:MC}
\end{table}
We have already assessed  the error from our Monte Carlo integration  in Section \ref{sub:null}, directly propagating the
(assumed correct) Monte Carlo integration error into our fit.   To comprehensively demonstrate the impact of Monte Carlo
integration error, we repeat our entire analysis reported in Figure \ref{fig:ILE-Ex}  multiple times.  Figure \ref{fig:MC} shows our directly comparable results; Table \ref{tab:MC} reports quantitative measures of how these distributions change.  Based on these quantities, we conclude the error introduced by our Mont Carlo is negligible.  Our results are consistent with Section \ref{sub:null}.

\subsection{Error budget for waveform extraction}
\label{sub:rextr}

\begin{figure*}
\includegraphics[width= \columnwidth]{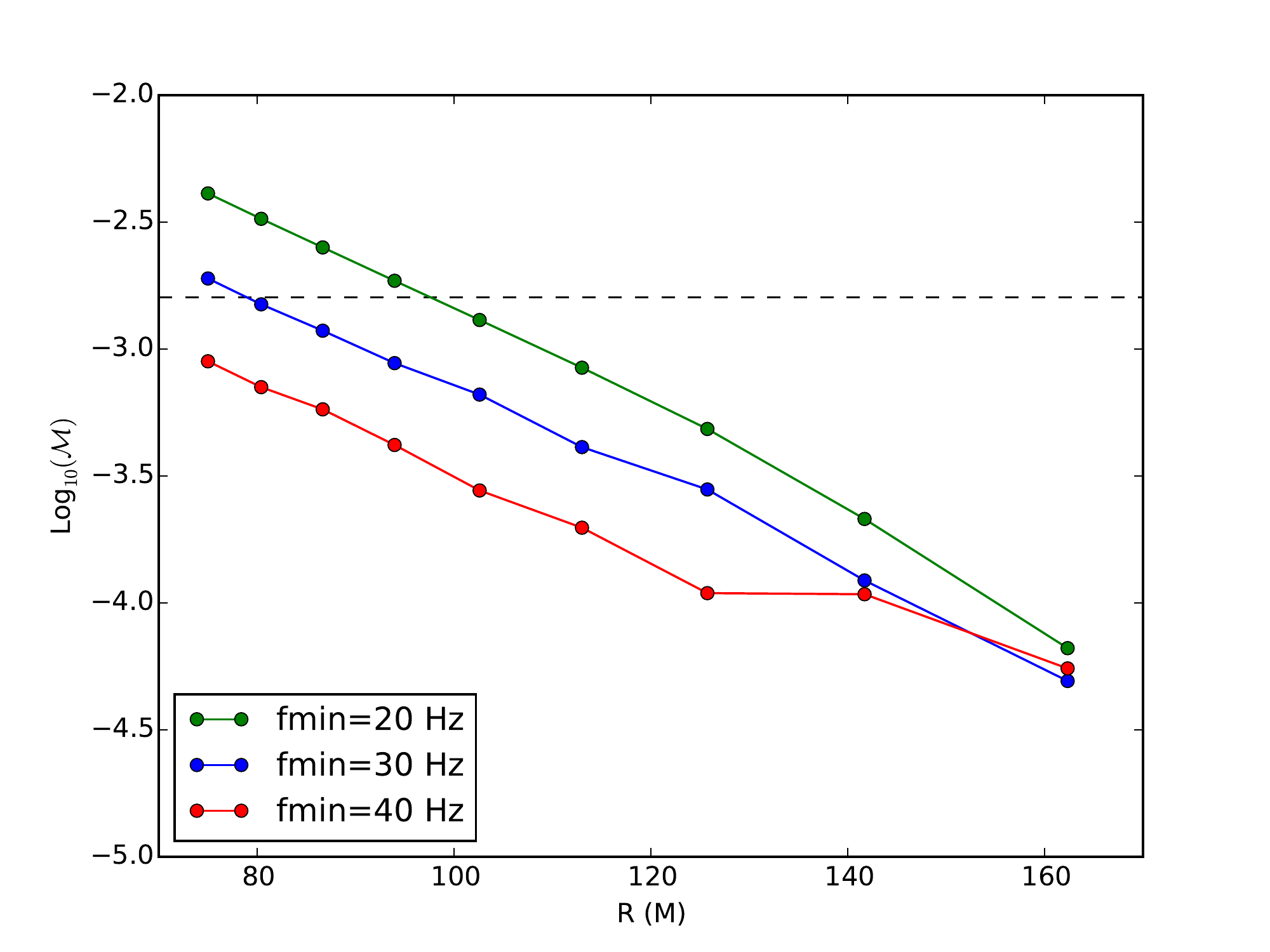}
\includegraphics[width= \columnwidth]{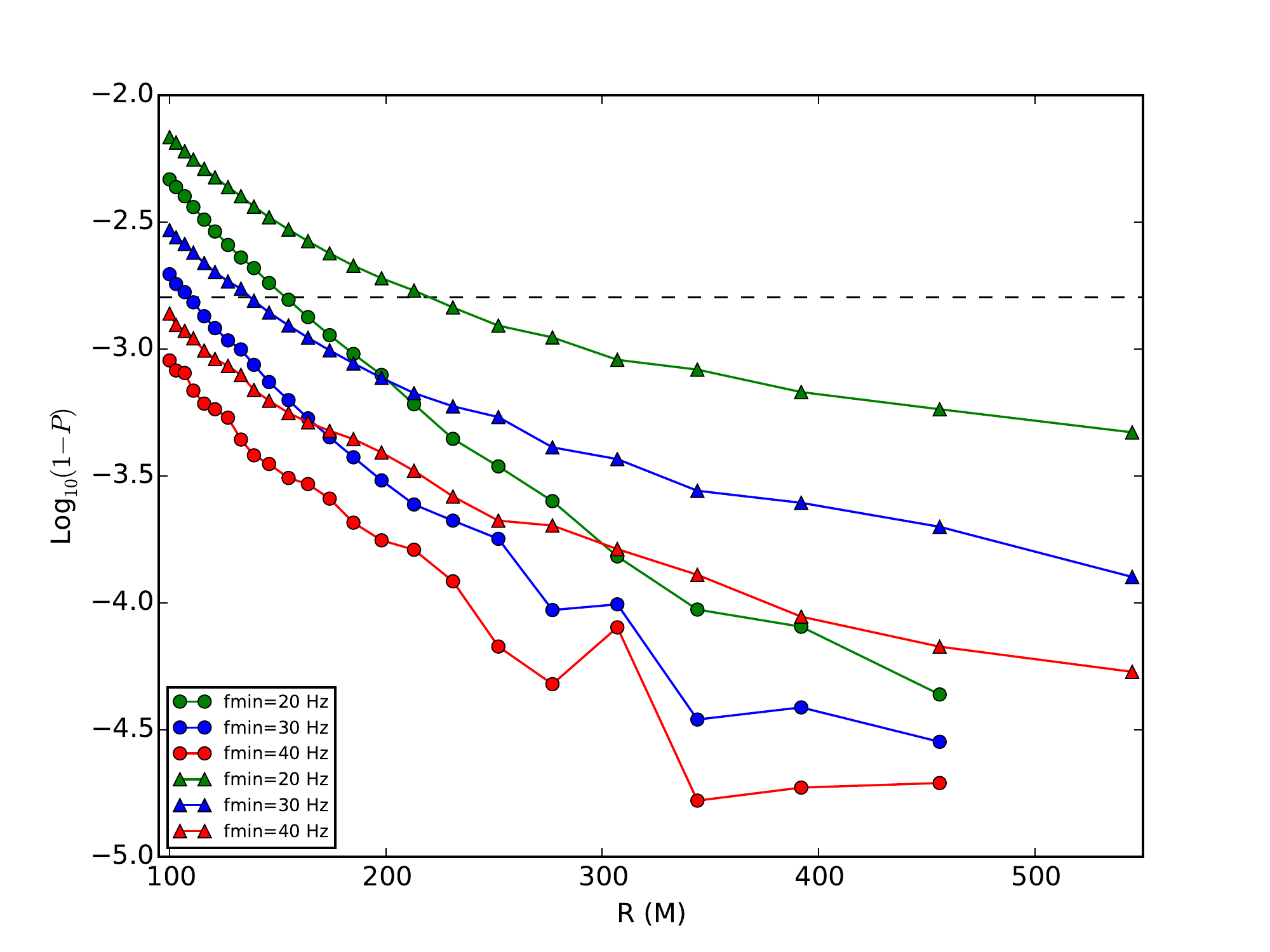}
\caption{\textbf{Mismatch between waveforms at different extraction radii using different NR groups and
      extraction techniques}:  Both panels show the mismatch between the radiation extracted from RIT-1a (left panel)
    and SXS-0233 (right panel) as a function of the extraction radius $r$.  All calculations are performed using  the same
    configurations as Figures \ref{fig:Ex1} and \ref{fig:Ex2}: a total mass of $70 M_\odot$ and an
    inclination $\iota=0.785$. %
In both panels, the green, blue, and red colors represent different choices of low frequency cutoff:  $f_{\rm min}=20, 30,
40 \unit{Hz}$ respectively.  For context and motivated by Eq. (\ref{eq:accuracy}), the dashed line denotes the mismatch
threshold implied by $\rho=25$ (i.e., $\log_{10}(1/25^{2})$).  
\emph{Left panel}: Mismatch calculations comparing a waveform perturbatively extracted at $r=190 M$
with a waveform that is perturbatively extracted at other extraction radii, [see Eq. (\ref{eq:PTExtract})].  
\emph{Right panel}: Circles correspond to results using a reference waveform extracted at $r=545 M$ via
perturbative extraction from their $\psi_4$ data; triangles denote calculations using a reference waveform evaluated
using the strain provided by SXS (i.e., using a polynomial extrapolation with $N=2$).  In both cases, the reference
waveform is compared to a waveform constructed via perturbative extraction using $\psi_4$ data at the specified radius.
\label{fig:rextr1}
}
\end{figure*}
While gravitational waves are defined at null infinity, the finite size of typical NR computational domains implies a
computational technique must identify the appropriate asymptotic radiation from the simulation 
\cite{Bishop2016}.  This method generally has error, often associated with systematic neglect of near-field physics in
the asymptotic expansion used to extract the wave (i.e., truncation error).  Our perturbative extrapolation method
shares this limitation.  As a result, if we decrease the radius at which we extract the asymptotic strain, we increase
the error in our approximation.  In other words, the mismatch between the waveform extracted at $r$ and some large
radius generally decreases with $r$; the trend of match versus $r$ provides clues into the reliability of our results.

Figure \ref{fig:rextr1} shows an example of a mismatch between two estimates of the strain: one evaluated at finite, largest possible radius
and one at smaller (and variable) radius.  For context, we show the nominal accuracy requirement
corresponding to a SNR=25 [see Eq. (\ref{eq:accuracy})] as a  black dotted line.  
First and foremost, this figure shows that, at sufficiently high extraction radius, the error introduced by mismatch errors is substantially
below our fiducial threshold for all choices of: cutoff frequency, waveform extraction location, and waveform
extraction technique; see also \cite{LIGO-Puerrer-NR-LI-Systematics}.   
Second, the second panel shows our perturbative extraction method is reasonably consistent with an
entirely independent approach to waveform extraction.
Agreement is far from perfect: our study also indicates a noticeable discrepancy between the results of our perturbative extraction technique and
the SXS strain extraction method.  Due to the good agreement reported elsewhere \cite{gwastro-mergers-nr-SXS_RIT-2016},
we suspect these residual disagreements arise from coordinate effects unique to our interpretation of SXS data; we will
assess this issue at greater depth in subsequent work.
Third and finally, as expected, comparisons that employ more of the NR signals are more discriminating: calculations with a smaller
$f_{\rm min}$ generally find a higher (i.e., worse) mismatch.   Nonetheless, our mismatch calculations significantly improve at large extraction
radius, when perturbative extrapolation is carried out well outside the near zone.

\begin{figure*}
\includegraphics[width=\columnwidth]{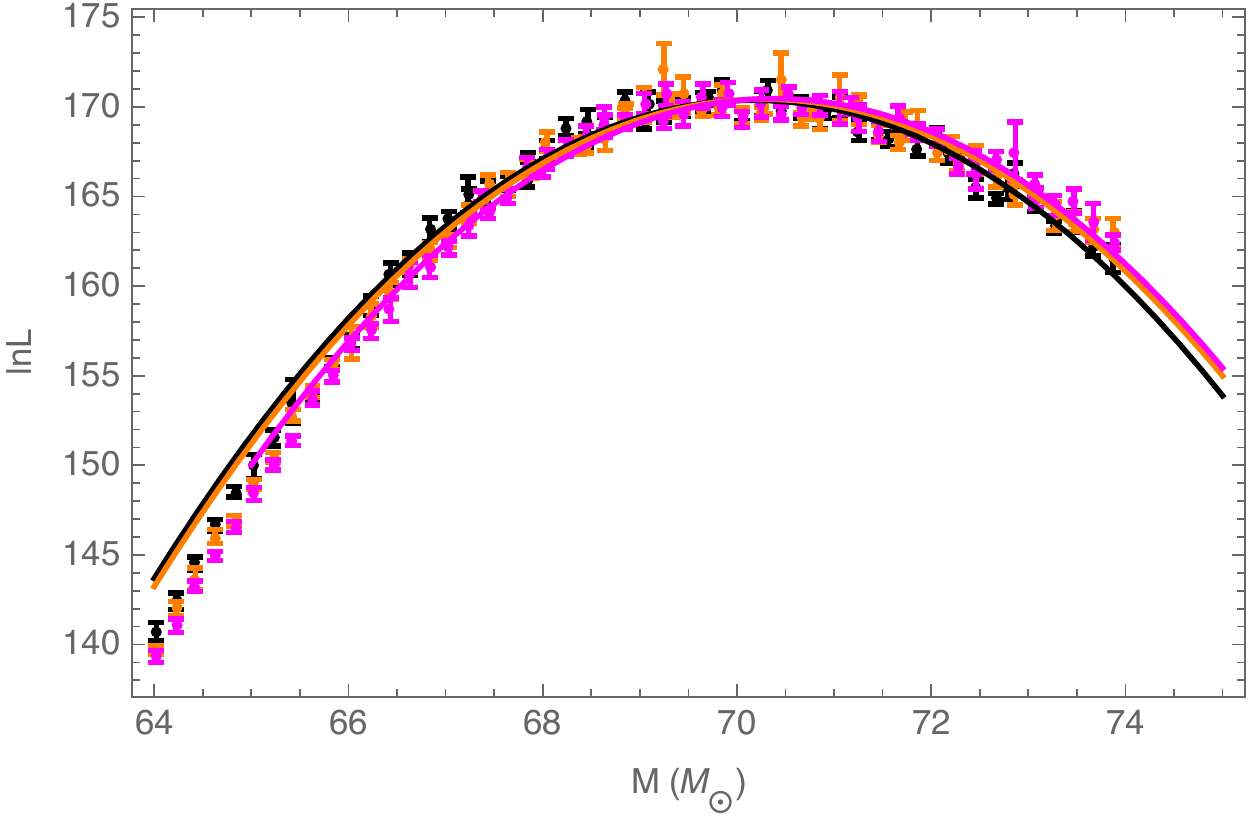}
\includegraphics[width=\columnwidth]{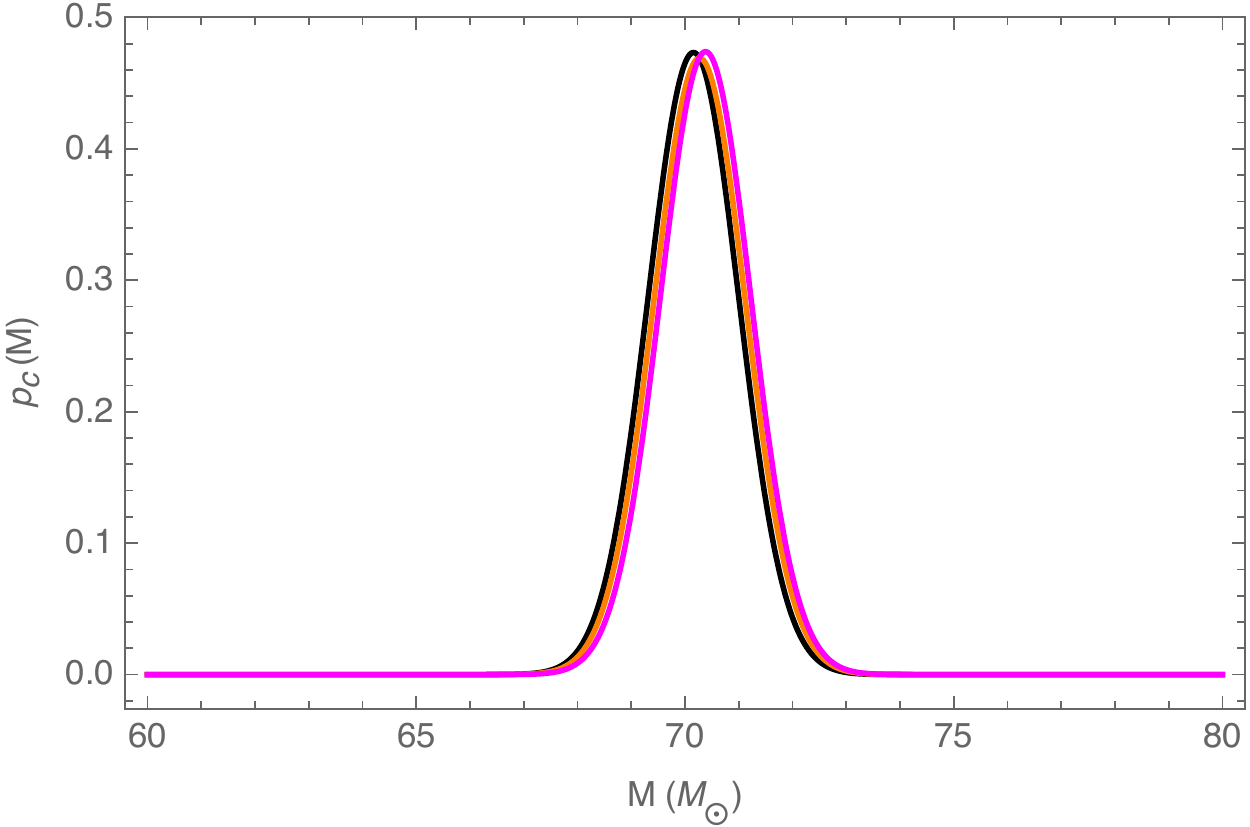}
\caption{\label{fig:rextr2}
\textbf{Propagating systematic error from finite extraction radius into posterior distributions }:  This figure shows
how small systematic errors from finite NR extraction radius propagate into parameter estimation posterior
distributions, by concrete example.
\emph{Left panel}: A plot of $\lnLmarg$ versus total mass.  In all cases, the source is RIT-1a at $r=190 M$; the
templates are also RIT-1a, using different extraction radii as templates.   Here,  magenta is $r=141.71M$, orange is $r=162.34 M$,
and black is $r=190 M$. We focus our search on only the last few extraction radii to avoid clutter.
 The error is relatively small but bigger than what our match study naively suggests (i.e., changes in
$\lnL$ of order $10^{-4}\rho^2/2 \simeq 2\times 10^{-2}$, though this result only applies to the change in the peak
value, which is indeed changes by less than than amount).  
\emph{Right panel}: One-dimensional posterior distributions $p_c(M)$  of each individual fit derived
from the three plots [see Eq. (\ref{eq:1d})].  Even though there are small differences, these PDFs are virtually
identical.
}
\end{figure*}

\begin{table}
\centering
\begin{tabular}{l|lccc}
Extraction Radius $(M)$ & $D_{KL}$ & CI (90\%)\\
190M/190M & 0 & (68.8 - 71.5)\\
162.34/190M & 9.3e-3 & (68.9 - 71.5)\\
141.71/190M & 3.6e-2 & (69.0 - 71.8)\\
\end{tabular}
\caption{\textbf{KL Divergence and 90\% CI between PDFs with different extraction radii}: This table shows the $D_{KL}$, calculated using Eq. (\ref{eq:dkl}) and 90\% CI for PDFs with three different extraction radii.  The $D_{KL}$ was calculated comparing the 1D distributions to the PDF with $r=190 M$ (notice its $D_{KL}$ is zero i.e. they're identical).  The CI also given to show the change between them.  Based on the $D_{KL}$ results, the 1D posteriors show some differences but are very similar.}
\label{tab:rextr}
\end{table}
To assess the observational impact of waveform extraction systematics, we evaluate $\lnLmarg(M)$ and $p_c(M)$ using
waveform estimates produced using different extraction radii.  Specifically, we take a simulation; use its large-radius
perturbative estimate as a source; and follow the procedures used in Figures \ref{fig:Ex1} and \ref{fig:Ex2} to produce
$\lnLmarg(M)$ and $p_c(M)$.   Figure \ref{fig:rextr2} shows our results; for clarity, we include only the last three
extraction radii  ($r=190M, 162M, 141M$).
The errors here are relatively small but bigger than expected from our match study; however, the error shown in the match only applies to changes in the peak value $\lnLmarg$, which can be seen in the left panel.
To again quantify these small differences, we use $D_{KL}$ and CI, as reported in Table
\ref{tab:rextr}.  As this table shows, the error introduced is insignificant as long as we pick a relative large extraction radius. This is
almost always the case for the current simulations available.  Some of the GT simulations require us to chose a lower
extraction radius due to an increase in the error as  the extraction radius increases beyond a certain point, but this does not affect our overall results.
\begin{figure*}[!ht]
\includegraphics[width=\columnwidth]{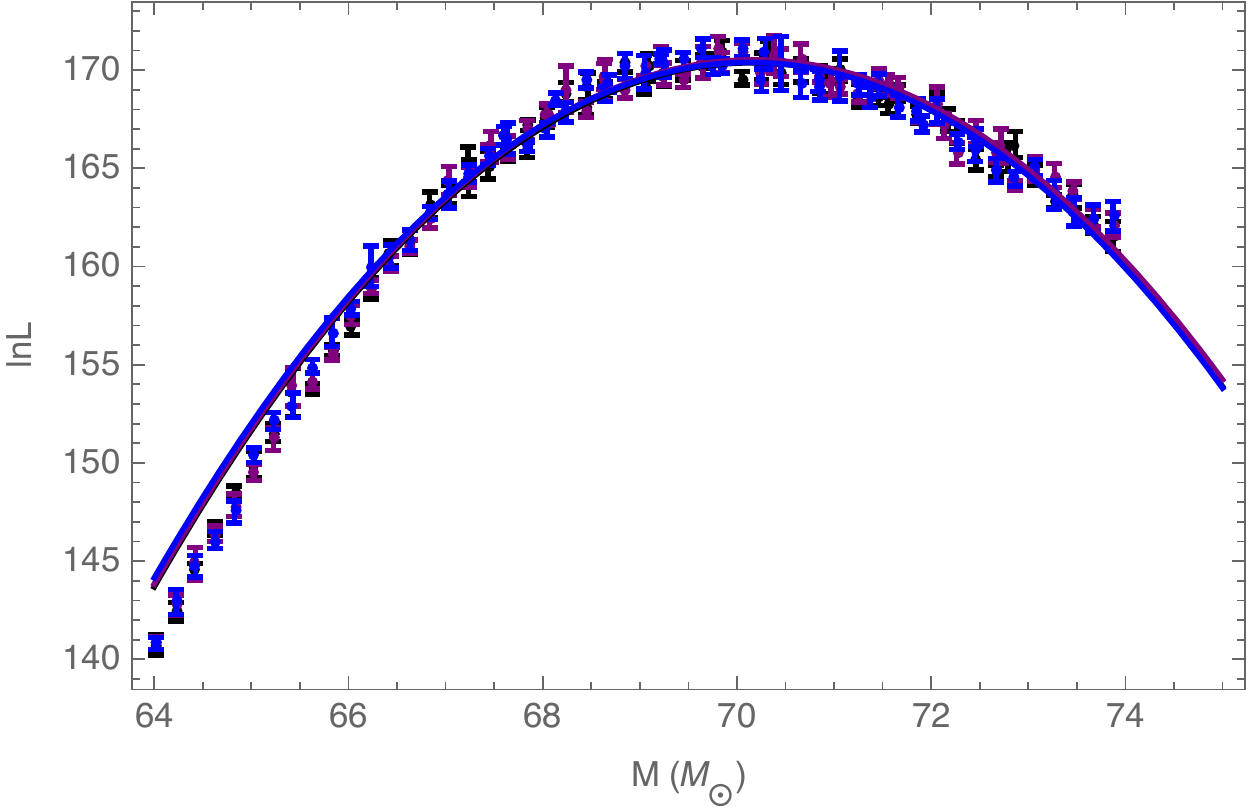}
\includegraphics[width=\columnwidth]{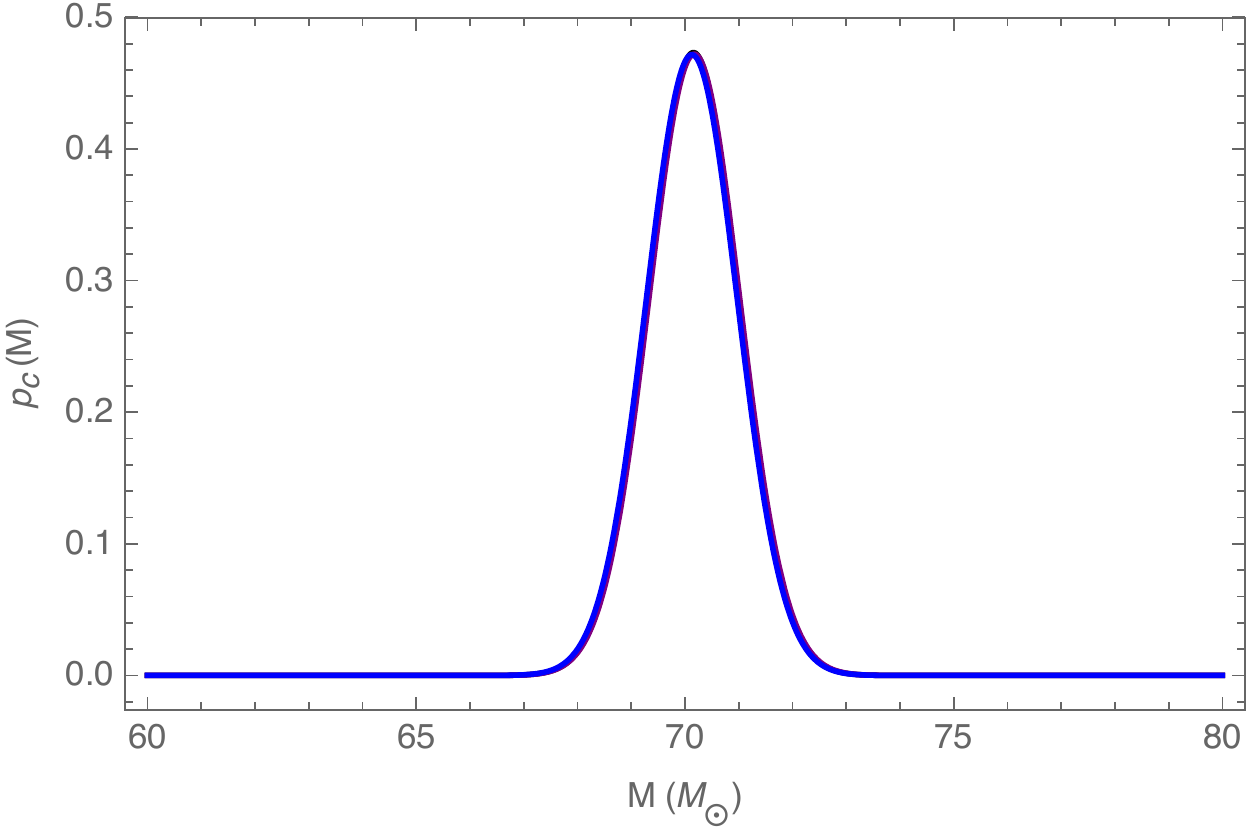}
\caption{\textbf{Single runs of \textit{ILE} with changing resolution and their corresponding PDFs}: The left panel consists of $\ln {\cal L}$ vs total mass curves with different numerical resolution.  Here we use RIT-1a as the source and compare it to simulations with the same parameters at different resolutions, specifically RIT-1b and RIT-1c.  The results were evaluated with $f_{\rm min}=30 \unit{Hz}$ at a total mass $M=70 M_{\odot}$ with a inclination $\imath=0.785$.  Here black is n120, purple is n110, and blue is n100.  Even though the error is clearly minuscule, we convert the fits to a PDFs for completeness.  The right panel shows the PDFs for the three different resolutions [see Eq. (\ref{eq:1d})]. It is clear that these are all the same PDFs, and the error introduced by different resolutions is irrelevant.}
\label{fig:res}
\end{figure*}

\subsection{Impact of simulation resolution}

\begin{table}
\centering
\begin{tabular}{l|lccc}
NR Label & Resolution & Mismatch\\
RIT-1a/RIT-1a & n120/n120 & 0.0\\
RIT-1b/RIT-1a & n110/n120 & 3.90e-5\\
RIT-1c/RIT-1a & n100/n120 & 5.27e-5\\
\end{tabular}
\caption{\textbf{Mismatch between waveforms with different numerical resolutions}: Here is a mismatch study between the different resolutions for one NR simulation.  Specifically RIT-1a vs RIT-1a, RIT-1a vs RIT-1b, and RIT-1a vs RIT-1c. The results were evaluated at $M=70 M_{\odot}$ and $\imath=0.785$. The mismatch between the different resolution is very small and is much smaller than our accuracy requirement.  We therefore expect the error introduced to be negligible.}
\label{tab:res}
\end{table}

Here we analyze errors introduced by different numerical resolutions.  Higher resolutions simulations take longer to run and computationally cost more than lower resolution ones.  If the effects of different resolutions are insignificant, numerical relativist will be able to run at a lower resolution while not introducing any systematic errors.  Table \ref{tab:res} shows a match comparison between the highest resolution RIT-1a and the two lower ones, RIT-1b and RIT-1c.  The mismatches are orders of magnitudes better than our accuracy requirement ($\sim10^{-2.8}$), and therefore introduce errors that are negligible.

Using $\lnLmarg$ as our diagnostic to compare these three simulations, we draw similar conclusions; see Figure
\ref{fig:res}.    We again see a error so small that changes between the three curves are
almost impossible to see, even far from the peak.  Table \ref{tab:res2} quantifies these extremely small differences.
In short,  different resolutions have no noticeable impact on our conclusions.  While this resolution study was only done
for a aligned RIT simulation, similar conclusions are expected when a wider range of simulations are used.

\begin{table}
\centering
\begin{tabular}{l|lccc}
Resolution $(M)$ & $D_{KL}$ & CI (90\%)\\
n120/n120 & 0 & (68.8 - 71.5)\\
n110/n120 & 2.0e-4 & (68.8 - 71.6)\\
n100/n120 & 6.5e-4 & (68.7 - 71.5)\\
\end{tabular}
\caption{\textbf{KL Divergence and 90\% CI between PDFs with different numerical resolution}: This table shows the $D_{KL}$, calculated using Eq. (\ref{eq:dkl}), and 90\% CI for PDFs with the three different resolutions for RIT-1a.  The $D_{KL}$ was calculated comparing the 1D distributions to the PDF with n120 (notice its $D_{KL}$ is zero i.e. they're identical).  The confidence intervals also given to show the change between them.  Based on the $D_{KL}$ results, the 1D posteriors are identical.}
\label{tab:res2}
\end{table}

Even though in this case the mismatch and \textit{ILE} studies show conclusively the minimal impact the numerical
resolution has on the waveform, we generate 1D distributions from the fits for completeness.  It is not surprising to
see in the right panel of Figure \ref{fig:res} the posteriors from the three fits match almost exactly.  To quantify
this similarity, we calculate $D_{KL}$ as well as the CI for the corresponding PDFs.  Based on the
$D_{KL}$, these distributions are clearly identical and using different resolutions does not effect the waveform in any
significant way.  This resolution study was only done for an aligned RIT simulation;  while extraction radius studies
have been performed for SXS for other extraction procedures \cite{2015arXiv150100918C}, a similar resolution investigation
needs to be done for SXS simulations for this extraction method. We hypothesize that this effect will also be minimal. 

\subsection{Impact of low frequency content and simulation duration}
\label{sub:DurationAndLowF}

\begin{figure*}[!ht]
\includegraphics[width=\columnwidth]{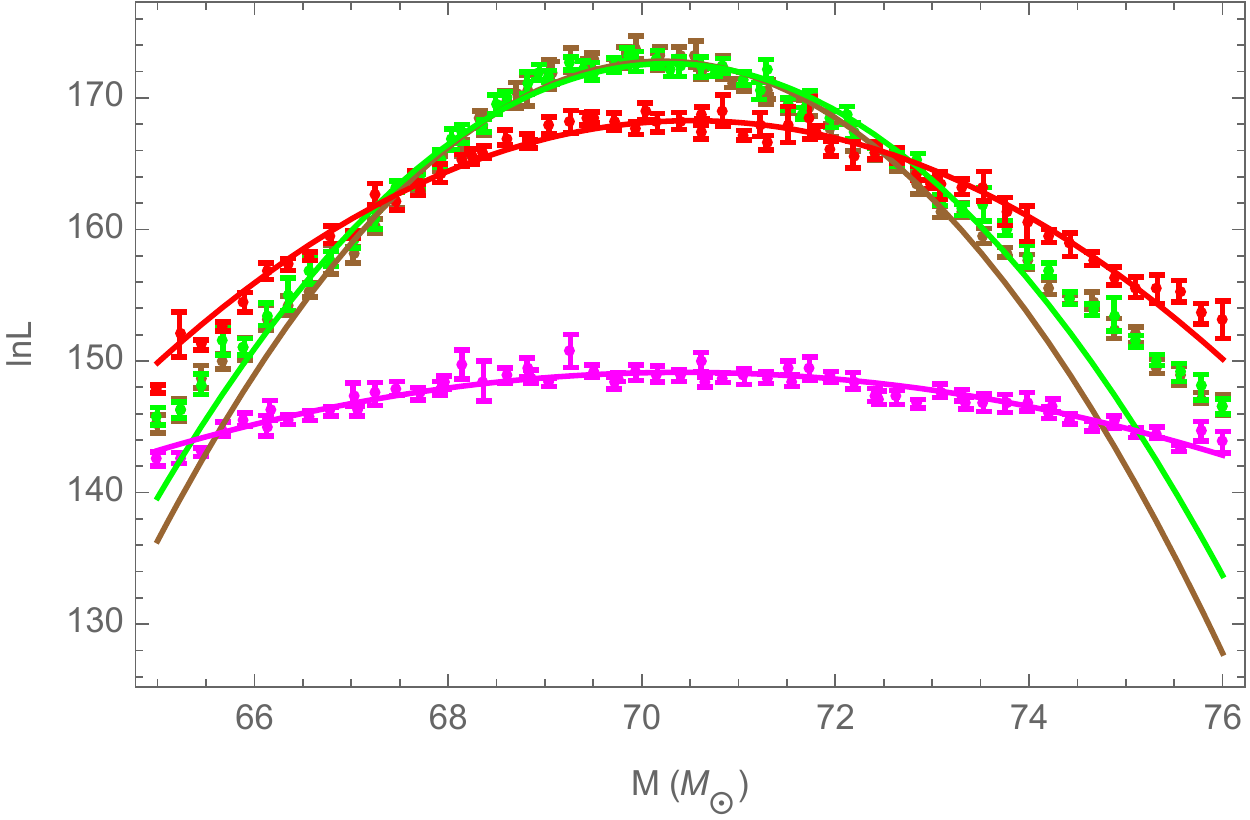}
\includegraphics[width=\columnwidth]{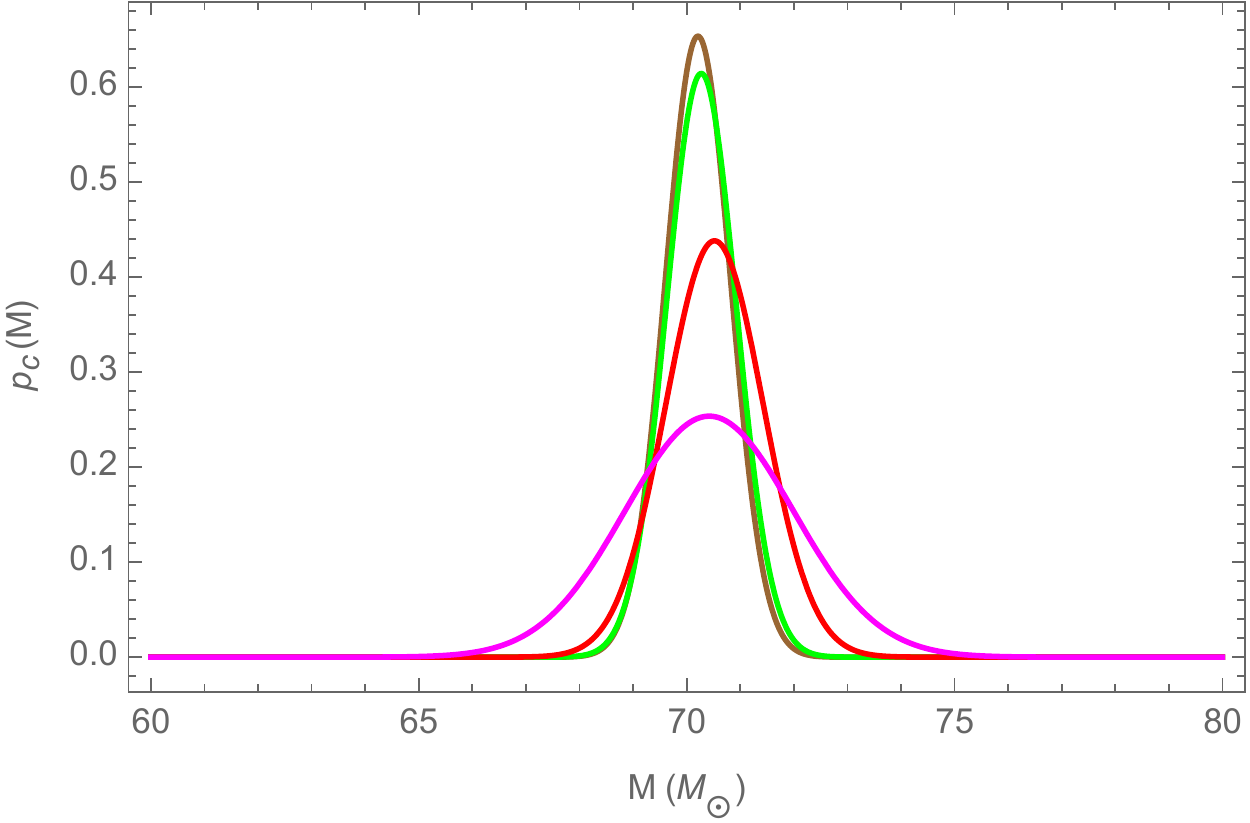}
\includegraphics[width=\columnwidth]{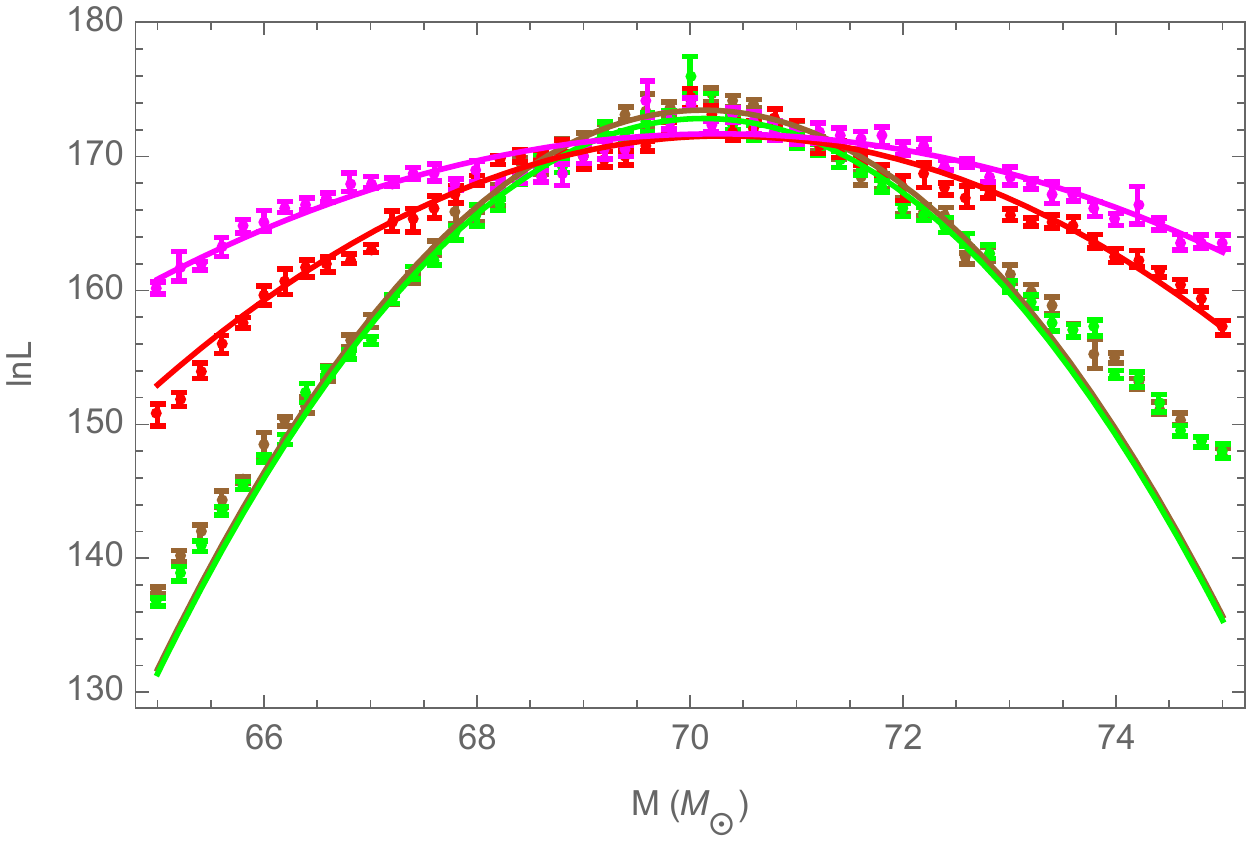}
\includegraphics[width=\columnwidth]{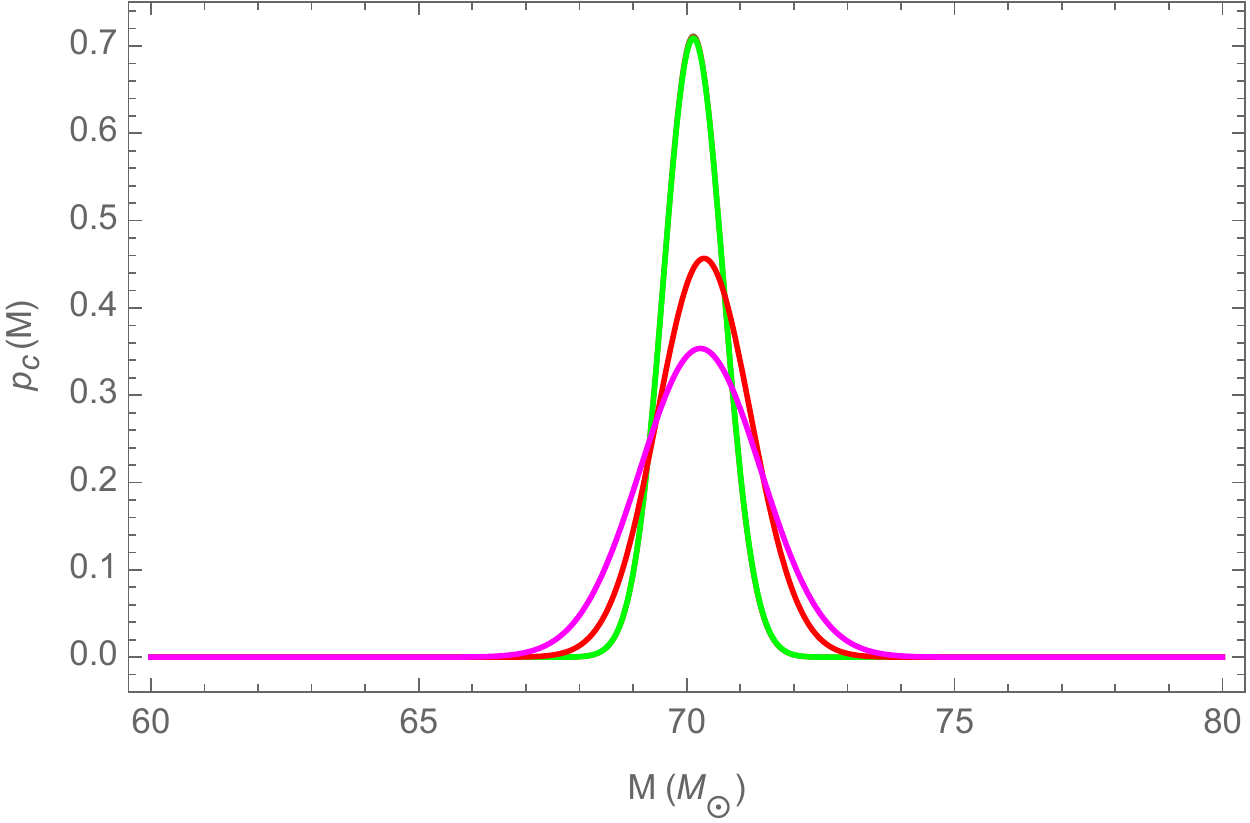}
\caption{\textbf{Assessing the impact of low frequency cutoff 2: Consistent cutoff choices}: 
  Revisiting the investigations shown in Figure \ref{fig:Ex3}, this figure uses (in the top panels) comparisons of an NR
  simulation  to itself as a method to isolate the impact of $f_{\rm min}$ (the low-frequency cutoff appearing in the
  likelihood).  The bottom panels repeat a comparable analysis using SEOB.
\emph{Top left panel}: Plots  different $\lnLmarg$ vs total mass curves with different $f_{\rm min}$.  Here we compared a RIT-4
source with a duration of 5.0 Hz source compared to itself at different $f_{\rm min}$ values.  Specifically brown has a
$f_{\rm min}=10$, green has a $f_{\rm min}=20$, red has a $f_{\rm min}=30$, and magenta has a $f_{\rm min}=40$.  These results are
similar to the SEOBNRv2 case in Figure \ref{fig:Ex3}.  As the cutoff increases, our $\lnL$ curve becomes wider, and the
peak value $\lnLmarg$ is lower. 
\emph{Top right panel}: One-dimensional posteriors $p_c(M)$ [Eq. (\ref{eq:1d})].  
This figure qualitatively resembles Figure \ref{fig:Ex3}; however, unlike the previous analysis, while the posterior is
wider (i.e., less informative), no significant bias is introduced by the low-frequency cutoff. 
\emph{Bottom left panel}: Similar to prior figures, a plot of $\lnLmarg(M)$, evaluated using SEOBNRv2.  In this comparison,
the  SEOBNRv2 source with a certain duration was compared to a SEOBNRv2 template with the same $f_{\rm min}$.  Specifically
brown has a $f_{\rm min}=10$, green has a $f_{\rm min}=20$, red has a $f_{\rm min}=30$, and magenta has a $f_{\rm min}=40$.   As the
cutoff increases, our $\lnL$ curve becomes wider.
\emph{Bottom right panel}:  The  corresponding PDFs to the fits [see Eq. (\ref{eq:1d})].  We again see similarities between this case and Figure \ref{fig:Ex3} minus the shift in total mass with increasing $f_{\rm min}$.
}
\label{fig:FF}
\label{fig:SEOB}
\end{figure*}

\begin{table}
\centering
\begin{tabular}{l|lccc}
$f_{\rm min}$ for \textit{ILE} run (Hz) & $D_{KL}$ & CI (90\%)\\
10/10 & 0.0 & (69.2 - 71.2)\\
20/10 & 9.2e-3 & (69.2 - 71.3)\\
30/10 & 0.34 & (69.0 - 72.0)\\
40/10 & 1.9 & (67.8 - 73.0)\\
\end{tabular}
\caption{\textbf{KL Divergence and 90\% CI of PDFs derived from RIT-4 sources with different low frequency cutoffs}: This table shows the $D_{KL}$ and 90\% CI for the four different configurations using a RIT-4 source with a set duration of $5 \unit{Hz}$ and compared against RIT-4 templates with different low frequency cutoffs.  The $D_{KL}$ was calculated comparing the 1D distributions to the $f_{\rm min}=10 \unit{Hz}$ case (notice its $D_{KL}$ is zero i.e. they're identical).  The CI also given to show the change between them.  Based on the $D_{KL}$ results, the 1D posteriors of $f_{\rm min}=10, 20 \unit{Hz}$ seem to be the same distribution; however, they differ significantly to $f_{\rm min}=30, 40 \unit{Hz}$.}
\label{tab:FF}
\end{table}

As demonstrated by Example 3 in Section \ref{sub:Ex3} above, the available frequency content provided by each simulation and used to the interpret the
data can significantly impact our interpretation of results.  In this section, we perform a more systematic analysis of
simulation duration and frequency content, again using the semi-analytic SEOBNRv2 model as a concrete waveform available
at all necessary durations.  Before we begin, we first carefully distinguish between two unrelated ``minimum frequencies'' that naturally
  show up in our analysis.  It is easy to get confused between the low frequency cutoff (in this work called $f_{\rm
    min}$) and simulation duration (or initial orbital frequency $M\omega_{0}$).  The simulation duration is the true
  duration of the simulation, which is a property of the binary and can be drastically different over many NR
  simulations.  The low frequency cutoff is an artificial cut to the signal that allows us to normalize the signal
  duration of all our waveforms.    As a result, with a lower $f_{\rm min}$, more of the NR simulation enters into our
  analysis. %

The top panels of Figure \ref{fig:FF} shows the result of compare a RIT-4 source with a duration of 5.0 Hz to itself with changing $f_{\rm min}$.  As $f_{\rm min}$ increases, a smaller portion of the simulation waveform is being used to analyze the data.  When $f_{\rm min}$ is high, we end up cutting off more of the waveform.  This results in a sharp decline in  $\lnLmarg$ since one is now comparing less of the waveform to itself.  In this panel it is clear that $f_{\rm min}\sim10-20\unit{Hz}$ seems to not significantly affect $\lnLmarg$; however, the curve changes drastically when $f_{\rm min}=30-40\unit{Hz}$.  For completeness Table \ref{tab:FF} shows the corresponding $D_{KL}$ and CI for different $f_{\rm min}$, again showing the similarities between the $f_{\rm min}=10,20 \unit{Hz}$ frequencies and the differences of the higher frequencies.  Hybrid NR waveforms will nullify this source of error by allowing us to compare more of the waveform while at the same time allowing us to standardize durations.

\begin{table}
\centering
\begin{tabular}{l|lccc}
$f_{\rm min}$ for \textit{ILE} run (Hz) & $D_{KL}$ & CI (90\%)\\
10/10 & 0.0 & (69.2 - 71.0)\\
20/10 & 1.7e-5 & (69.2 - 71.1)\\
30/10 & 0.33 & (68.9 - 71.8)\\
40/10 & 0.85 & (68.4 - 72.1)\\
\end{tabular}
\caption{\textbf{KL Divergence and 90\% CI of PDFs derived from SEOB sources}: This table shows the $D_{KL}$ and 90\% CI for the four different configurations using a SEOB source compared against SEOB templates with the same duration/$f_{\rm min}$ (i.e. if the source has a duration of 10 Hz, the template has a $f_{\rm min}=10\unit{Hz}$).  The $D_{KL}$ was calculated comparing the 1D distributions to the $f_{\rm min}=10 \unit{Hz}$ case (notice its $D_{KL}$ is zero i.e. they're identical).  The CI also given to show the change between them.  Based on the $D_{KL}$ results, the 1D posteriors of $f_{\rm min}=10, 20 \unit{Hz}$ seem to be the same distribution; however, they differ significantly to $f_{\rm min}=30, 40 \unit{Hz}$.}
\label{tab:SEOB}
\end{table}
To investigate the shift in mass seen in Figure \ref{fig:Ex3} further, we compare a SEOBNRv2 source to a SEOBNRv2 template with the same duration/$f_{\rm min}$ (i.e. the source has a duration of 10 Hz therefore the template has a $f_{\rm min}=10 \unit{Hz}$).  This was done to investigate the shift in total mass seen in Figure \ref{fig:Ex3} for a SEOBNRv2 source with a fixed duration compared to a SEOBNRv2 template with different low frequency cutoffs.  As the bottom panels of Figure \ref{fig:SEOB} now show, this shift was a product of comparing a source and templates with different signal lengths.  When we now set the same duration for the source and $f_{\rm min}$ for the template, the \textit{ILE} results and their corresponding PDFs peak around the same mass point.  We still see a widening of the curves with increasing $f_{\rm min}$; this corresponds to a wider and shorter PDF.  We calculate $D_{KL}$ and CI for this case as well, see Table \ref{tab:SEOB}.  These values shows that $f_{\rm min}=10,20\unit{Hz}$ are relatively similar while the higher frequencies are significantly different.

\section{ Reconstructing properties of synthetic data I: Zero, Aligned, and Precessing spin}
\label{sec:aligned}
This section is dedicated to end-to-end demonstrations of this parameter estimation technique.   Unless otherwise specified, we adopt a total binary mass of $M=70 M_\odot$ and use the fiducial early-O1 PSD \cite{PEPaper} to qualitatively reproduce the characteristic features of data analysis for GW150914.   Without loss of generality and consistent with common practice, we adopt a ``zero noise" realization (i.e., the data used for each instrument is equal to its expected response to our synthetic source).  Table \ref{tab:simulations} is a list of simulations we have used as sources in our end-to-end runs; these include zero, aligned, and precessing systems all at different inclinations. Here we start with a end-to-end demonstration with zero spin from SXS.

\subsection{Zero Spin: A fiducial example demonstrating the method's validity}

\begin{figure*}
\includegraphics[width=\columnwidth]{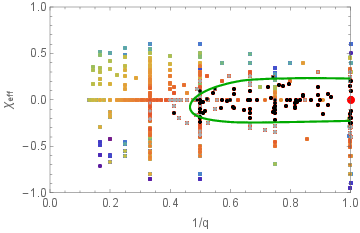}
\includegraphics[width=\columnwidth]{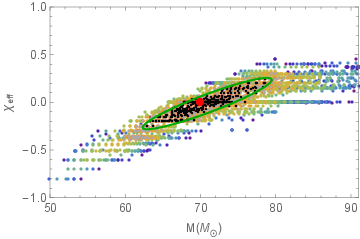}
\caption{\label{fig:ZeroSpinBasicResults-I}\textbf{Parameter recovery for zero spin equal mass binary I}: Each point
  represents a NR simulation and a particular total mass compared against a SXS-1 source.  The left panel shows $\chi_{\rm eff}$ vs 1/q with q=$m_{1}/m_{2}$
  and $\chi_{\rm eff}$ defined in Eq. (\ref{eq:chieff}), and the right panel shows $\chi_{\rm eff}$ vs $M$.  The gray points
  represent points that fall between $\lnLmarg=130$ and $\lnLmarg=127$.  The black points represent points that fall in
  $\lnLmarg>130$, i.e. templates that best match the source.  The peak value with this run was $\lnLmarg=134$.  These intervals were determined using the inverse $\chi^{2}$ distribution (see Eq. \ref{eq:chi}).  The rest of the colors represent all the points $\lnLmarg<127$ with the red represent the highest in the region.  The green contour is the 90\% CI derived using the quadratic fit to $\lnLmarg$ for nonprecessing systems only.  The big red dot represents the true parameters of the source.  We are able to recover the 2D posterior distribution that is consistent with the distributions with $\lnLmarg>130$ (black points).}
\end{figure*}

\begin{figure*}
\includegraphics[width=\columnwidth]{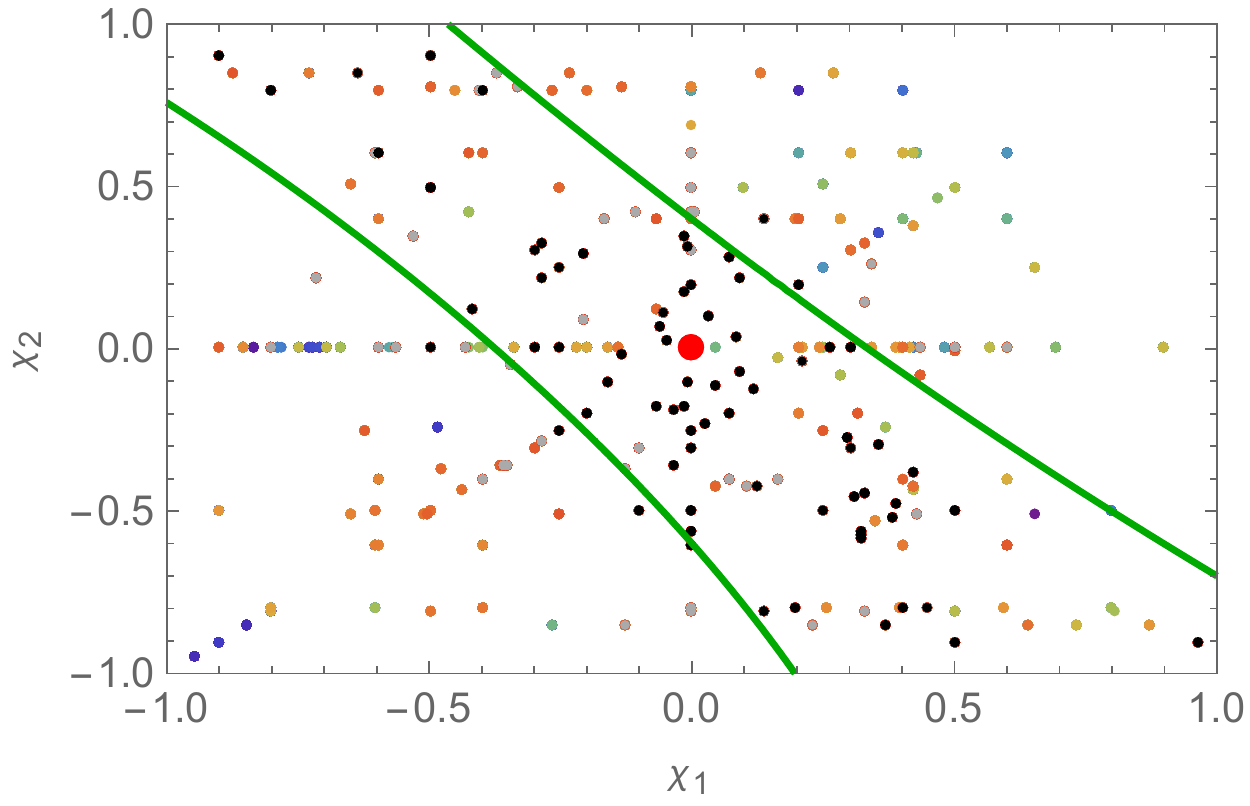}
\includegraphics[width=\columnwidth]{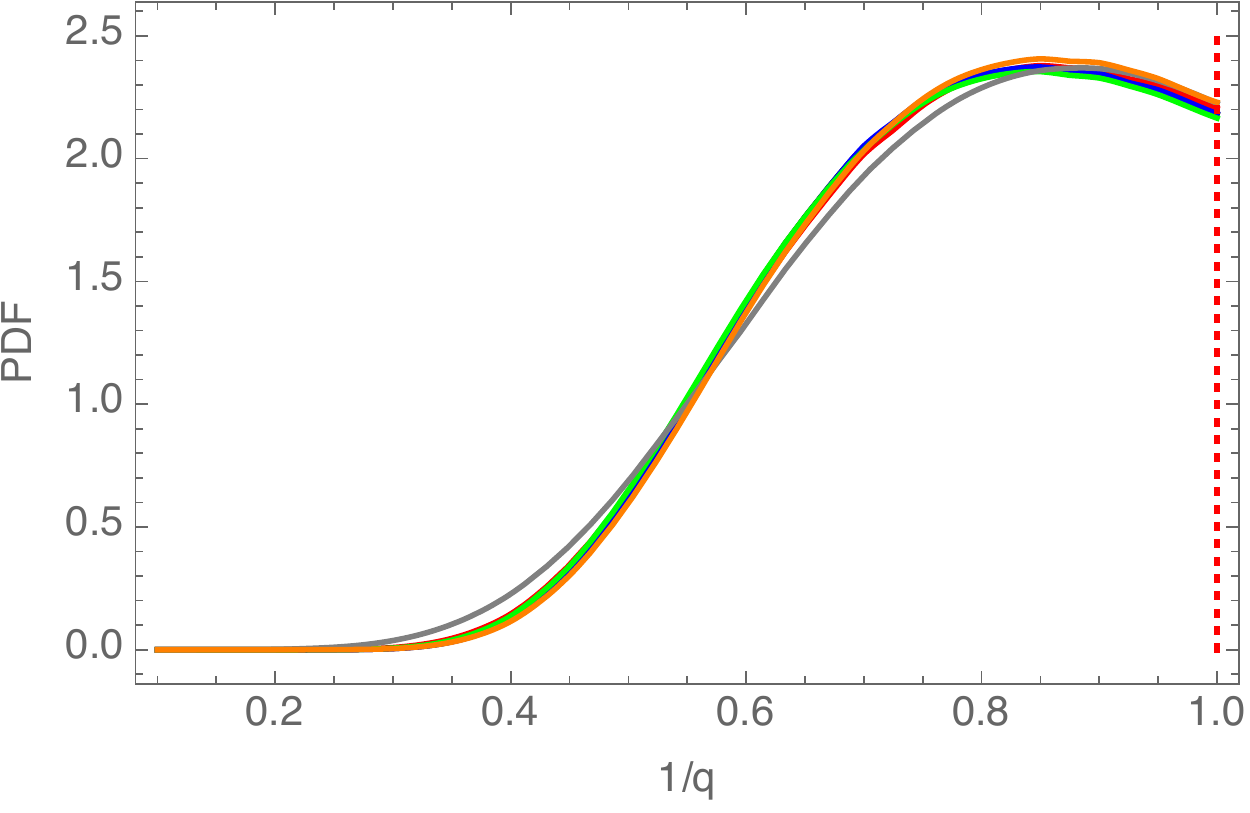}
\caption{\label{fig:ZeroSpinBasicResults-II}\textbf{Parameter recovery for zero spin equal mass binary II}: The left panel shows the $\lnLmarg$ as a
  function of $\chi_{1z}$ and $\chi_{2z}$.  The rainbow, gray, and black points represent the same intervals as in Figure
  \ref{fig:ZeroSpinBasicResults-I}.  The green contour also represents the same CI as Figure
  \ref{fig:ZeroSpinBasicResults-I}.  The right panel shows the 1D posterior distribution for $1/q$.  This 1D posterior
  was derived from the quadratic fit of to $\lnLmarg$ for nonprecessing systems only.  Here we show results for six
  inclinations: $\imath=0.0$ (black), $\imath=0.5$ (red), $\imath=0.785$ (blue), $\imath=1.0$ (green), $\imath=1.5$
  (gray), $\imath=2.35$ (orange).  We see that the results from all the inclinations are the same,
  i.e. no more information can be obtained with higher order modes.}
\end{figure*}
We first illustrate the simplest possible and most-well-studied scenario: a compact binary with zero spin and equal mass, as represented here by SXS-1.  To enable comparison with other cases where higher-order modes will be more significant, we adopt inclinations $\imath=0,0.5,0.785,1.0,1.5,2.35$.  For the purposes of illustration, we present our end-to-end plots using an inclination $\imath=0$.

The left panel of Figure \ref{fig:ZeroSpinBasicResults-I} shows $\chi_{\rm eff}$ vs 1/q; the points represent the maximum log likelihood $\lnLmarg$ of all the different \textit{ILE} runs across parameter space.  The green contour is the 90\% CI derived using the quadratic fit to $\lnLmarg$ for nonprecessing systems only.  The colored points represent points that fall in  $\lnLmarg<127$ region with the red points representing higher $\lnLmarg$ and violet represent lower $\lnLmarg$.  The gray points represent points that fall between $\lnLmarg=130$ and $\lnLmarg=127$.  The black points represent points that fall in $\lnLmarg>130$.  These intervals were determined using the inverse $\chi^{2}$ distribution [see Eq. (\ref{eq:chi})] adopting $d=4$ (two masses with aligned spin) for the black points and $d=8$ (two masses with precessing spins).  This CI is consistent with the point distribution $\lnLmarg>130$ (i.e. black points), which represents the points closest to the maximum.  The right panel of Figure \ref{fig:ZeroSpinBasicResults-I} shows the $\chi_{\rm eff}$ vs $M$ with the same green contour and black point distribution.  As with the left panel, the green contour is consistent with the black point distribution.  Both plots recover the true parameters (indicated by the big red dot) with regards to the confidence interval and the black point distributions.

The left panel of Figure \ref{fig:ZeroSpinBasicResults-II} shows the $\chi_{1z}$ vs $\chi_{2z}$ where $\chi_{1z,2z}$ is
the z component of the dimensionless spin [see Eq. (\ref{eq:chii})].  All the colors here represent the same as in Figure \ref{fig:ZeroSpinBasicResults-I}.  We again see that the green contour is consistent with the black point distribution.  The right panel of Figure \ref{fig:ZeroSpinBasicResults-II} shows the 1D posteriors for $1/q$ for six different inclinations.
These produce distributions we expect to see; all the curves from the different inclinations lie on top of each other.
This implies that higher order modes for this particular case are not expected to provide any extra information.  By construction, this source needs no higher order modes to completely recover the parameters.  Since all inclinations have the same distribution shape, the results here are independent of inclination at a fixed SNR.

\subsection{Nonprecessing binaries: unequal mass ratios and aligned spin}

\begin{figure*}
\includegraphics[width=\columnwidth]{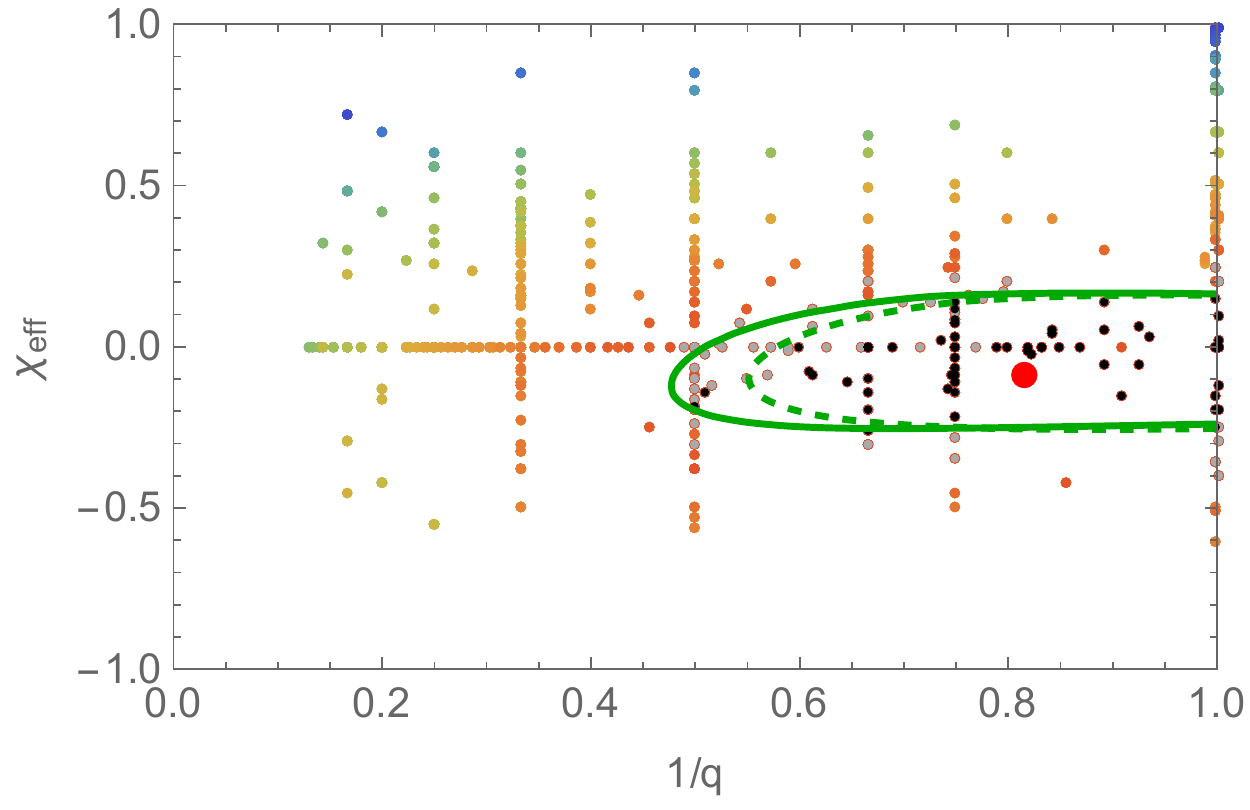}
\includegraphics[width=\columnwidth]{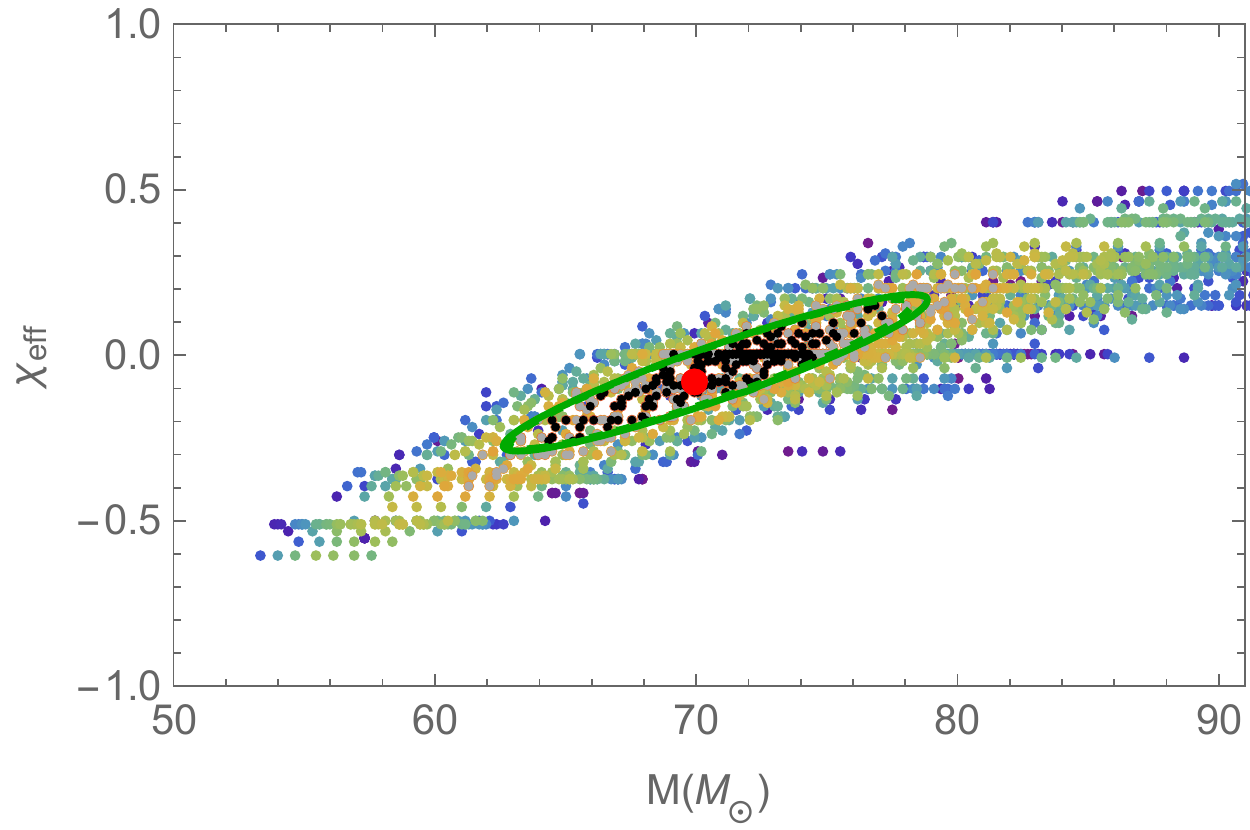}
\caption{\label{fig:0233BasicResults-I}\textbf{Parameter recovery for an aligned, GW150914-like unequal mass binary I}:
  Each point represents a NR simulation and a particular total mass compared against a SXS-0233 source.  The left panel shows $\chi_{\rm eff}$ vs 1/q with
  q=$m_{1}/m_{2}$, and the right panel shows $\chi_{\rm eff}$ vs M with $\chi_{\rm eff}$ defined in
  Eq. (\ref{eq:chieff}).  The gray points represent points that fall between $\lnLmarg=167$ and $\lnLmarg=165$.  The
  black points represent points that fall in $\lnLmarg>167$, i.e. templates that best match the source.  The rest of the
  colors represent all the points $\lnLmarg<165$ with the red represent the highest in the region.  The green contours
  are the 90\% CI derived using the quadratic fit to $\lnLmarg$ for nonprecessing systems only.  The
  dash line is the CI for $l\le 3$, and the solid line is the CI for $l\le2$.  The big
  red dot represents the true parameters of the source.  We are able to better constrain the posterior by using higher modes for this system.}
\end{figure*}

\begin{figure*}
\includegraphics[width=\columnwidth]{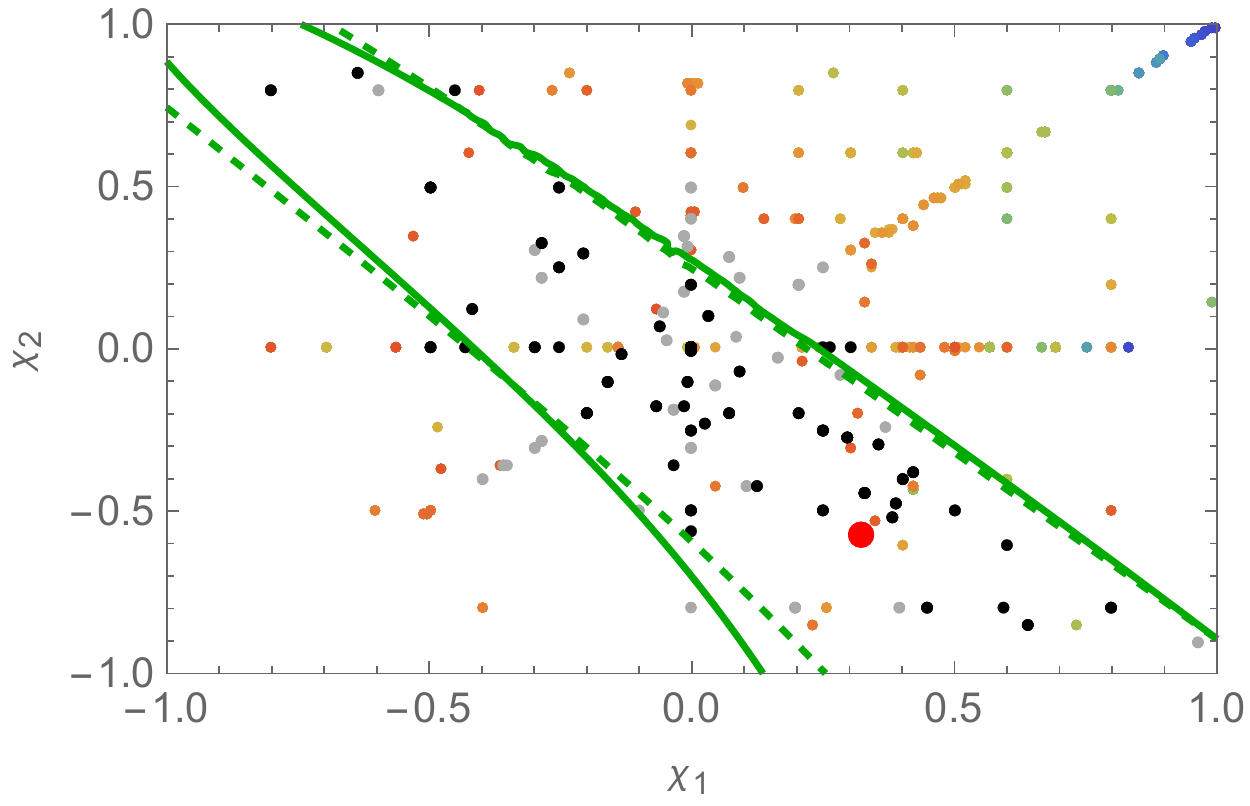}
\includegraphics[width=\columnwidth]{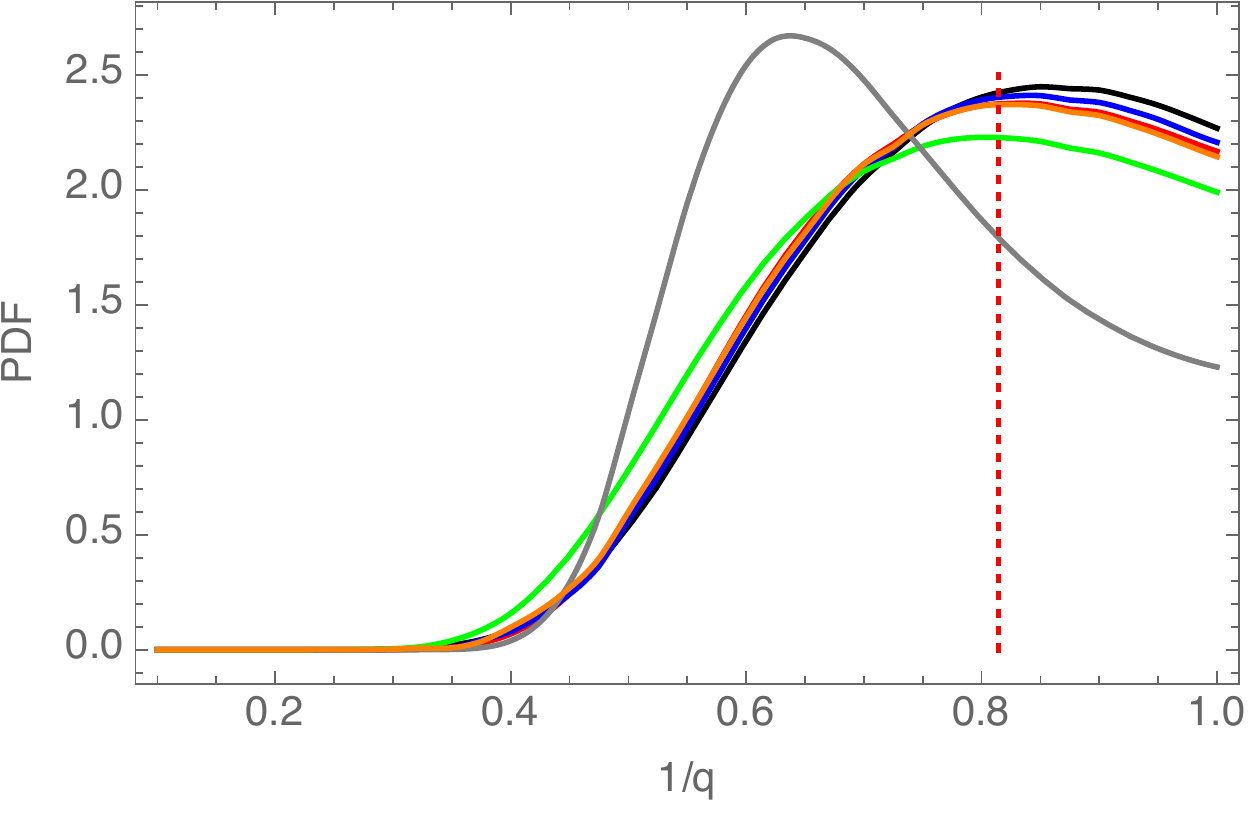}
\caption{\label{fig:0233BasicResults-II}\textbf{Parameter recovery for an aligned, GW150914-like unequal binary II}: The
  left panel shows the $\lnLmarg$ as a function of $\chi_{1z}$ and $\chi_{2z}$.  The colored, gray, and black points represent the
  same intervals as in Figure \ref{fig:0233BasicResults-I}. The green contours also represents the same CI as Figure \ref{fig:ZeroSpinBasicResults-I}.  The big red dot represents the true parameters of the source.  The right panel shows the 1D posterior distribution for 1/q.  This 1D posterior was derived from the quadratic fit of to $\lnLmarg$ for nonprecessing systems only.  Here we show results for six inclinations all represented by the same colors as the zero spin case, see Figure \ref{fig:ZeroSpinBasicResults-II}.  In this case, we see significant differences between the curves implying that higher order modes could important for accurate analysis of this source.}
\end{figure*}

In the previous zero spin case, the higher order modes had a minimal impact.  Now we introduce an aligned spin
GW150914-like simulation as the source, SXS-0233.  For our total mass of $M=70 M_{\odot}$, we expect that the impact of higher order modes border on being significant.  Because of this, we did 2 end-to-end runs
with SXS-0233: one with $l\le2$ and the other with $l\le 3$.  The panels in Figure \ref{fig:0233BasicResults-I} are the same
type of plots as in the previous case; however, we have also included a contour representing the 90\% CI for $l\le 3$ (green dashed line).  In the left panel of Figure \ref{fig:0233BasicResults-I}, the posterior
corresponding to $l\le3$ better constrains the mass ratio than that of the posterior corresponding to $l\le2$.  In this
case, including higher order modes provides more information about the mass ratio, allowing us to constrain it more tightly. The right
panel of Figure \ref{fig:0233BasicResults-I} is the same type of plot as the bottom panel of Figure \ref{fig:ZeroSpinBasicResults-I}; however, this includes the results from the $l\le3$ runs.  Since the $\lnLmarg$ was higher, the number of black and gray points slightly decreased.  It is clear from these
two plots that higher order modes are significant and need to be included for this source to get the best possible
constrains on the parameters.  The right panel in Figure \ref{fig:0233BasicResults-I} shows the $\chi_{\rm eff}$ vs $M$; these show little difference between the $l\le2$ and the $l\le 3$ contours.  The contours agree very well with each as well as the black points' distribution in both panels of Figure \ref{fig:0233BasicResults-I}.  We recover the true parameters in both plots and with $l\le2$ and $l\le3$; however, we can better constrain q with
higher order modes.

As with the zero spin case, we plot  $\lnLmarg$ as a function of $\chi_{1z}$ and $\chi_{2z}$ in the left panel in Figure
\ref{fig:0233BasicResults-II}.  Here again the dashed and solid green contour represents the confidence interval for
$l\le2$ and $l\le3$ respectively and are largely consistent with each other.  The right panel of Figure \ref{fig:0233BasicResults-II} shows the 1D distributions for 1/q for different inclination values.  The difference in the curves here could be explained by higher order modes; however, more needs to be done to corroborate this hypothesis.

In this particular case, higher order modes have a relatively modest impact on the
  posterior.  The minimal impact is by design: moving away from zero spin and equal mass within the posterior of
  GW150914, we have explicitly selected a point in parameter space where higher-order
  modes have just become marginally significant.   Even remaining within the posterior of GW150914, as we move towards
  more extreme antisymmetric spins and mass ratios, higher-order modes can play an increasingly significant role.  We
  will address this issue further in subsequent work.

\subsection{Precessing binaries: unequal mass ratios and precessing spin, but short duration}

\begin{figure*}
\includegraphics[width=\columnwidth]{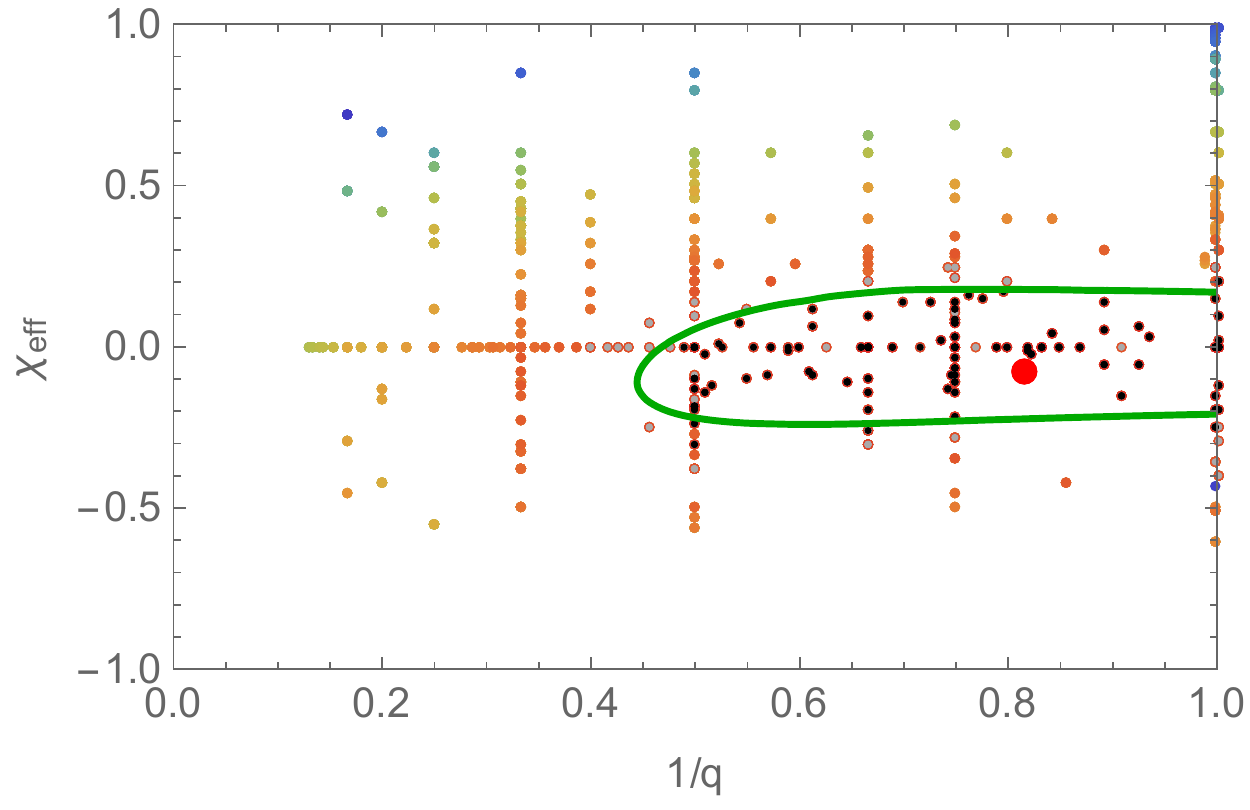}
\includegraphics[width=\columnwidth]{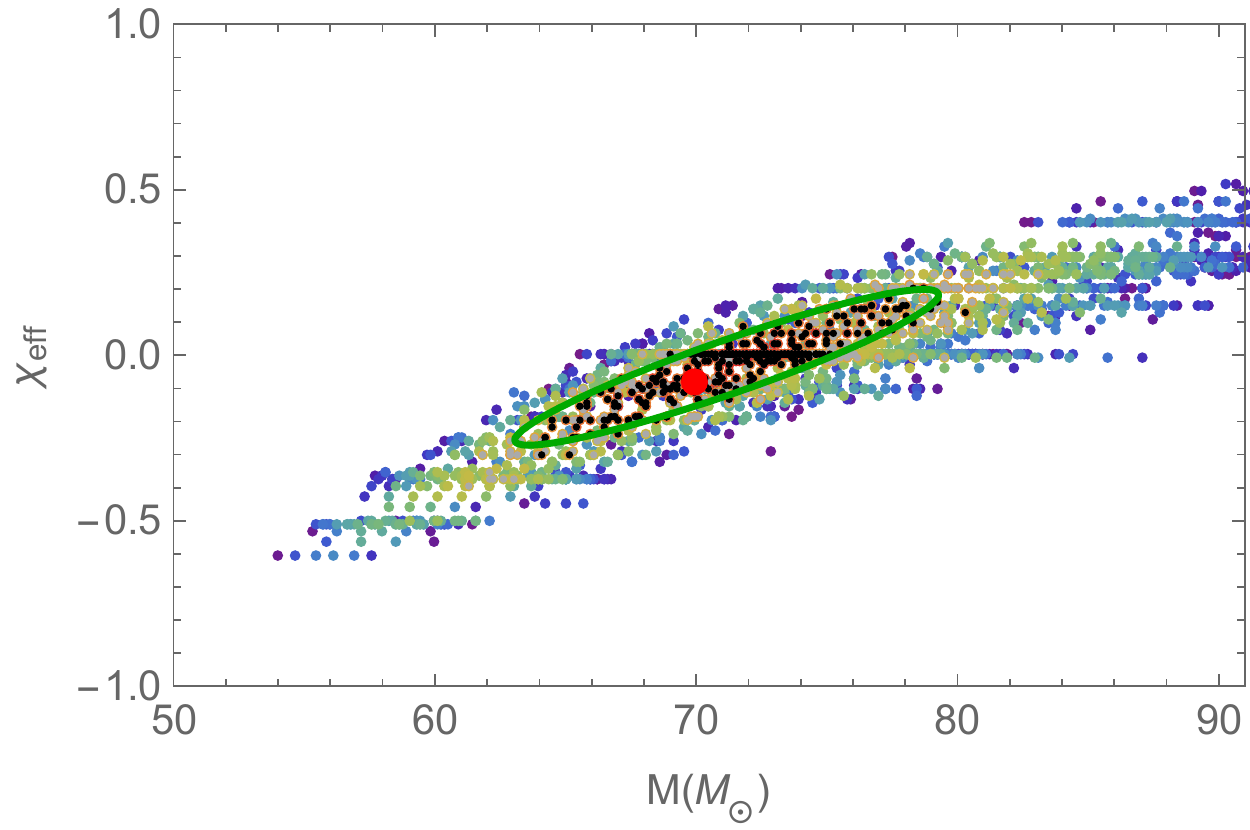}
\caption{\label{fig:0234v2BasicResults-I}\textbf{Parameter recovery for an precessing, short, unequal mass binary I}: Each point represents a NR simulation and a particular total mass compared against a SXS-0234v2 source with $l\le2$ modes.  The left panel shows the $\chi_{\rm eff}$ vs 1/q with q=$m_{1}/m_{2}$ and $\chi_{\rm eff}$ defined in Eq. (\ref{eq:chieff}), and the right panel shows the $\chi_{\rm eff}$ vs $M$.  The gray points represent points that fall between $\lnLmarg=165$ and $\lnLmarg=163$.  The black points represent points that fall in $\lnLmarg>165$, i.e. templates that best match the source.  The rest of the colors represent all the points $\lnLmarg<163$ with the red represent the highest in the region.  The green contour is the 90\% CI derived using the quadratic fit to $\lnLmarg$ for nonprecessing systems only.  The big red dot represents the true parameters of the source.  We are able to recover the 2D posterior distribution that is consistent with the distributions with $\lnLmarg>165$ (black points).}
\end{figure*}

\begin{figure*}
\includegraphics[width=\columnwidth]{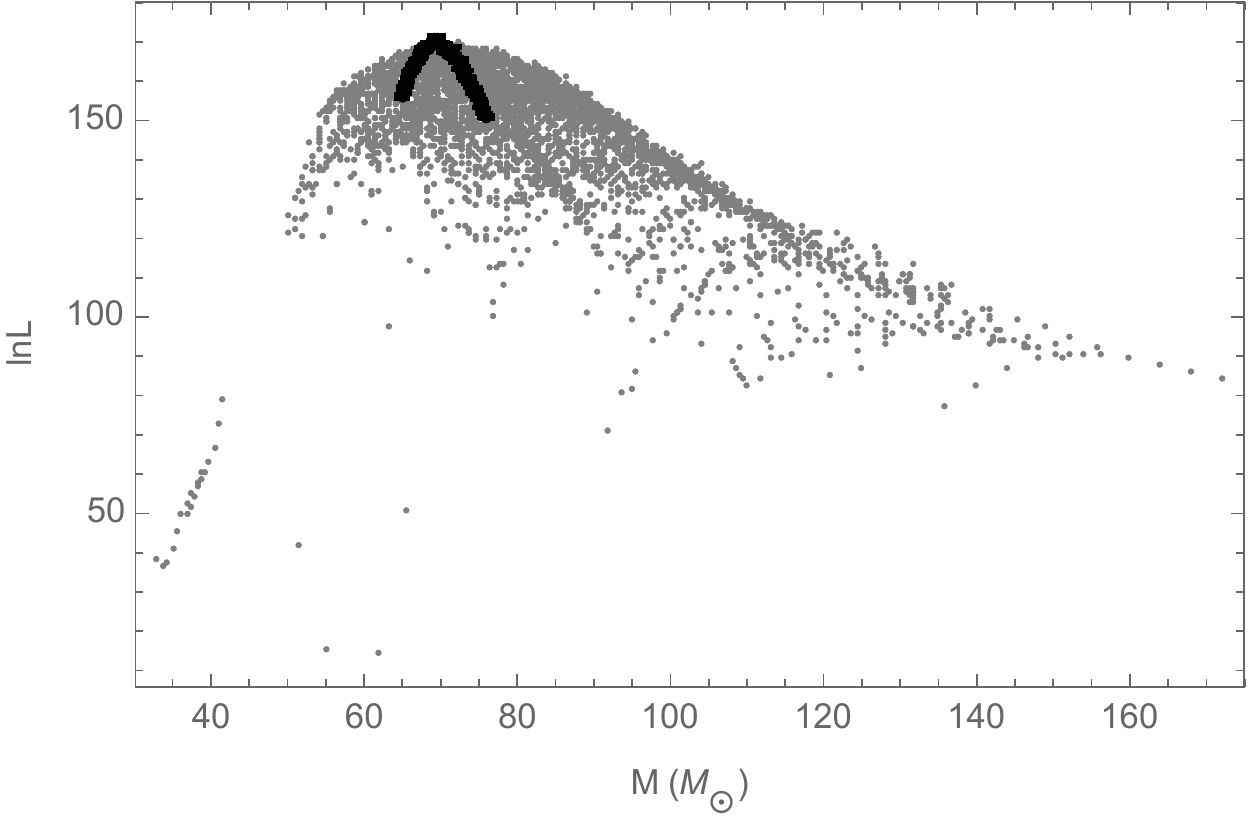}
\includegraphics[width=\columnwidth]{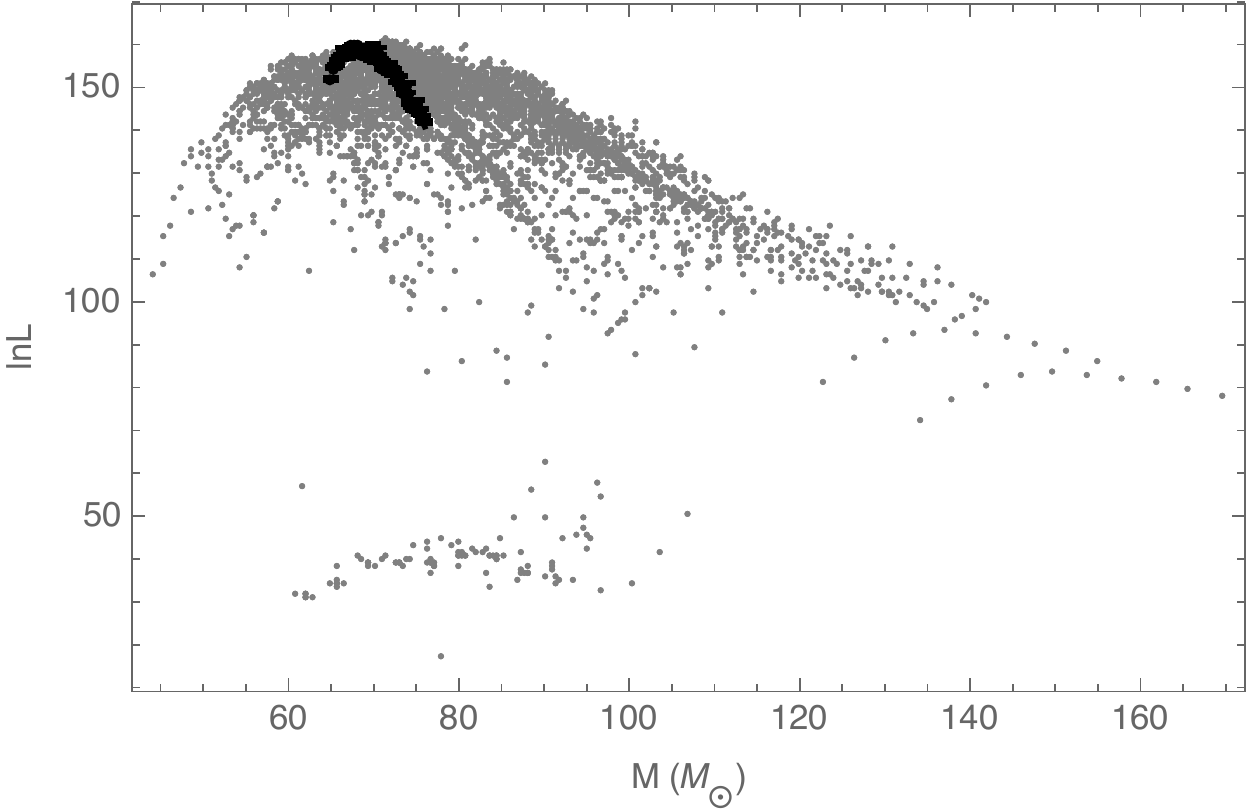}
\caption{\label{fig:0234v2Proof}\textbf{Proof of parameter recovery for an precessing, short, unequal mass binary}: Here is $\lnLmarg(M)$ of a single \textit{ILE} null run comparing SXS-0234v2 with itself (black) and the $\lnLmarg(M)$ for the full end-to-end run with SXS-0234v2 as its source (gray).  The left panel represent runs with a source with $\imath=0.0$, and the right panel represent runs with a source with $\imath=1.5$.  The gray points only include the nonprecessing templates.  If we take the difference between the $\lnLmarg$ from the whole end-to-end run and the $\lnLmarg$ from the null run, we get a $\Delta\ln L\sim0.97$ for $\imath=0.0$ and $\Delta\ln L\sim1.8$ for $\imath=1.5$.  Even if we were to include the best template in our end-to-end runs (which is itself), we only get a slight increase in the $\lnLmarg$ for the face-on inclination.  However, the edge-on case change seems significant; see Figure \ref{fig:ILE-all} for an investigation focusing on the peak values.}
\end{figure*}

Since all the fits in this study have only used the nonprecessing binaries, one might come to the conclusion that this
limits us to analyzing only zero spin and aligned source.  We can potentially recover parameters of precessing sources
if the duration of these sources are short enough; this translates to only a few cycles and therefore little to no precession before merger, see before Eq. (9) in \cite{PEPaper}.  Figure
\ref{fig:0234v2BasicResults-I}  are the same type of plots as in Figure \ref{fig:ZeroSpinBasicResults-I}. Here the gray
points represent points that fall between $\lnLmarg=165$ and $\lnLmarg=163$, and the black points represent points that
fall in $\lnLmarg>165$.  The colored points represent the points that fall in the region $\lnLmarg<163$ with the red points represent the higher $\lnLmarg$ values.  As with the previous cases, these
intervals were determined using the inverse $\chi^{2}$ distribution [see Eq. (\ref{eq:chi})] adopting
$d=4$ (two masses with aligned spin) for the black points and $d=8$ (two masses with precessing spins) for the gray
points.  As we expected, the short duration of this source allows us to recover the parameters with a fit that only uses
the nonprecessing cases as shown in the left panel of Figure \ref{fig:0234v2Proof}. Here we plot the $\lnLmarg(M)$ of a
single null run of \textit{ILE} comparing SXS-0234v2 with itself (black) and  the whole end-to-end $\lnLmarg(M)$ using
SXS-0234v2 as the source.  By construction, the $\lnLmarg$ from the null run of SXS-0234v2 is the highest $\lnLmarg(M)$
possible. If the maximum $\lnLmarg$ from the whole end-to-end run is close ($\Delta\ln L\le1$), we can recover the
parameters of the simulations without fitting with the precessing systems. In this case, the $\Delta\ln L=0.97$.  We can
therefore accurately recover the parameters of this precessing  system as evident by Figure
\ref{fig:0234v2BasicResults-I}.\footnote{When interpreting the above statement, however, it is important to note our analysis by construction uses only
  information $f>30 \unit{Hz}$.  If we had access to a wider range of long simulations, we could have access to
  information from precession cycles between $10-30\unit{Hz}$, even for sources of this kind and in this data.  More
  work is needed to assess the prospects for recovery for longer, more generic sources.}

\begin{figure*}
\includegraphics[width=\columnwidth]{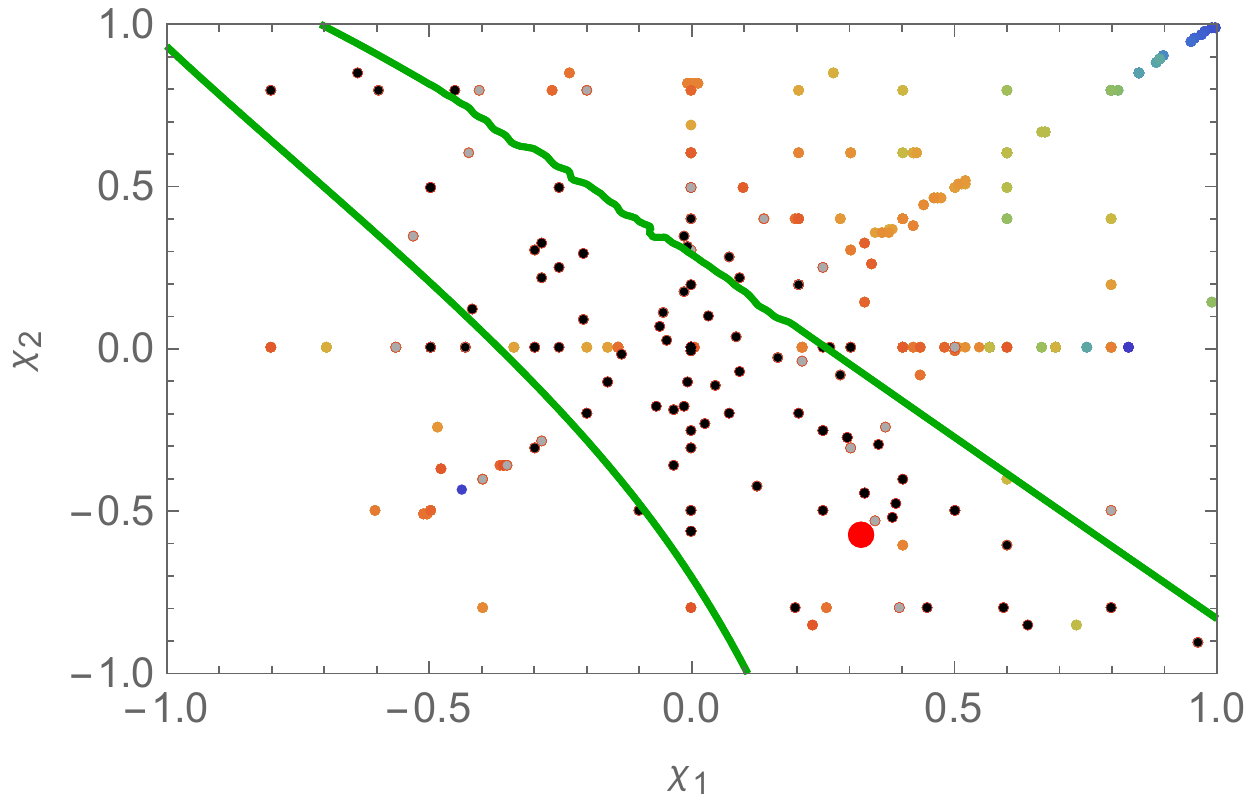}
\includegraphics[width=\columnwidth]{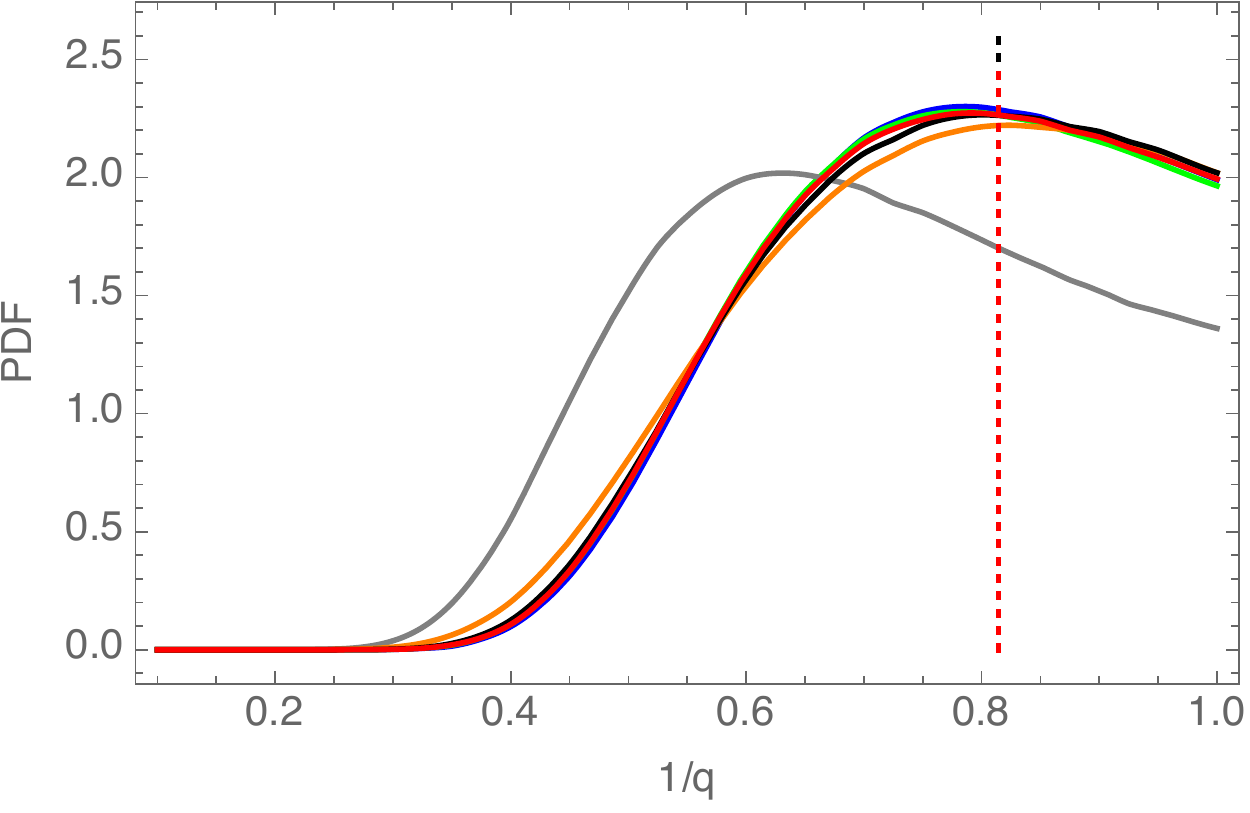}
\caption{\label{fig:0234v2BasicResults-II}\small\textbf{Parameter recovery for an precessing, short, unequal mass binary II}: The left panel shows the $\lnLmarg$ as a function of $\chi_{1z}$ and $\chi_{2z}$.  The gray, black, and other color points represent the same intervals as in Figure \ref{fig:0234v2BasicResults-I}.  The green contour represents the same contour as in Figure \ref{fig:0234v2BasicResults-I}.  The big red dot represents the true parameters of the source.  The green contour is consistent with the black point distribution.  The right panel shows the 1D posterior distribution for 1/q.  This 1D posterior was derived from the quadratic fit of to $\lnLmarg$ for nonprecessing systems only.  Here we show results for the same 6 inclinations all represented by the same colors as the zero spin case, see Figure \ref{fig:ZeroSpinBasicResults-II}.  In this case, we see significant differences between the curves implying that higher order modes could important for accurate analysis of this source.  He also see a large discrepancies between the $\imath=1.5$ distribution and the other inclinations.  See Figure \ref{fig:0234v2Proof} and Figure \ref{fig:ILE-all} for further analyses.}
\end{figure*}
\begin{figure}
\includegraphics[width=\columnwidth]{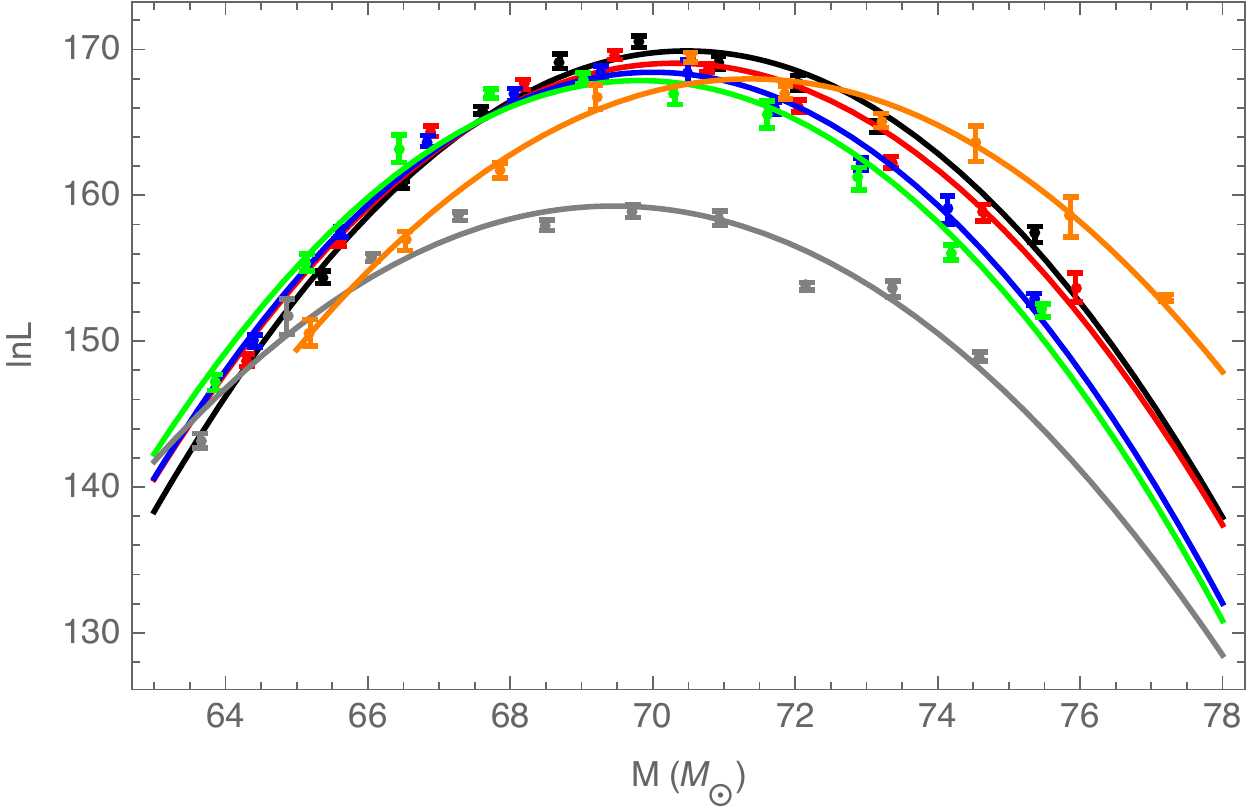}
\caption{\label{fig:ILE-all}\textbf{Discrepancy in $\lnLmarg(M)$ for $\imath=1.5$}: This is a plot of multiple $\lnLmarg(M)$ comparing SXS-0234v2 with itself at different inclinations.  Here $\imath=0.0$ is black, $\imath=0.5$ is red, $\imath=0.785$ is blue, $\imath=1.0$ is green, $\imath=1.5$ is gray, and $\imath=2.35$ is orange.  The edge-on case is clearly different than the rest of inclinations; more needs to be done to discover the origin of this discrepancy; however, this could be due to many significant higher modes.}
\end{figure}

We again show $\lnLmarg$ as a function of $\chi_{1z}$ and $\chi_{2z}$ in the left panel of Figure \ref{fig:0234v2BasicResults-II} with all the
colors and contours representing the as in Figure \ref{fig:ZeroSpinBasicResults-II}.  The green contour are consistent with the black point distribution.  We again plot the 1D distribution for 1/q for different inclinations in the right panel of Figure \ref{fig:0234v2BasicResults-II} with all the colors
corresponding to the same inclinations as in the right panel of Figure \ref{fig:ZeroSpinBasicResults-II}.  Here we see relative
consistency between the different inclinations, with a consistent trend towards extracting marginally more information
as the inclination increases. We have an outlier for $\imath=1.5$: a nearly edge-on line of sight.  For such a
line of sight, keeping in mind we tune the source distance to fix the network SNR, precession-induced modulations are
amplified; this outlier \emph{could} and probably does represent the impact of precession. To investigate this further, we again plot
$\lnLmarg(M)$ of a single null run of \textit{ILE} comparing SXS-0234v2 with itself (black) and  the whole end-to-end
$\lnLmarg(M)$ using SXS-0234v2 with $\imath=1.5$ as the source, see the right panel of Figure \ref{fig:0234v2Proof}.
By construction, the $\lnLmarg$ from the null run of SXS-0234v2 is the highest $\lnLmarg(M)$ possible.  Here we find a
bigger difference between $\lnLmarg$ of the null run and $\lnLmarg$ of the entire end-to-end run: $\Delta\ln L\sim1.8$.
We then take all the individual runs from the end-to-end runs that compared 0234v2 to itself and plot $\lnLmarg(M)$ for
each inclination.  As evident in Figure \ref{fig:ILE-all}, the $\imath=1.5$ curve lies well below the rest of the
inclinations.  More investigations are needed to be done to figure out this discrepancy; however, this could imply SXS-0234v2 has many modes that are relevant, reflecting precession-induced modulation most apparent perpendicular to $\bar{J}$ the total angular momentum vector.  In future work, where we attempt to recover all spin degrees of freedom for precessing sources, we will focus in particular on edge-on lines of sight like this.

\section{Conclusions}
\label{sec:conclusion}
We have presented and assessed  a method to directly interpret real gravitational wave data by comparison to numerical
solutions of Einstein's equations.  This method can employ existing harmonics and physics that has been or can be
modeled.  While any other method can do so as well if suitable models have been developed and calibrated, this method skips the step of
translating NR results into model improvements, circumventing the effort and potential biases introduced in doing so. %

We also provided a detailed systematic study of the potential errors introduced in our method.  
We first used the overlap or mismatch to assess the difference between different simulations along fiducial lines of sight.  As noted
in Eq. (\ref{eq:snr-lnL}), we expect that  $\lnL$ is approximately proportional to the mismatch by an
overall constant.  We demonstrate this relationship explicitly, using NR sources and synthetic data.
  Once we obtained $\lnLmarg$, we fitted with a simple quadratic and derived a PDF using Eq. (\ref{eq:1d}) with its corresponding 90\% CI. Using the PDFs, we can graphically see any errors that would have been propagated through.  To quantify this change, we calculated a KL Divergence between two PDFs [see Eq. (\ref{eq:dkl})].  By using these diagnostics,  we addressed and quantified systematic errors that could affect our parameter estimation results.

Our validation studies systematically assessed the impact of (a)  Monte Carlo error, (b) waveform extraction error, (c) simulation
resolution, and (d) low frequency cutoff/signal duration via our diagnostics.  
\begin{itemize}
\item (a) Based on our results from our examples, we
were confident that the error from our Monte Carlo integration would be small.  To quantify the results that seem
apparent by eye,  we applied our diagnostics (omitting the mismatch) and found the $D_{KL}$ between the PDFs (i.e. $D_{KL}$(v1,v1), $D_{KL}$(v1,v2), $D_{KL}$(v1,v3)) to be all $D_{KL}\sim10^{-5}$. 
\item (b)  In a similar fashion, we applied our diagnostics to GW150914-like
simulations from the SXS and RIT NR groups.  We validated the utility of the perturbative extraction technique but noted
some differences between the strain provided by SXS and perturbative extraction applied to their $\psi_4$
  data.  Based on excellent agreement between RIT (with perturbative extraction) and SXS provided strain, we expect the
  discrepancies relate to improper assumptions regarding SXS coordinates.  More needs to be done to discover the origin
of this disparity.   From our match study, we determined that the impact of the error due to waveform extraction is insignificant at a large
 enough extraction radius.  This was validated via the $D_{KL}$ between three PDFs with the highest possible extraction
 radii, which were all around $10^{-2}-10^{-3}$.
\item (c) 
When using our mismatch study to assess the impact of resolution
 error, it was determined that the mismatch for all the different resolution was $\mathcal{M}\sim10^{-5}$.  This
 seemingly small difference in the waveform was then reaffirmed by the corresponding $D_{KL}\sim10^{-4}-10^{-5}$.  From
 our diagnostics, it was clear that the error introduced by numerical resolution was negligible.  
\item (d) We finally used our diagnostics to the assess impact of low frequency cutoffs and signal duration.  For both NR and
  analytic models, the available frequency content provided can significantly affect our results.  After deriving our
  PDFs and calculating the $D_{KL}$, we found the lower $f_{\rm min} (10,20 \unit{Hz})$ were very similar with a narrow PDF
  and a high peak while the higher $f_{\rm min} (30,40 \unit{Hz})$ produced a wider PDF with a lower peak.  We stress the
  importance of the hybridization of the NR waveforms to allow for a low $f_{\rm min}$ to standardization NR waveforms while
  providing the longest waveform possible.
\end{itemize}

We also provided three end-to-end examples with three different types of sources.  First, we used a simple example -- zero spin equal mass, where no significant higher order modes complicate our interpretation --  to show
our method works.  Second, we examined an aligned, GW150914-like, unequal mass source.  Though the
  leading-order quadruple radiation from such a source is nearly degenerate with an equal mass, zero spin system, this
  binary has asymmetries which produce higher order modes.  We used our method with the $l\le2$ as well as the $l\le 3$ modes
and found we could better constrain q using higher modes.  We also found significant differences between the 1D
probability distributions for 1/q; this implied that higher modes were significant.  Third, we used our method on a
precessing but short unequal mass source.  Due to its short duration of the observationally accessible signal,
this comparable-mass binary has little to no time to precess in band.  This allows us to recover the parameters of the binary
even though we construct a fit based on the nonprecessing binaries.  Even though the recovery of parameters was
possible, the edge-on case for our 1D distributions were significantly different than the rest.  For this line
  of sight, precession-induced modulations are most significant; the simplifying approximation that allowed success for
  the other lines of sight break down.  Even though we suspect this is also due to higher order modes, more needs to be
done to validate this claim.  In the future, we will extend this  strategy to recover parameters of generic precessing sources.
The method presented here relies on interpolation between existing simulations of quasi-circular black hole binary
mergers.  For nonprecessing binaries, this three-dimensional space has been reasonably well-explored.   For generic
quasi-circular mergers, however, substantially more simulations may be required to fill the seven-dimensional parameter
space sufficiently for this method.   Fortunately,  targeted followup numerical simulations of heavy binary black holes are
 always possible.  These simulations will be incredibly valuable to validate any inferences about binary black hole
 mergers, from this or any other method.    For this method in particular, followup simulations can be used to directly
 assess our estimates, and revise them.  We will outline followup strategies and iterative fitting procedures
 in subsequent work.

\begin{acknowledgements}
The RIT authors gratefully acknowledge the NSF for financial support from Grants: No. PHY-1505629, No. AST-1664362
No. PHY-1607520, No. ACI-1550436, No. AST-1516150, and No. ACI-1516125.
Computational resources were provided by XSEDE allocation
TG-PHY060027N, and by NewHorizons and BlueSky Clusters 
at Rochester Institute of Technology, which were supported
by NSF grant No. PHY-0722703, DMS-0820923, AST-1028087, and PHY-1229173.
This research was also part of the Blue Waters sustained-petascale computing
NSF projects ACI-0832606, ACI-1238993, and OCI-1515969, OCI-0725070. 

The SXS collaboration authors gratefully acknowledge the NSF for financial support from Grants: No. PHY-1307489, No. PHY-1606522, PHY-1606654, and AST- 1333129. They also  gratefully acknowledge support for this research at CITA from NSERC 
of Canada, the Ontario Early Researcher Awards Program, the Canada
Research Chairs Program, and the Canadian Institute for Advanced 
Research.  Calculations were done on the ORCA computer cluster, supported by NSF grant PHY-1429873, the Research Corporation for Science Advancement, CSU Fullerton, the GPC supercomputer at the SciNet HPC Consortium~\cite{scinet}; SciNet is funded by: the Canada Foundation for Innovation (CFI) under the
auspices of Compute Canada; the Government of Ontario; Ontario Research Fund (ORF) -- Research Excellence; and the University of Toronto. Further calculations were performed on the Briar\'ee cluster at Sherbrooke University, managed by Calcul Qu\'ebec and Compute Canada and with operation funded by the Canada Foundation for Innovation (CFI), Minist\'ere de l'\'Economie, de l'Innovation et des Exportations du Quebec (MEIE), RMGA and the Fonds de recherche du Qu\'ebec - Nature et Technologies (FRQ-NT).

The GT authors gratefully acknowledge the NSF for financial support from Grants: No. ACI-1550461 and No. PHY-1505824. Computational resources were provided by XSEDE and the Georgia Tech Cygnus Cluster.

Finally, the authors are grateful for computational resources used for the parameter estimation runs provided by the Leonard E Parker
Center for Gravitation, Cosmology and Astrophysics at the University of
Wisconsin-Milwaukee; the Albert Einstein Institute at Hanover, Germany; and the California Institute of Technology at Pasadena, California.
\end{acknowledgements}

\bibliography{references,paperexport,references2}
\bibliographystyle{unsrt}
\appendix

\begin{figure*}
\includegraphics[width=\columnwidth]{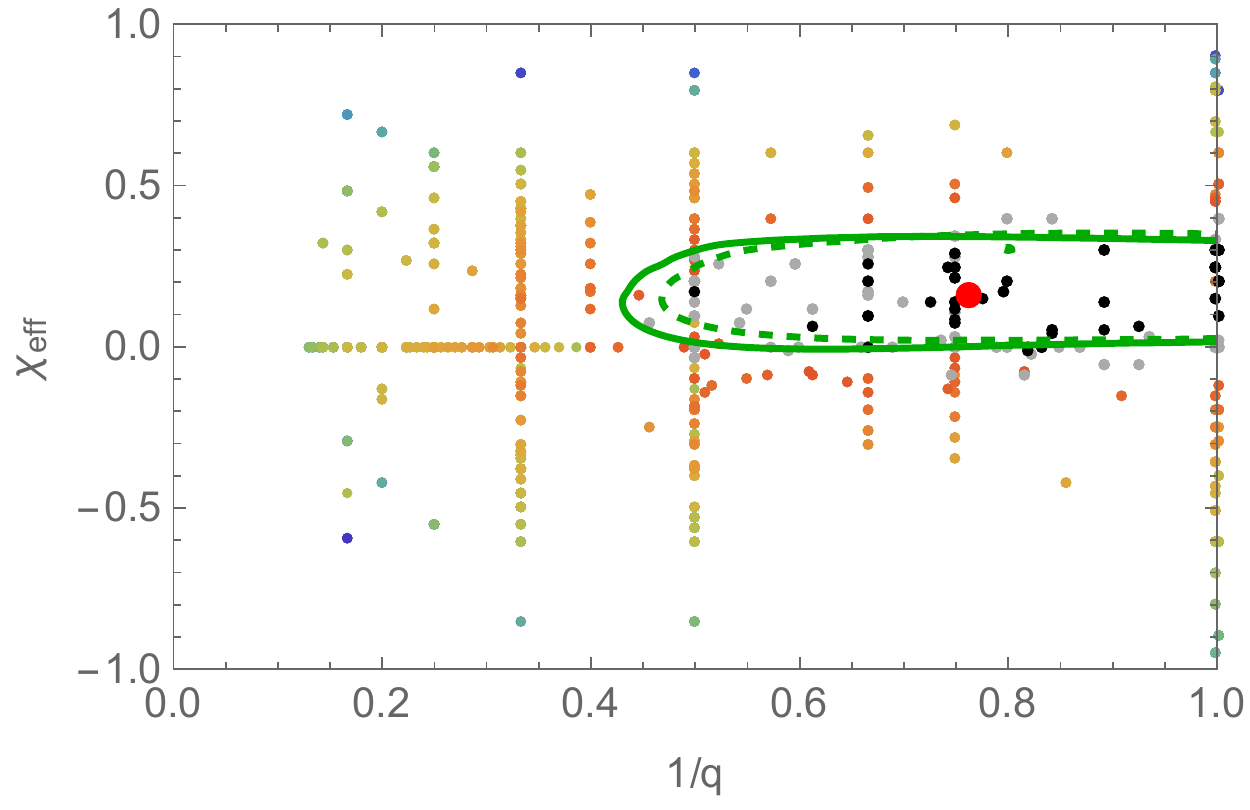}
\includegraphics[width=\columnwidth]{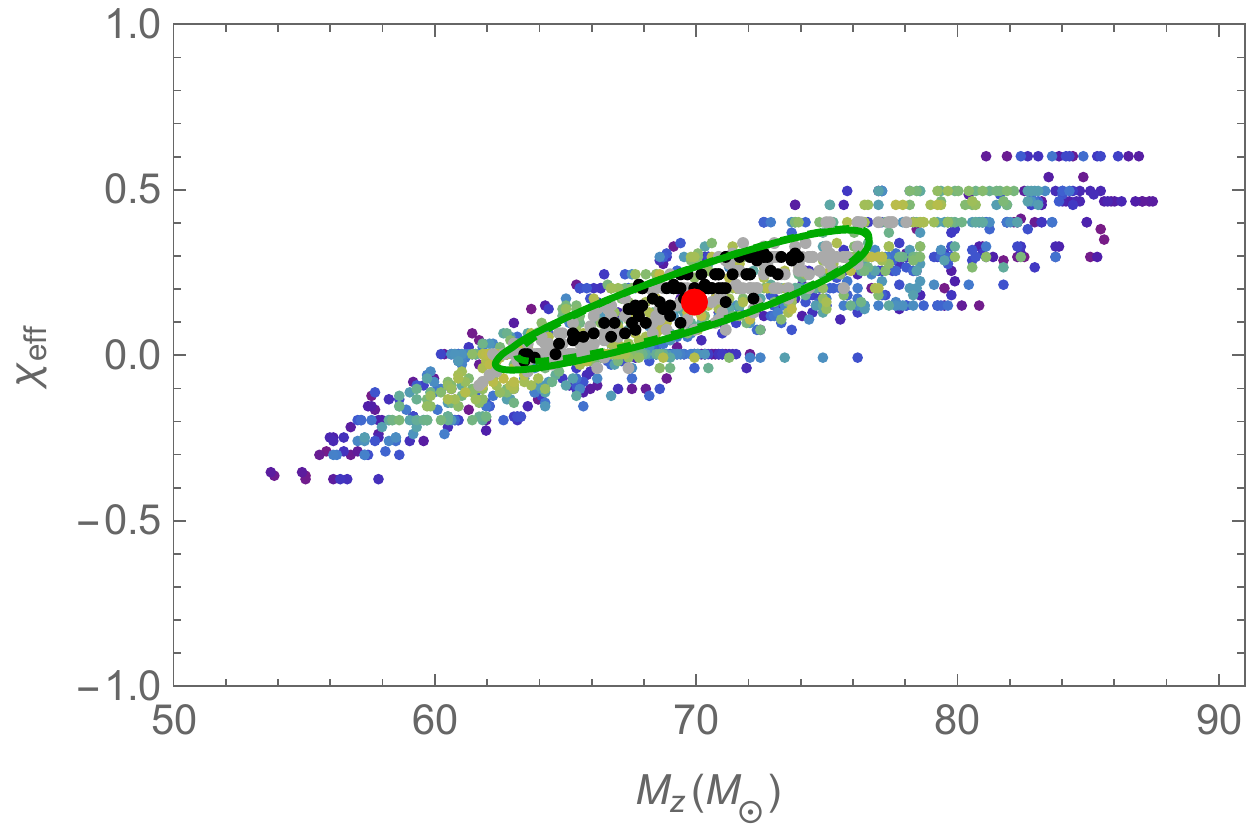}
\includegraphics[width=\columnwidth]{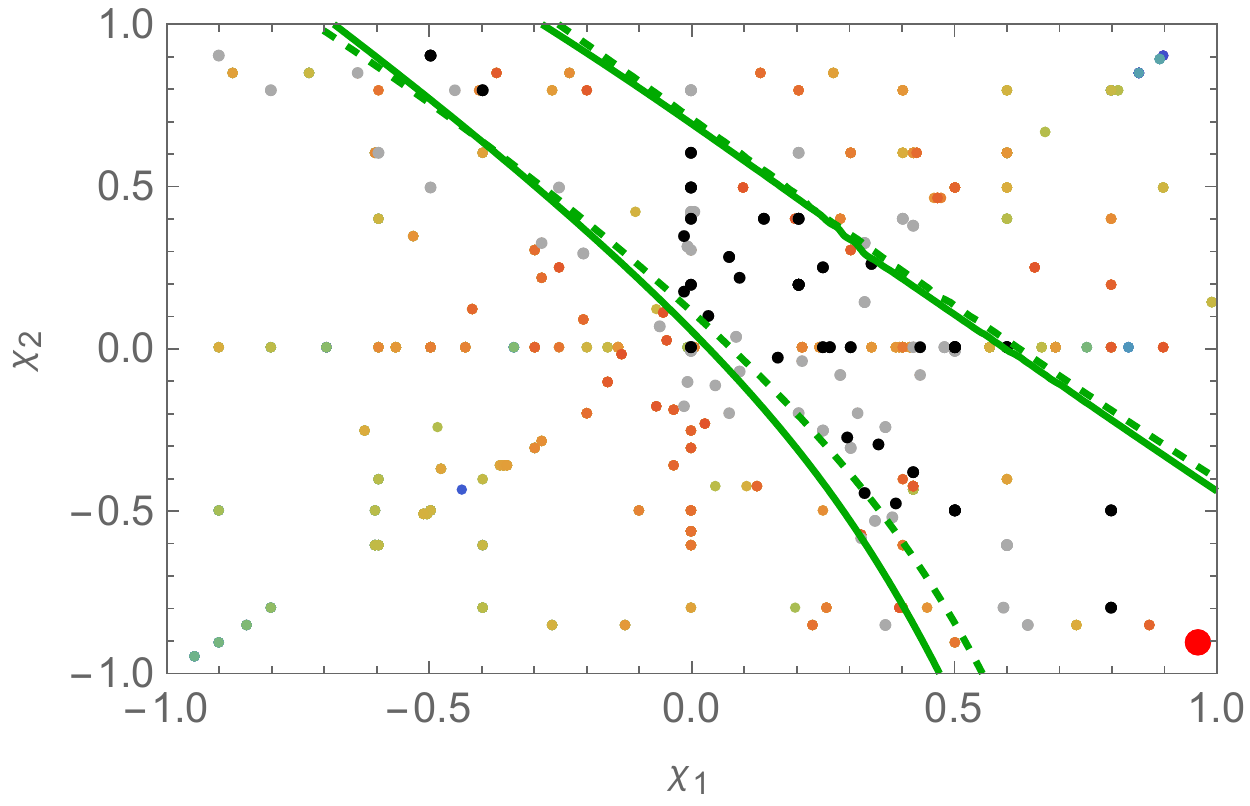}
\caption{\label{fig:SXS-0.9}\textbf{Parameter recovery for a high, anti-aligned spin $q=1.31$ binary}: Each point
  represents a NR simulation and a particular total mass compared against a SXS-high-antispin source.  The top left panel shows the $\chi_{\rm eff}$ vs 1/q with
  q=$m_{1}/m_{2}$ and $\chi_{\rm eff}$ defined in Eq. (\ref{eq:chieff}), the top right panel shows the $\chi_{\rm eff}$
  vs $M$, and the bottom panel shows the $\chi_1$ vs $\chi_2$.  The gray points represent points that fall between
  $\lnLmarg=167$ and $\lnLmarg=164$.  The black points represent points that fall in $\lnLmarg>167$, i.e. templates that
  best match the source.  The rest of the colors represent all the points $\lnLmarg<164$ with the red represent the
  highest in the region.  The green contour is the 90\% CI derived using the quadratic fit to
  $\lnLmarg$ for nonprecessing systems only.  The dash line is the CI for $l\le 3$, and the solid line is the CI for $l\le2$.  The big red dot represents the true parameters of the source.  We are able to recover the 2D posterior distribution that is consistent with the distributions with $\lnLmarg>167$ (black points).}
\end{figure*}

\begin{figure*}
\includegraphics[width=\columnwidth]{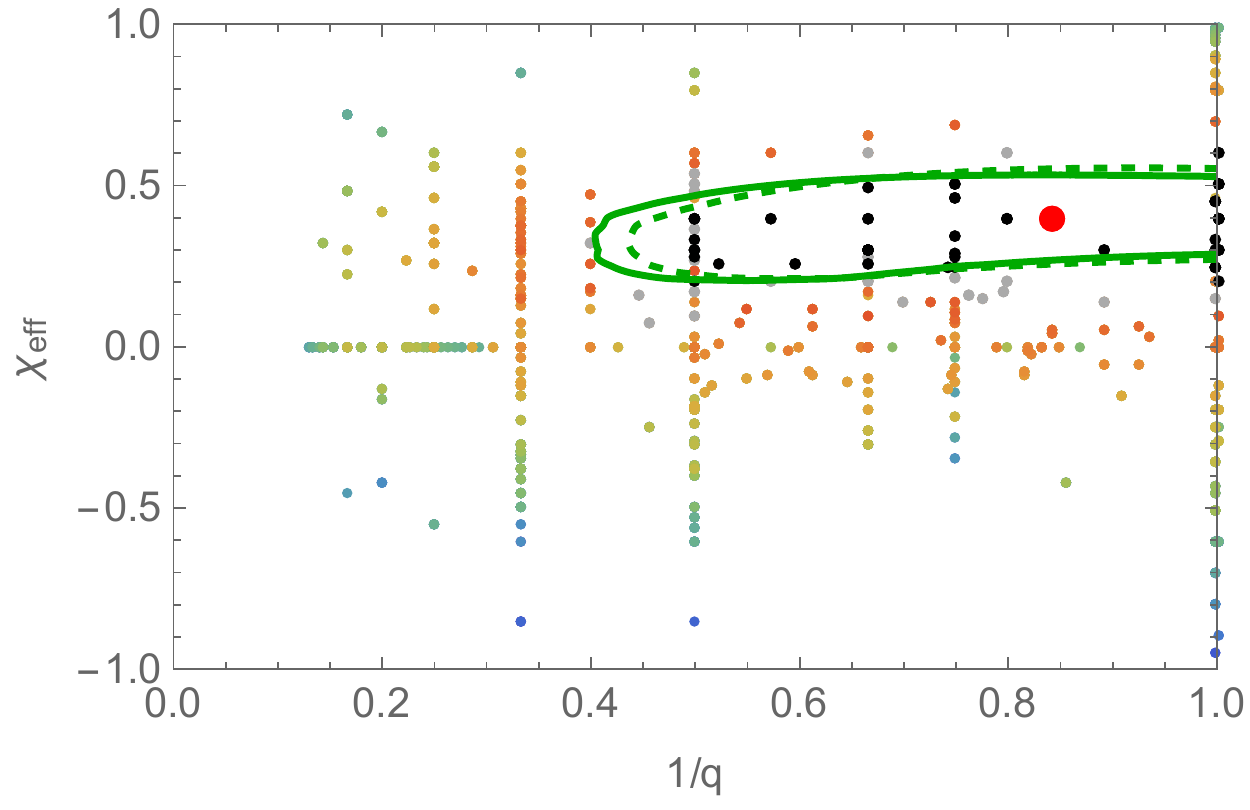}
\includegraphics[width=\columnwidth]{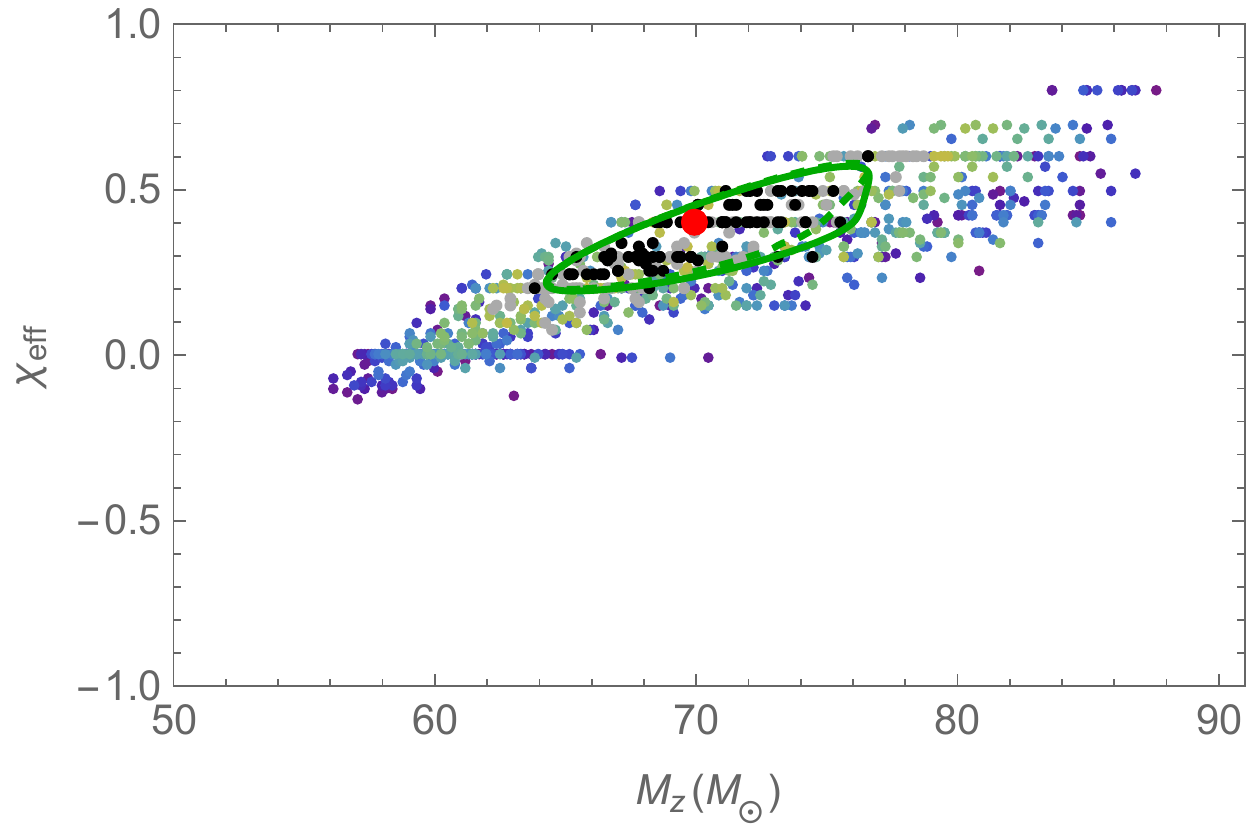}
\includegraphics[width=\columnwidth]{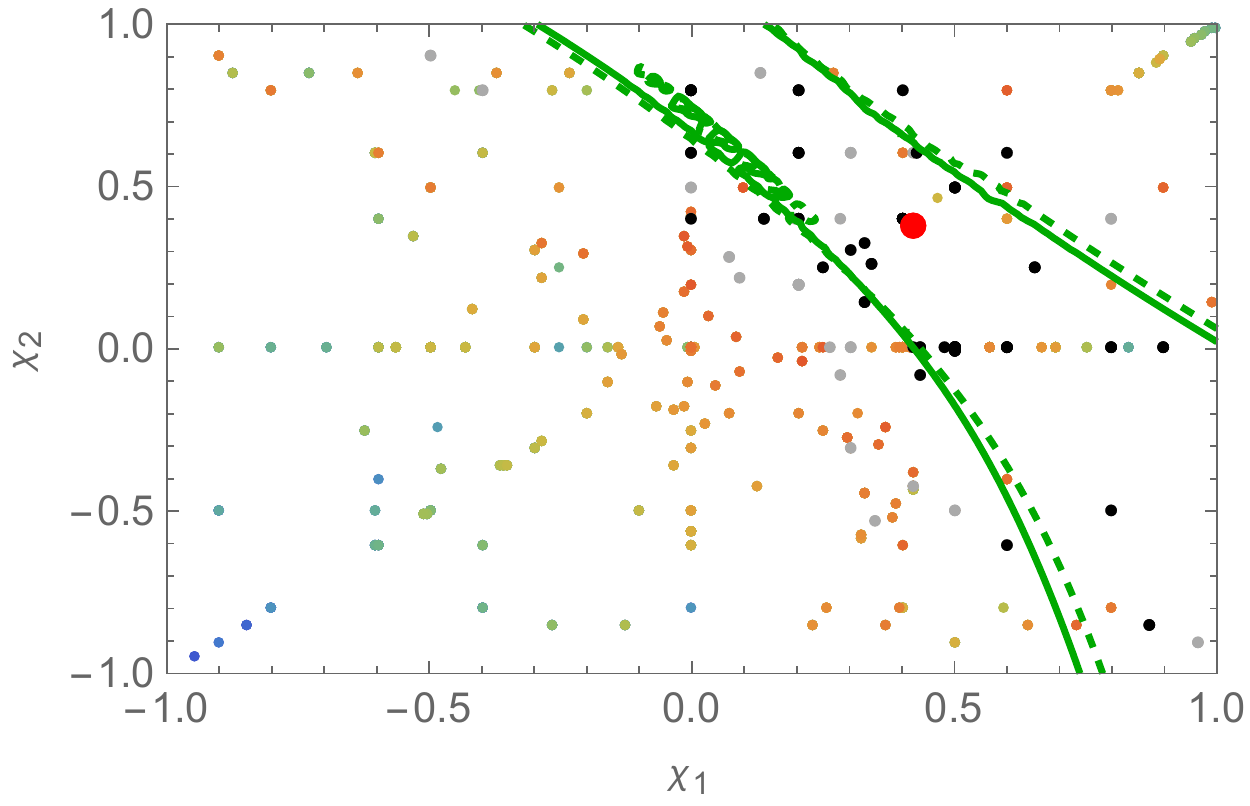}
\caption{\label{fig:SXS-chi_eff0.4}\textbf{Parameter recovery for a $\chi_{\rm eff}=0.4$ spin $q=1.19$ binary}: Each
  point represents a NR simulation and a particular total mass compared against a SXS-$\chi_{\rm eff}$0.4 source.  The top left panel shows the $\chi_{\rm eff}$ vs 1/q with q=$m_{1}/m_{2}$ and $\chi_{\rm eff}$ defined in Eq. (\ref{eq:chieff}), the top right panel shows the $\chi_{\rm
    eff}$ vs $M$, and the bottom panel shows the $\chi_1$ vs $\chi_2$.  The gray points represent points that fall
  between $\lnLmarg=167$ and $\lnLmarg=164$.  The black points represent points that fall in $\lnLmarg>167$,
  i.e. templates that best match the source.  The rest of the colors represent all the points $\lnLmarg<164$ with the
  red represent the highest in the region.  The green contour is the 90\% CI derived using the
  quadratic fit to $\lnLmarg$ for nonprecessing systems only.  The dash line is the CI for $l\le 3$, and the solid line is the CI for $l\le2$.  The big red dot represents the true parameters of the source.  We are able to recover the 2D posterior distribution that is consistent with the distributions with $\lnLmarg>167$ (black points).}
\end{figure*}

\begin{figure*}
\includegraphics[width=\columnwidth]{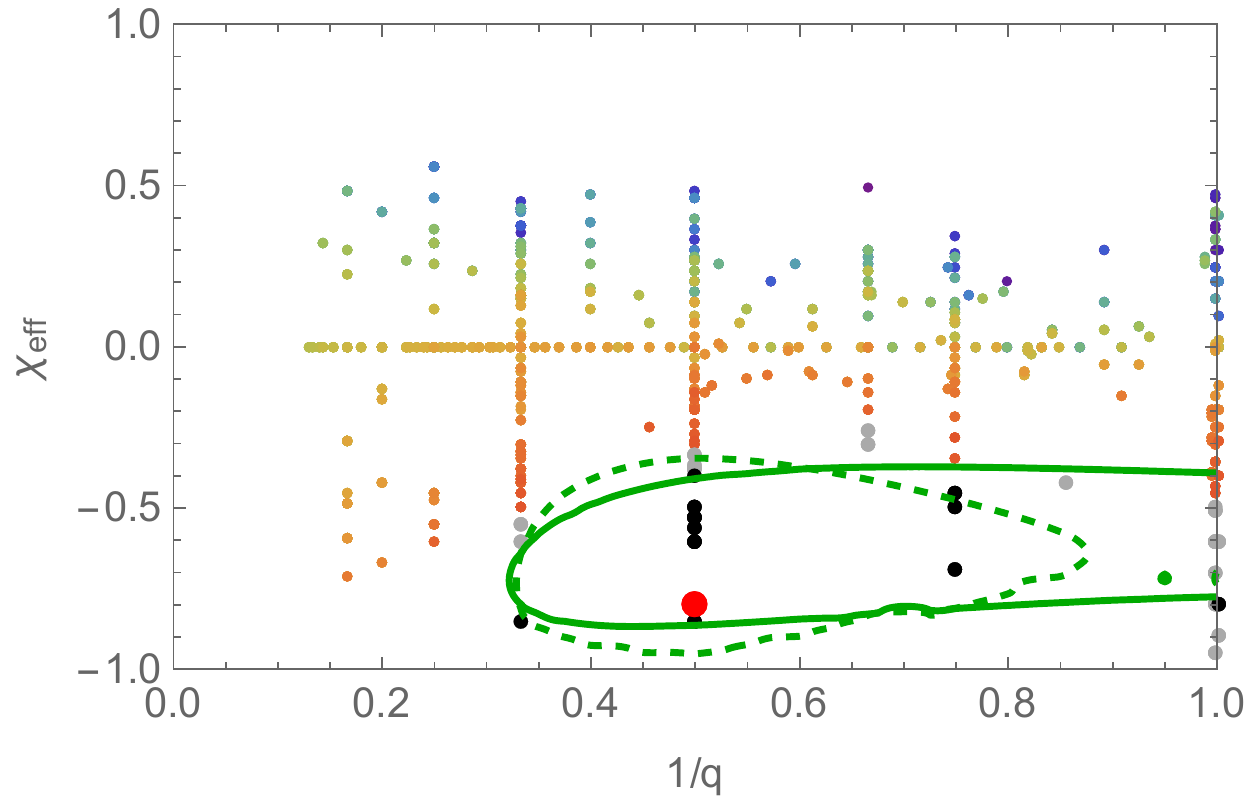}
\includegraphics[width=\columnwidth]{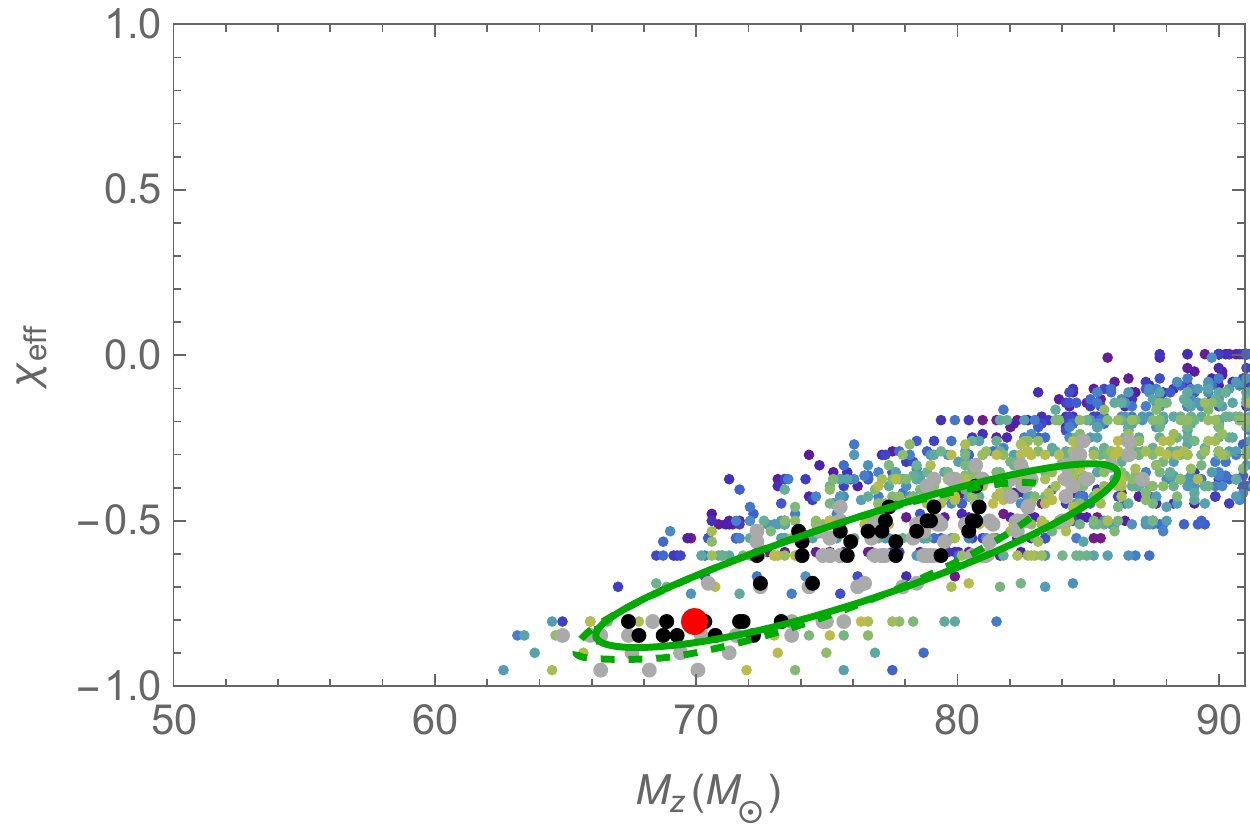}
\includegraphics[width=\columnwidth]{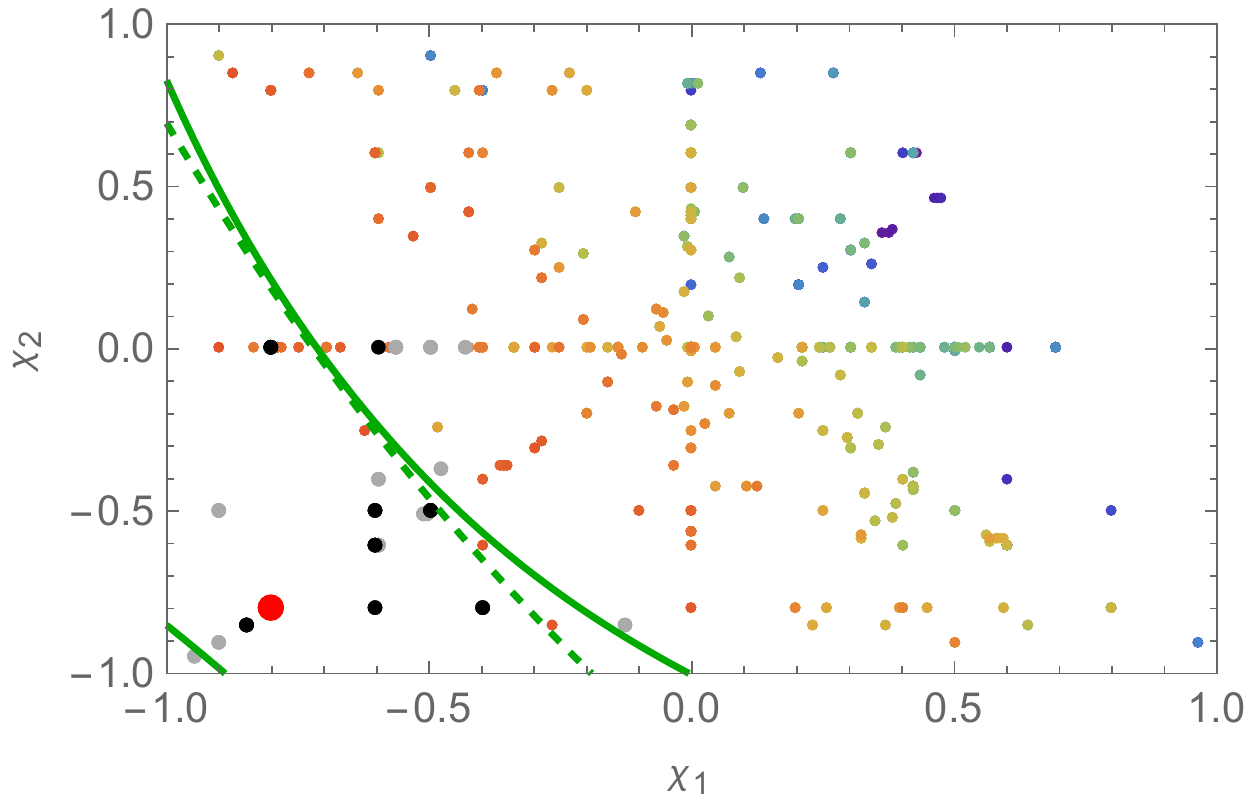}
\caption{\label{fig:RIT-q2-a0.8}\textbf{Parameter recovery for a $\chi_1=\chi_2=-0.8$ spin $q=2.0$ binary}: Each point
  represents a NR simulation and a particular total mass compared against a RIT-2 source.  The top left panel shows the $\chi_{\rm eff}$ vs 1/q with
  q=$m_{1}/m_{2}$ and $\chi_{\rm eff}$ defined in Eq. (\ref{eq:chieff}), the top right panel shows the $\chi_{\rm eff}$
  vs $M$, and the bottom panel shows the $\chi_1$ vs $\chi_2$.  The gray points represent points that fall between
  $\lnLmarg=165$ and $\lnLmarg=162$.  The black points represent points that fall in $\lnLmarg>165$, i.e. templates that
  best match the source.  The rest of the colors represent all the points $\lnLmarg<162$ with the red represent the
  highest in the region.  The green contour is the 90\% CI derived using the quadratic fit to
  $\lnLmarg$ for nonprecessing systems only.  The dash line is the CI for $l\le 3$, and the solid line is the CI for $l\le2$.  The big red dot represents the true parameters of the source.  We are able to recover the 2D posterior distribution that is consistent with the distributions with $\lnLmarg>165$ (black points).}
\end{figure*}

\clearpage
\begin{widetext}
\section{Exploring the parameter space}
\label{sec:appendixA}
In this appendix, we provide additional examples of our method using numerical relativity simulations in different
regions of parameter space.  We demonstrate our method works reliably for extreme black hole spins (Figure
\ref{fig:SXS-0.9}) as well as  in regions where few  simulations with comparable parameters are available  (Figures
\ref{fig:SXS-chi_eff0.4} and \ref{fig:RIT-q2-a0.8}). For the parameters of each source, see the following source labels (in order as they appear) in Table \ref{tab:simulations}: RIT-5, SXS-high-antispin, SXS-$\chi_{\rm eff}$0.4, and RIT-2.
\end{widetext}

\label{sec:appendixA}

\begin{figure*}
\includegraphics[width=\columnwidth]{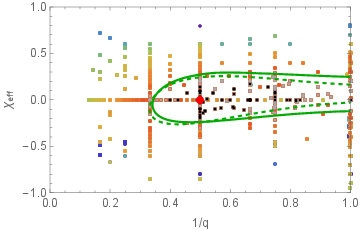}
\includegraphics[width=\columnwidth]{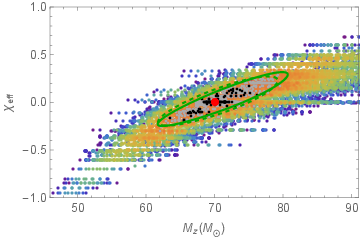}
\includegraphics[width=\columnwidth]{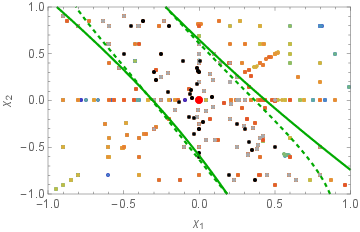}
\caption{\label{fig:q2-a0}\textbf{Parameter recovery for a zero spin, $q=2$ binary}: Each point represents a NR
  simulation and a particular total mass compared against a RIT-5 source.  The top left panel shows the $\chi_{\rm eff}$ vs 1/q with q=$m_{1}/m_{2}$ and
  $\chi_{\rm eff}$ defined in Eq. (\ref{eq:chieff}), the top right panel shows the $\chi_{\rm eff}$ vs $M$, and the
  bottom panel shows the $\chi_1$ vs $\chi_2$.  The gray points represent points that fall between $\lnLmarg=166$ and
  $\lnLmarg=164$.  The black points represent points that fall in $\lnLmarg>167$, i.e. templates that best match the
  source.  The rest of the colors represent all the points $\lnLmarg<164$ with the red represent the highest in the
  region.  The green contour is the 90\% CI derived using the quadratic fit to $\lnLmarg$ for
  nonprecessing systems only.  The dash line is the CI for $l\le 3$, and the solid line is the CI for $l\le2$.  The big red dot represents the true parameters of the source.  We are able to recover the 2D posterior distribution that is consistent with the distributions with $\lnLmarg>167$ (black points).}
\end{figure*}

\end{document}